\documentclass[a4paper,11pt]{article}
\pdfoutput=1 

\usepackage{jheppub} 

\usepackage[T1]{fontenc} 

\usepackage{hyperref}

\def\beq{\begin{equation}}
\def\eeq{\end{equation}}

\newcommand{\eref}[1]{\eqref{#1}}

\newcommand{\cref}[1]{Corollary \ref{#1}}

\newcommand{\co}{\ , \ \ \ \ \ \ }

\newcommand{\te}{\textrm}
\newcommand{\ap}{\alpha'}

\newcommand{\ee}{\textrm{e}}

\newcommand{\newt}{\mathbb T}

\newcommand{\cor}[1]{\langle #1 \rangle}
\newcommand{\scor}[1]{\langle \! \langle #1 \rangle \! \rangle}

\newcommand{\ga}{\gamma}
\newcommand{\de}{\delta}

\newcommand{\ssum}{{\cal W}}
\newcommand{\ssuma}[2]{\ssum \left[ \begin{array}{c}
		\! #1 \! \\
		#2
	\end{array}\right]}

\newcommand{\tht}[1]{\theta_{#1}}

\title{\boldmath Fermionic one-loop amplitudes of the RNS superstring}

\author[a]{Seungjin Lee}
\author[a,b]{Oliver Schlotterer}

\affiliation[a]{Max--Planck--Institut f\"ur Gravitationsphysik, Albert--Einstein--Institut,\\14476 Potsdam, Germany}
\affiliation[b]{Perimeter Institute for Theoretical Physics,\\Waterloo, ON N2L 2Y5, Canada}

\emailAdd{seungjin.lee@aei.mpg.de}
\emailAdd{olivers@aei.mpg.de}


\abstract{We investigate massless $n$-point one-loop amplitudes of the open RNS superstring with two external fermions and determine their worldsheet integrands. The contributing correlation functions involving spin-$1/2$ and spin-$3/2$ operators from the fermion vertices are evaluated to any multiplicity. Moreover, we introduce techniques to sum these correlators over the spin structures of the worldsheet fermions such as to manifest all cancellations due to spacetime supersymmetry. These spin sums require generalizations of the Riemann identities among Jacobi theta functions, and the results can be expressed in terms of doubly-periodic functions known from the mathematics literature on elliptic multiple zeta values. On the boundary of moduli space, our spin-summed correlators specialize to compact representations of fermionic one-loop integrands for ambitwistor strings.
}

\arxivnumber{1710.07353}

\begin{document} 
\maketitle
\flushbottom
\section{Introduction}\label{sec;intro}

Scattering amplitudes in field and string theories are usually significantly simpler than traditional methods of computation suggest. A classic example concerns the tremendous cancellations between bosonic and fermionic loop corrections in supersymmetric amplitudes. One-loop amplitudes of maximally supersymmetric gauge theory and gravity have been firstly determined from the point-particle limit of superstrings in 1982 \cite{Green:1982sw}, and the ease of that computation exemplifies the general virtue of string amplitudes to study field theories from a new perspective. Since then, a variety of formalisms have been developed for string amplitudes, including the manifestly supersymmetric pure-spinor description \cite{Berkovits:2000fe, Berkovits:2015yra, Berkovits:2016xnb} of the superstring and the more recent ambitwistor strings \cite{Mason:2013sva, Berkovits:2013xba, Adamo:2013tsa, Adamo:2015hoa} which directly compute $D$-dimensional field-theory amplitudes.

The traditional approach to superstring amplitudes through the Ramond--Neveu--Schwarz formalism (RNS) is based on worldsheet spinors which allow for different boundary conditions or {\it spin structures} on Riemann surfaces of genus $g\geq 1$ \cite{Ramond:1971gb, Neveu:1971rx, Neveu:1971iv}. Spacetime supersymmetry is hidden the RNS formalism since external fermions are described through spin fields \cite{Cohn1986}, and the supersymmetry cancellations within loop amplitudes originate from the interplay of different spin structures.

At one loop, the summation over spin structures is well understood for RNS amplitudes with any number of external bosons \cite{Tsuchiya:1988va, Stieberger:2002wk, Broedel:2014vla}: The supersymmetry cancellations stem from generalizations of the Riemann identities among Jacobi theta functions and neatly connect \cite{Broedel:2014vla} with the mathematics of iterated integrals on an elliptic curve \cite{Brown2011,eMZV}. In this work, we extend the simplification of spin sums to one-loop amplitudes with two external fermions and any number of external bosons, and the general strategy is expected to apply to any number of fermions. The relevant genus-one correlation functions involving spin fields will be determined in a systematic manner, and the combinations of Jacobi theta functions in their spin sums are explicitly shown to conspire to the doubly-periodic functions of \cite{  Brown2011,eMZV,Broedel:2014vla}.

The pure-spinor formalism \cite{Berkovits:2000fe, Berkovits:2005bt}, on the other hand, automatically leads to superfield representations of scattering amplitudes which simultaneously address external bosons and fermions. This framework significantly extended the computational reach for superstring amplitudes, see \cite{Mafra:2011nv} for the $n$-point amplitude at tree level and \cite{Berkovits:2004px, Berkovits:2005df, Berkovits:2005ng, Gomez:2010ad, Gomez:2013sla, Gomez:2015uha} for multiloop results. However, the composite nature of the b-ghost \cite{Berkovits:2005bt} currently poses difficulties in the {\it direct} evaluation of loop amplitudes with six and more external legs. Still, indirect methods have been successfully applied to pinpoint the complete one-loop six-point result \cite{Mafra:2016nwr}, and further developments towards higher multiplicity (and ultimately loop order) are in progress \cite{Mafra:2017ioj, workinprogress}.

In this context, one major motivation for this work is to advance the RNS methods for cases where the pure-spinor tools for explicit evaluation of amplitudes are still being developed. Our results will make two-fermion-$n$-boson amplitudes at one loop completely accessible and allow for comparison with the respective superspace components of the pure-spinor expressions \cite{Mafra:2012kh,Mafra:2016nwr}. It has the two-fold potential to either assist in the refinement of pure-spinor methods or to check the equivalence between the RNS and pure-spinor formalism in more advanced contexts. Such explicit tests of equivalence would complement recent work \cite{Berkovits:2015yra, Berkovits:2016xnb} on common origins on the pure-spinor, Green-Schwarz and RNS formalisms.

We will compute the spin-summed worldsheet integrands for $n$-point open- and closed-string one-loop amplitudes with two fermions and manifest all the supersymmetry cancellations. While our results apply to generic points in the moduli space of the genus-one worldsheet of torus, cylinder or Moebius-strip topology, their degeneration limits for a pinched A-cycle can be exported to the ambitwistor setup: Cachazo--He--Yuan (CHY) formulae \cite{Cachazo:2013gna, Cachazo:2013hca, Cachazo:2013iea} for loop amplitudes boil down to correlators on a nodal Riemann sphere \cite{Geyer:2015bja, Geyer:2015jch, Geyer:2016wjx}, and the $\tau \rightarrow i\infty$ limit of the subsequent spin sums yields CHY formulae involving two fermions. For external bosons, the analogous spin sums have been recently applied to convert superstring correlators at genus one \cite{He:2017spx} into new representations of field-theory amplitudes which obey the Bern--Carrasco--Johansson (BCJ) duality between color and kinematics \cite{Bern:2008qj, Bern:2010ue}. In other words, the spin sums of this work encode one-loop BCJ numerators for maximally supersymmetric field-theory amplitudes with two fermions and $n$ bosons in $D$ spacetime dimensions\footnote{The dimensional reduction of the ten-dimensional Weyl fermions in the expressions of this work yields both lower-dimensional chiralities in different representations of the R symmetry.}.

This work is organized as follows: We start by reviewing aspects of the genus-one amplitude prescription of the RNS superstring, known results on the contributing one-loop correlators as well as a convenient system of elliptic functions in section \ref{sec;sa}. A novel class of genus-one correlators involving excited spin fields -- spin-$\frac{3}{2}$ operators from the fermionic vertices -- is presented in section \ref{sec;2}, the main result being the $n$-point function in subsection \ref{sec;loopX}. Section \ref{sec;5} is dedicated to summing the correlators of the earlier sections over the spin structures of the worldsheet spinors. We identify universal theta-function building blocks and describe their manipulations which allow for an algorithmic simplification of spin-summed correlators, see subsection \ref{sec;53} and appendix \ref{sec;moremoremore} for a summary of the final results. Various representations of higher-point correlators and detailed derivations of certain theta-function identities can be found in several appendices.


\section{Review}\label{sec;sa}

\subsection{Massless vertex operators in the RNS formalism}\label{sec;sa1}

In the RNS formulation of the open superstring, massless bosons and fermions with momenta $p$ enter the amplitude prescription through their vertex operators
\begin{subequations}\label{eq;ver}
	\begin{align}
	U_{b}^{(-1)}(e,p,z) & = e_{\mu}  \, \psi^{\mu} \, \ee^{-\phi} \, \ee^{ipX} (z)
	\\
	U^{(-1/2)}_f(\chi,p,z) & =  \chi^a \, S_a \,\ee^{-\phi/2} \, \ee^{ipX}(z) \  .
	\end{align}
\end{subequations}
They are conformal primary fields of weight $h=1$ on a worldsheet with complex coordinate $z$. The on-shell degrees of freedom are encoded in transverse polarization vectors $p_\mu e^\mu=0$ and Majorana--Weyl spinors $\chi^a$ subject to the gauge-equivalence $e_\mu \cong e_\mu+p_\mu$ as well as the massless Dirac equation $p_\mu \gamma^\mu_{ab}\chi^b=0$, respectively. Vector and spinor indices of $SO(1,9)$ are taken from the Greek and Latin alphabet as $\mu,\nu,\lambda,\ldots=0,1,\ldots,9$ and $a,b,\ldots=1,2,\ldots,16$, respectively, and $\gamma^\mu_{ab}=\gamma^\mu_{ba}$ denote the $16{\times}16$ Pauli matrices\footnote{Antisymmetrized products are normalized as $
	\gamma^{\mu_1 \mu_2\ldots \mu_p} \equiv \frac{1}{p!} \sum_{\sigma \in S_n} {\rm sgn}(\sigma) \gamma^{\mu_{\sigma(1)}}\gamma^{\mu_{\sigma(2)}}\ldots  \gamma^{\mu_{\sigma(p)}}
	$, e.g.\ $\gamma^{\mu \nu}=\frac{1}{2}(\gamma^\mu \gamma^\nu-\gamma^\nu \gamma^\mu)$.}. Subscript indices as in $S_a$ refer to right-handed Weyl spinors while superscripts as in $\chi^a$ are used for left-handed ones.

The matter sector of the RNS model is furnished by the worldsheet scalars $X^\mu$ and worldsheet spinors $\psi^\mu$ in (\ref{eq;ver}). The fermionic vertex additionally involves the spin field $S_a$ of weight $h=\frac{5}{8}$ that interchanges periodic and antiperiodic boundary conditions of $\psi^\mu$ \cite{Cohn1986}. Moreover, the superscripts of $U_{b,f}^{(q)}$ refer to the superghost charge $q$ of the bosonized representative $e^{q\phi}$ of the $\beta$-$\gamma$ system \cite{Friedan:1985ey, Friedan:1985ge}. For open strings, the vertex-operator insertion is integrated over worldsheet boundaries, and the Chan--Paton generators for the color degrees of freedom are suppressed for ease of notation.

Apart from the {\it canonical} superghost pictures in (\ref{eq;ver}), gluon and gluino vertices exist at higher superghost charge $q=0$ and $q=\frac{1}{2}$, respectively\footnote{Another term $\sim b e^{3\phi/2}$ of $U_f^{(+\frac12)}$ with a different ghost structure \cite{Blumenhagen:2013fgp} is suppressed since it cannot contribute to tree-level and one-loop amplitudes by ghost-charge conservation.},
\begin{subequations}\label{eq;vers}
	\begin{align}\label{eq;ver2}
	U_b^{(0)}(e,p,z) &= e^{\mu} \, \big[ (p\cdot \psi)\psi_{\mu} + i\partial X_{\mu} \big] \, \ee^{ipX}(z)
	\\
	U_f^{(+\frac12)}(\chi,p,z) &= \chi^a \, \big[\tfrac{1}{\sqrt{2}}i \partial X_{\mu} \gamma^{\mu}_{ab}S^b + p_{\mu}S^{\mu}_a \big] \,   \ee^{+\phi/2}\, \ee^{ipX}(z) \ ,
	\label{eq;ver2a}
	\end{align}
\end{subequations}
setting $2\ap=1$.
The second term of the fermion vertex involves excited spin fields\footnote{In \cite{AlanKostelecky1987, Koh:1987hm}, excited spin fields are combined with the descendant fields $\partial S_a$ in the composite operators $S_{a}^{\mu} \propto \psi^{\mu}\psi_{\nu}\gamma^{\nu}_{ab} S^b$.} $S^{\mu}_a$, conformal primaries of weight $h=\frac{13}{8}$ in the spin-$\frac{3}{2}$ representation of $SO(1,9)$ which appear at the subleading order of the OPE \cite{Feng:2012bb}
\beq
\psi^\mu(z) S_a(0) \sim \frac{ \gamma^\mu_{ab} S^b(0)}{z^{1/2}} + z^{1/2} \big[
S^\mu_a(0) + \tfrac{\sqrt{2}}{5} \gamma^\mu_{ab} \partial S^b(0)
\big] + {\cal O}(z^{3/2})
\label{psiSOPE}
\eeq
and satisfy the irreducibility constraint
\beq
S^\mu_a \gamma_\mu^{ab} = 0 \ .
\label{notrace}
\eeq
We are using the shorthand $\partial = \frac{ \partial }{\partial z}$ in (\ref{eq;vers}), 
(\ref{psiSOPE}) and later equations of this work.

\subsection{One-loop superstring amplitudes with two fermions}\label{sec;sa2}

One-loop amplitudes of the open RNS superstring are computed from correlation functions with vanishing overall superghost charge, that is why most of the massless states enter in the superghost pictures of (\ref{eq;vers}). For $n$ external bosons and two external fermions, we have
\begin{align}
&{\cal A}^{\te{1-loop}}(1,2,\ldots,n,A,B) = \int_{{\cal M}_{1;n+2}} \sum_{\nu=1}^{4} (-1)^{\nu + 1}  \label{eq;sa} \\
& \ \ \ \ \times  \langle  \prod_{j=1}^n U^{(0)}_b(e_j,p_j,z_j)  U^{(-1/2)}_f(\chi,p_A,z_A) U^{(+1/2)}_f(\bar\chi,p_B,z_B)  \rangle_\nu \ ,
\notag
\end{align}
where the moduli space ${\cal M}_{1;n+2}$ of the $(n{+}2)$-punctured genus-one surface has to be adjusted to the color-order\footnote{For single-trace amplitudes $\sim {\rm Tr}\{t^1 t^2 \ldots t^k \}$, the insertion points of the vertex operators are integrated over a single worldsheet boundary with $z_1<z_2<\ldots < z_k$. Double-trace amplitudes $\sim {\rm Tr}\{t^1 t^2 \ldots t^j \}{\rm Tr}\{t^{j+1} \ldots t^k \}$ stem from punctures on both boundaries of a cylinder worldsheet such that their cyclic ordering reflects the trace structure \cite{Green:1987mn}.} of ${\cal A}^{\te{1-loop}}(1,2,\ldots,n,A,B)$. The four spin structures indexed by $\nu=1,2,3,4$ refer to the four boundary conditions for the worldsheet spinors $\psi^\mu$ which may be independently chosen as periodic or antiperiodic under translations around the A- and B-cycle. The correlation function in (\ref{eq;sa}) factorizes into contributions from the decoupled CFT sectors of the superghost fields $\ee^{q\phi}$, the $\{\psi^\mu,S_a, S^\mu_b\}$ system as well as the bosons $X^\mu$,
\begin{align}
&\langle  \prod_{j=1}^n U^{(0)}_b(e_j,p_j,z_j)  U^{(-1/2)}_f(\chi,p_A,z_A) U^{(+1/2)}_f(\bar\chi,p_B,z_B)  \rangle_\nu = \chi^a \bar \chi^b  \cor{\ee^{-\phi/2}(z_A) \ee^{\phi/2}(z_B)}_{\nu} \notag \\
&\ \times \sum_{12\ldots n \atop{=P\cup Q}} \bigg\{ p^B_\mu \langle\prod_{l \in Q} p_{\lambda_l} e_{\rho_l}  \psi^{\lambda_l} \psi^{\rho_l}(z_l) S_a(z_A) S^\mu_b(z_B)\rangle_\nu   \langle \prod_{j \in P} e_j \! \cdot \! i \partial X(z_j) \! \! \prod_{k=A,B, \atop{1,2,\ldots,n}} \! \! \ee^{ip_k X(z_k)} \rangle \label{totalcorr} \\
& \ + \frac{ \gamma^\mu_{bc} }{\sqrt{2}} \langle\prod_{l \in Q} p_{\lambda_l} e_{\rho_l}  \psi^{\lambda_l} \psi^{\rho_l}(z_l) S_a(z_A) S^c(z_B)\rangle_\nu    \langle i\partial X_\mu(z_B) \prod_{j \in P} e_j \! \cdot \! i \partial X(z_j) \! \! \prod_{k=A,B, \atop{1,2,\ldots,n}} \! \! \ee^{ip_k X(z_k)} \rangle
\bigg\} \ , \notag
\end{align}
and only the former two depend on the spin structure $\nu$. The sum over $2^n$ partitions $12\ldots n  =P\cup Q$ combines the two terms of the bosonic vertex (\ref{eq;ver2}). For ease of notation, the dependence of the correlators and their building blocks on the modular parameter $\tau$ of the genus-one surface is often suppressed in (\ref{totalcorr}) and later equations.

In the RNS formalism, the two-fermion amplitudes (\ref{eq;sa}) with $n=2$ bosons has firstly been computed in \cite{Atick:1986rs} (also see \cite{Lin:1986ia, Lin:1988xb} for work on $n=3$). For higher numbers of bosons, however, the challenges from the correlators in (\ref{totalcorr}) and the sum over spin structures have never been addressed so far.

The superghost pictures of the above vertex operators partially depart from the prescription 
of \cite{Witten:2012bh, Sen:2014pia} on the distribution of superghost charges near a worldsheet 
degeneration. This can be balanced by relocating the superghost pictures which will generically 
introduce boundary terms in moduli space (see appendix A of \cite{Mafra:2016nwr} for explicit 
examples in the pure-spinor formalism). Such boundary terms are likely to vanish with the large 
amount of supersymmetry in ten-dimensional
flat spacetime, but they might play a role in compactifications with reduced supersymmetry. It would 
be interesting to pinpoint the onset of such boundary terms.

In the following subsections, we will briefly review the genus-one correlators of the worldsheet bosons, the superghosts and the combination of Lorentz currents $ \psi^{\lambda} \psi^{\rho}$ with unexcited spin fields $S_a(z_A) S^b(z_B)$. New results to be given in sections \ref{sec;2} and \ref{sec;5} include the genus-one correlators involving excited spin fields as well as the spin sum over the $\nu$-dependent correlators of (\ref{totalcorr}).

\subsection{Structure of the correlators}\label{sec;sa3}

The genus-one correlation functions relevant to the integrand (\ref{totalcorr}) of one-loop amplitudes can be expressed in terms of Jacobi theta functions $\theta_\nu(z,\tau)$. According to the spin structures in (\ref{eq;sa}), there is one odd instance at $\nu=1$
\beq
\tht{1}(z,\tau) = -i \sum_{n=-\infty}^{\infty} (-1)^{n} q^{\frac{1}{2}(n+\frac{1}{2})^2} \ee^{2\pi i(n+\frac{1}{2})z}
\label{thetaZ}
\eeq
and three even instances at $\nu=2,3,4$,
\begin{align}
\tht{2}(z,\tau) &=  \sum_{n=-\infty}^{\infty} q^{\frac{1}{2}(n+\frac{1}{2})^2}  \ee^{2\pi i(n+\frac{1}{2})z} \notag \\
\tht{3}(z,\tau) &=  \sum_{n=-\infty}^{\infty}  q^{\frac{1}{2}n^2} \ee^{2\pi inz}  \label{thetaY} \\
\tht{4}(z,\tau) &=  \sum_{n=-\infty}^{\infty}  (-1)^nq^{\frac{1}{2}n^2} \ee^{2\pi inz} \ , \notag
\end{align}
where the dependence on the second argument via $q \equiv e^{2\pi i \tau}$ will often be suppressed in the subsequent. Bosonic correlators of the free fields $X^\mu$ can be straightforwardly computed from the two-point function on the torus
\beq
\langle iX^\mu(z) iX^\lambda(0)  \rangle = \eta^{\mu \lambda} \bigg[ \log \left|  \frac{ \theta_1(z)}{\theta_1'(0)} \right|^2 - \frac{2\pi }{\Im(\tau)} [\Im(z)]^2 \bigg] \equiv \eta^{\mu \lambda} G(z)
\label{thetaX}
\eeq
via Wick-contractions, e.g.\footnote{In order to obtain a double-copy representation of closed-string correlators, one can follow the prescription of chiral factorization \cite{DHoker:1988pdl} and exclude the contributions from the joint zero modes of $\partial X^\mu$ and $\bar \partial X^\mu$ from the Wick contractions. This simplifies the contractions in (\ref{thetaW}) to the meromorphic expression $\partial X^\mu(z_1) \ee^{ip_j X(z_j)} \rightarrow p_j^\mu \partial \log \theta_1(z_{1j}) \ee^{ip_j X(z_j)}$.}
\beq
\langle i \partial X^\mu(z_1) \prod_{j=1}^{N} \ee^{ip_j X(z_j)} \rangle = \sum_{l=2}^N p_l^\mu \Big( \partial \log \theta_1(z_{1l}) + 2\pi i \frac{ \Im(z_{1l}) }{\Im(\tau)}  \Big) \prod_{i<j}^N \ee^{p_i \cdot p_jG(z_{ij})}
\label{thetaW}
\eeq
with $z_{ij} \equiv z_i - z_j$ and additional Wick contractions $i\partial X^\mu(z) i\partial X^\lambda(0) \sim \eta^{\mu \lambda} \partial^2 G(z)$ between multiple insertions of $\partial X^\mu$.

The CFT sectors which are sensitive to the spin structure involve the prime form
\beq
E(z,w) = \frac{\tht{1}(z-w)}{\tht{1}'(0)} \ ,
\label{thetaV}
\eeq
raised to some fractional powers. By design of the GSO projection, the powers of the prime form always conspire to integers when combining the individual correlators of the superghost system \cite{Atick1987c}
\begin{align}\label{eq;exams}
\cor{ \ee^{-\phi/2}(z_A) \ee^{+\phi/2}(z_B) }_\nu = \frac{\tht{1}'(0) E(z_A,z_B)^{1/4}}{\tht{\nu}(\frac12(z_B-z_A))}
\end{align}
and the $\{\psi^\mu,S_a, S^\mu_b\}$ system, starting with the two-point function of the spin field~\cite{Atick1987c}
\beq
\cor{ S_a(z_A) S^b(z_B) }_\nu =  \frac{\delta_a^b\,  \theta_\nu(\frac{1}{2}(z_B-z_A))^5 }{\theta'_1(0)^5E(z_A,z_B)^{5/4}} \ .
\label{thetaT}
\eeq
We will often use the notation
\beq
\scor{\psi^\mu(z_1) \ldots S_a(z_2)}_\nu = \cor{\ee^{-\phi/2}(z_A) \ee^{+\phi/2}(z_B) }_{\nu}  \cor{\psi^\mu(z_1) \ldots S_a(z_2)}_\nu
\label{thetaS}
\eeq
for the single-valued combinations relevant to (\ref{totalcorr}), where the ellipsis refers to an arbitrary combination of field insertions $\{\psi^\mu,S_a, S^\mu_b\}$.

In slight abuse of notation, the combined partition function $( \frac{ \theta_\nu(0)}{\theta_1'(0)})^4$ of ten worldsheet bosons and fermions as well as the respective ghosts has been absorbed into the normalization of the $\nu$-dependent correlators in (\ref{eq;exams}), (\ref{thetaT}) and (\ref{thetaS}). This is useful for a unified treatment of the odd spin structure $\nu=1$ and the even ones $\nu=2,3,4$ such that their contributions to the amplitude (\ref{eq;sa}) can be efficiently combined. In particular, this choice of normalization
bypasses indeterminates of the form $\frac{0}{0}$ from the formally vanishing $\theta_\nu(0)$ in the partition function of the odd spin structure.

\subsection{Multiparticle correlators involving spin fields}\label{sec;sa4}

The spin-field correlators in the last line of (\ref{totalcorr}) can be assembled from the results of \cite{Hartl2011} for any number of $\psi^{\lambda} \psi^{\rho}$ insertions.

\subsubsection{Lower-point example}

In the notation of (\ref{thetaS}), the simplest generalization of the two-point function
\beq
\scor{ S_a(z_A) S^b(z_B) }_\nu = \frac{\delta_a^b \, \theta_\nu(\frac{1}{2}(z_B-z_A))^4 }{\theta'_1(0)^4E(z_A,z_B)}
\label{thetaR}
\eeq
is given by
\begin{align}
\scor{\psi^{\lambda} \psi^{\rho}(z_1) S_a(z_A) S^b(z_B) }_\nu &=  \frac{\gamma^{\lambda \rho}{}_a{}^b \, \tht{\nu}(\tfrac{1}{2}z_{AB})^2 \tht{\nu}(\tfrac{1}{2}(z_{A1}{+}z_{B1}))^2 }{2\, \theta'_1(0)^4 \, E_{1A} E_{1B}} \ ,
\label{thetaQ1}
\end{align}
where we used the shorthands
\beq
z_{ij} \equiv z_i - z_j  \ , \ \ \ \ \ \ E_{ij} \equiv E(z_i,z_j)\ . \label{shand}
\eeq
In the subsequent cases with multiple insertions of $\psi^{\lambda} \psi^{\rho}$, it is convenient to introduce the notation
\beq
T_{ij}^{\nu} \equiv \frac{E_{iA}E_{jB}\tht{\nu}(z_{ij}{+}\frac{1}{2}z_{AB}) + E_{jA}E_{iB}\tht{\nu}(z_{ji}{+}\frac{1}{2}z_{AB})}{E_{ij}E_{AB}\tht{\nu}(\frac12z_{AB})}  \ , \ \ \ \
t_{i}^{\nu} \equiv \frac{ \tht{\nu}( \tfrac{1}{2} (z_{Ai}+z_{Bi})) }{ \tht{\nu}( \tfrac{1}{2}z_{AB})}
\label{defT}
\eeq
for the coefficients of the tensor structures:
\begin{align}
& \scor{\psi^{\mu_1} \psi^{\nu_1}(z_1) \psi^{\mu_2} \psi^{\nu_2}(z_2) S_a(z_A) S^b(z_B) }_\nu =  \frac{ E_{AB}\tht{\nu}(\tfrac{1}{2}z_{AB})^4 }{4 \, \theta'_{1}(0)^4\, E_{1A} E_{1B} E_{2A} E_{2B}} \label{thetaQ2} \\
& \ \times \Big[ \gamma^{\mu_1 \nu_1 \mu_2\nu_2}{}_a{}^b \, (t_{1}^{\nu} t_{2}^{\nu})^{2} + \eta^{\nu_1 [\mu_2}  \eta^{\nu_2]   \mu_1}  \delta_a^b\, (T_{12}^\nu)^2 + (\eta^{ \mu_2 [\nu_1} \gamma^{\mu_1] \nu_2}{}_a{}^b - \eta^{ \nu_2 [\nu_1} \gamma^{\mu_1] \mu_2}{}_a{}^b) \, T^\nu_{12} t^{\nu}_{1} t^{\nu}_{2}     \Big]
\notag
\end{align}
The relative signs in the second line depend on the conventions for the Clifford algebra, and we follow
\cite{Hartl2010, Schlotterer2010,Hartl2011} with a minus sign on the right-hand side of the anticommutator $\{ \gamma^\mu, \gamma^\nu \} = -2 \eta^{\mu \nu} $.

Given that vector indices are antisymmetrized with the normalization convention $ \eta^{\nu_1 [\mu_2}  \eta^{\nu_2] \mu_1} =  \eta^{\nu_1 \mu_2}  \eta^{\nu_2 \mu_1} - \eta^{\nu_1 \nu_2}  \eta^{\mu_2 \mu_1}  $, each tensor in the $[\ldots]$ bracket of (\ref{thetaQ2}) and the subsequent equation appears with a prefactor of $\pm 1$. We note that the building blocks in (\ref{defT}) are related via $t_i^\nu = T_{Bi}^\nu$.

The correlator with one more pair of $\psi^{\mu}$ is given by
\begin{align}
&\scor{\psi^{\mu_1} \psi^{\nu_1}(z_1)  \psi^{\mu_2} \psi^{\nu_2}(z_2)   \psi^{\mu_3} \psi^{\nu_3}(z_3) S_a(z_A) S^b(z_B) }_\nu = \frac{E^2_{AB}\tht{\nu}(\tfrac{1}{2}z_{AB})^4}{8\, \theta'_{1}(0)^4 \, \prod_{i=1}^{3}E_{iA} E_{iB}} \label{thetaQ3} \\
&\ \times \bigg[  \gamma^{\mu_1\nu_1\mu_2\nu_2\mu_3\nu_3}{}_a{}^b\,
(t^{\nu}_{1} t^{\nu}_{2} t^{\nu}_{3})^{2} + (\eta^{\nu_1[\mu_2} \eta^{\nu_2][\mu_3} \eta^{\nu_3] \mu_1}-\eta^{\mu_1[\mu_2} \eta^{\nu_2][\mu_3} \eta^{\nu_3] \nu_1}) \delta_{a}^b\,
T^{\nu}_{12}T^{\nu}_{23}T^{\nu}_{13} \notag \\
& \ \ \ \ \ + \Big\{ (\eta^{\mu_2[\nu_1} \gamma^{\mu_1]\nu_2\mu_3\nu_3}{}_a{}^b-\eta^{\nu_2[\nu_1} \gamma^{\mu_1]\mu_2\mu_3\nu_3}{}_a{}^b)\,
T^{\nu}_{12} t^{\nu}_{1} t^{\nu}_{2} (t^{\nu}_{3})^{2}  + \eta^{\nu_1[\mu_2}\eta^{\nu_2] \mu_1}\gamma^{\mu_3\nu_3}{}_a{}^b\,
(T^{\nu}_{12}t^{\nu}_{3})^{2}
\notag\\
& \ \ \ \ \ \ \ \ \ +  (\eta^{\mu_3[\nu_2} \eta^{\mu_2][\nu_1} \gamma^{\mu_1]\nu_3}{}_a{}^b- \eta^{\nu_3[\nu_2} \eta^{\mu_2][\nu_1} \gamma^{\mu_1]\mu_3}{}_a{}^b)\,
T^{\nu}_{12}T^{\nu}_{23} t^{\nu}_{1} t^{\nu}_{3}    + {\rm cyc}(1,2,3)  \Big\} \bigg]\ ,
\notag
\end{align}
where the cyclic sum in the curly bracket does not extend to the totally symmetric terms in the second line.


\subsubsection{The $n$-point function}\label{sec;nunex}

The above examples of spin-field correlators involving $S_a(z_A)$, $S^b(z_B)$ and $n {\leq} 3$ currents $\psi^{\mu_j}\psi^{\nu_j}$ point to the generalization to $n$ insertions of $\psi^{\mu_j} \psi^{\nu_j}(z_j)$ (which can be derived from the results of \cite{Hartl2011}). The structure of the results is captured by
\beq
\scor{\prod_{j=1}^n\psi^{\mu_j} \psi^{\nu_j}(z_j)   S_a(z_A) S^b(z_B) }_\nu =   \frac{ E_{AB}^{n-1} \theta_\nu(\tfrac{1}{2}z_{AB})^4 }{ 2^n \theta_1'(0)^4 \prod_{j=1}^n E_{jA}E_{jB} } \sum_{i}
(\ell_{(i)})^{[\mu_i \nu_i]}{}_{a}{}^{b}   \varphi_\nu^{(i)}(z)  \ ,
\label{eq;moregrule}
\eeq
where the sum over $i$ gathers Lorentz tensors $\ell_{(i)}$ with the index structure of the left-hand side along with spin-structure dependent functions of the $n{+}2$ punctures $\varphi_\nu^{(i)}(z)= \varphi_\nu^{(i)}(z_1,z_2,\ldots,z_n,z_A,z_B)$. The prefactors are $\pm1$ once the $\ell_{(i)}$ in (\ref{eq;moregrule}) are organized in terms of a single form $(\gamma^{\rho_1 \rho_2 \ldots \rho_{2k}})_{a}{}^{b}$ and products of $\eta^{\cdot \cdot }$ with antisymmetrizations in $\mu_i\leftrightarrow \nu_i$, cf.\ (\ref{thetaQ3}).

Most importantly, each Lorentz tensor $\ell_{(i)}$ in (\ref{eq;moregrule}) can be translated into its accompanying function $\varphi_\nu^{(i)}(z)$ through the following dictionary (where $\supset$ is understood as ``contains a factor of'')
\begin{subequations} \label{moresuperrules}
	\begin{align}
	\ell_{(i)} \supset (\gamma^{\ldots \mu_j \ldots})_{a}{}^{b} \ {\rm or} \ (\gamma^{\ldots \nu_j \ldots})_{a}{}^{b} \ \ &\Rightarrow  \ \ \varphi_\nu^{(i)}(z) \supset t^{\nu}_{j}
	\label{superrule4} \\
	\ell_{(i)} \supset \eta^{\mu_j \mu_k} \ {\rm or} \ \eta^{\nu_j \nu_k} \ {\rm or} \ \eta^{\mu_j \nu_k} \ \ &\Rightarrow  \ \ \varphi_\nu^{(i)}(z) \supset T^\nu_{jk} \, , \
	\label{superrule5}
	\end{align}
\end{subequations}
see (\ref{defT}) for the definitions of $t^{\nu}_{j}$ and $T^\nu_{jk}$. The summation range $\sum_i$ in (\ref{eq;moregrule}) involves all Lorentz tensors that can be obtained from partitions of the antisymmetrized pairs of indices $[\mu_1 \nu_1],[\mu_2 \nu_2],\ldots, [\mu_n \nu_n]$ into a form $(\gamma^{\ldots })_{a}{}^{b}$ and products of $\eta^{\cdot \cdot }$.

For each Lorentz tensor $\ell_{(i)}$, the relative prefactor $\pm 1$ can be read off by starting with the $2n$-form $\gamma^{\mu_1\nu_1 \mu_2 \nu_2 \mu_3 \nu_3 \dots \mu_n   \nu_n}$ and moving the pairs of indices entering the given $\eta^{\cdot \cdot }$ to neighboring position. The rule is that the indices in $\eta^{\mu_i \mu_j }$, $\eta^{\nu_i \nu_j }$, $\eta^{\mu_i \nu_j }$ with $i<j$ must be moved into the order $\gamma^{\ldots \mu_i \mu_j \ldots }$, $\gamma^{\ldots \nu_i \nu_j \ldots}$, $\gamma^{\ldots \mu_i \nu_j \ldots}$ and {\it not} the converse one (such as $\gamma^{\ldots \mu_j \mu_i \ldots }$). Then, the number of transpositions among the $\mu_i $ and $\nu_i$ required to attain the pairs of neighbors determines the sign of the Lorentz tensor $\ell_{(i)}$ according to the total antisymmetry of the $\gamma^{\ldots}$. The leftover indices of the form must be left in their order after transferring the neighboring pairs to the $\eta^{\cdot \cdot }$.

For instance, the negative sign of $ \eta^{\nu_1\mu_2} \eta^{\nu_2 \nu_3} \gamma^{\mu_1\mu_3}{}_a{}^b$ in (\ref{thetaQ3}) can be seen by rearranging $\gamma^{\mu_1\nu_1 \mu_2 \nu_2 \mu_3 \nu_3} =- \gamma^{\mu_1 \nu_1 \mu_2 \nu_2 \nu_3 \mu_3 }$ and then removing the neighboring pairs $\nu_1\mu_2 \rightarrow \eta^{\nu_1\mu_2}$ and $\nu_2 \nu_3\rightarrow \eta^{\nu_2 \nu_3}$, leaving $- \gamma^{\mu_1 \nu_1 \mu_2 \nu_2 \nu_3 \mu_3 }\rightarrow - \gamma^{\mu_1\mu_3 }$.

\subsection{Doubly-periodic functions \& bosonic one-loop amplitudes}\label{sec;sa5}

At fixed spin structure $\nu=2,3,4$, the combined correlators of $\{\psi^\mu,S_a, S^\mu_b,e^{q\phi}\}$ involve the even Jacobi theta functions (\ref{thetaY}) and look very different from bosonic correlators of $X^\mu$. After the spin sum in the one-loop amplitude (\ref{eq;sa}), however, the $\nu$-dependent correlator will collapse to a system of doubly-periodic functions that generalizes the singular function in (\ref{thetaW}) from contractions of $i\partial X^\mu$ ,
\beq
f^{(1)}(z) \equiv \partial \log \theta_1(z) + 2\pi i \frac{ \Im(z)}{\Im(\tau)} \ .
\label{thetaP}
\eeq
A system of doubly-periodic functions $\{f^{(n)}(z), n \in \mathbb N_0\}$ which is suitable to capture the results of spin sums over the above $\scor{\ldots}_\nu$ can be generated from a non-holomorphic extension of the Kronecker--Eisenstein series \cite{Kronecker,Brown2011}
\beq
F(z,\alpha) =\frac{ \theta_1'(0) \theta_1(z+\alpha)}{\theta_1(z) \theta_1(\alpha)}
\ , \ \ \ \ \ \
\Omega(z,\alpha) = \ee^{2\pi i \alpha \frac{ \Im(z)}{\Im(\tau)}}  F(z,\alpha) =\sum_{n =0}^{\infty} \alpha^{n-1} f^{(n)}(z) \ .
\label{thetaO}
\eeq
The simplest expansion coefficients besides (\ref{thetaP}) read
\beq
f^{(0)}(z)=1 \ , \ \ \ \
f^{(2)}(z) = \frac{1}{2} \Big\{
\Big( \partial \log \theta_1(z) + 2\pi i \frac{ \Im(z)}{\Im(\tau)} \Big)^2 + \partial^2 \log \theta_1(z) - \frac{   \theta'''_1(0)}{3  \theta'_1(0)}
\Big\} \ .
\label{thetaN}
\eeq
The functions $f^{(n)}$ in (\ref{thetaO}) can be used to generate homotopy invariant iterated integrals over an elliptic curve \cite{Brown2011} and therefore enter the definition of elliptic multiple zeta values \cite{eMZV}. The latter have been identified as a convenient language for the $\ap$-expansion of one-loop open-string amplitudes \cite{Broedel:2014vla}, including double-trace contributions \cite{Broedel:2017jdo}.

\subsubsection{Spin sums on bosonic one-loop amplitudes}

In order to exemplify the relevance of the doubly-periodic $f^{(n)}$ in (\ref{thetaO}) for spin sums, let us review their instances in the $N$-gluon amplitudes. From the $N$ vertex operators (\ref{eq;ver2}), we are led to products of the Szeg\"o kernels
\beq
P_\nu(z) \equiv \frac{ \theta_1'(0) \theta_\nu(z)}{\theta_1(z) \theta_\nu(0)}
\label{thetaM}
\eeq
which ultimately appear in the combinations
\beq
{\cal G}_n(z_1,z_2,\ldots,z_n)  \equiv \sum_{\nu=2}^4 (-1)^{\nu+1} \Big( \frac{ \theta_\nu(0)}{\theta'_1(0)} \Big)^4 P_\nu(z_1) P_{\nu}(z_2) \ldots P_\nu(z_n) \ , \ \ \ \ \ \ \sum_{j=1}^n z_j = 0
\label{thetaL}
\eeq
with $n\leq N$.
All-multiplicity techniques for the simplification of (\ref{thetaL}) have been given in \cite{Tsuchiya:1988va}, also see \cite{Stieberger:2002wk} for an alternative method. As pointed out in \cite{Broedel:2014vla}, the above $f^{(n)}$ functions together with holomorphic Eisenstein series
\beq
{\rm G}_k \equiv \sum_{m,n \in \mathbb Z \atop{(m,n) \neq (0,0)}} \frac{1}{(m\tau+n)^k} \ , \ \ \ \ \ \ k \geq 2
\label{thetaK}
\eeq
allow to compactly represent the spin sums (\ref{thetaL}), starting with \cite{Tsuchiya:1988va, Stieberger:2002wk, Broedel:2014vla}
\begin{align}
{\cal G}_0 &= {\cal G}_2(z_1,z_2) = {\cal G}_3(z_1,z_2,z_3) = 0 \ , \ \ \ \ \ \ {\cal G}_4(z_1,z_2,z_3,z_4) = 1 \label{thetaJ} \\
{\cal G}_5(z_1,\ldots,z_5) &= \sum_{j=1}^5 f^{(1)}(z_j) \ , \ \ \ \ \ \  {\cal G}_6(z_1,\ldots,z_6) = \sum_{j=1}^6 f^{(2)}(z_j) + \! \!\sum_{1=i<j}^6 \! \! f^{(1)}(z_i) f^{(1)}(z_j) \ .\notag
\end{align}
The patterns at higher multiplicity are conveniently captured by the elliptic functions
\beq
V_w(z_1,z_2,\ldots,z_n) \equiv \alpha^n \Omega(z_1,\alpha) \Omega(z_2,\alpha) \ldots \Omega(z_n,\alpha) \big|_{\alpha^w} \ , \ \ \ \ \ \ \sum_{j=1}^n z_j = 0 \ ,
\label{thetaH}
\eeq
starting with
\begin{align}
V_0(z_1,z_2,\ldots,z_n) &= 1 \ , \ \ \ \ \ \
V_1(z_1,z_2,\ldots,z_n) = \sum_{j=1}^n f^{(1)}(z_j) \\
V_2(z_1,z_2,\ldots,z_n) &= \sum_{j=1}^n f^{(2)}(z_j) + \sum_{1=i<j}^n f^{(1)}(z_i) f^{(1)}(z_j) \ .
\label{thetaG}
\end{align}
For instance the $n\leq 9$-point results of \cite{Tsuchiya:1988va} translate into \cite{Broedel:2014vla}
\begin{align}
{\cal G}_n(z_1,z_2,\ldots,z_n) &= V_{n-4}(z_1,z_2,\ldots,z_n) \co 4\leq n \leq 7 \notag \\
{\cal G}_8(z_1,z_2,\ldots,z_8) &= V_{4}(z_1,z_2,\ldots,z_8) + 3 {\rm G}_4 \label{thetaF} \\
{\cal G}_9(z_1,z_2,\ldots,z_9) &= V_{5}(z_1,z_2,\ldots,z_9) + 3 {\rm G}_4 V_{1}(z_1,z_2,\ldots,z_9) \ , \notag
\end{align}
where further simplifications arise in the degeneration limit $\tau \rightarrow i\infty$ \cite{He:2017spx}.

Hence, the worldsheet integrand for the $N$-gluon amplitude comprising spin sums (\ref{thetaL}) and correlators of $X^\mu$ can be entirely expressed in terms of $f^{(n)}$ functions in (\ref{thetaO}). This motivates to express the two-fermion amplitudes in (\ref{eq;sa}) which are related to external bosons by supersymmetry in the same language, also see \cite{Mafra:2016nwr} for the six-point one-loop amplitude in pure-spinor superspace involving $f^{(2)}_{ij}$ \& $f^{(1)}_{ij}f^{(1)}_{pq}$.

Note that the same techniques can be used for spin sums in bosonic one-loop $N$-point amplitudes in orbifold compactifications with reduced supersymmetry \cite{Berg:2016wux} (see \cite{Bianchi:2006nf, Bianchi:2015vsa} for earlier work on the four-point function).


\section{Correlators involving excited spin fields}\label{sec;2}

On top of the spin-field correlators reviewed in section \ref{sec;sa4}, the integrand (\ref{totalcorr}) for two-fermion amplitudes requires correlators of the form $\scor{\prod_j \psi^{\lambda_j} \psi^{\rho_j}(z_j) S_a(z_A) S^\mu_b(z_B)}_\nu$
with an excited spin field $S^\mu_b$. Following the techniques of \cite{Atick1987c,AlanKostelecky1987,Atick1987b,Hartl2011}, we will determine the structure of these genus-one correlators using bosonization techniques for any number of $\psi^{\lambda} \psi^{\rho}$ insertions.

\subsection{Bosonization}\label{sec;21}

The interacting nature of spin fields as reflected in their OPE (\ref{psiSOPE}) with the worldsheet spinors $\psi^\mu$ renders $SO(1,9)$-covariant correlation functions inaccessible to free-field methods. In other words, correlators cannot be obtained from a naive sum over Wick contractions as in (\ref{thetaW}), and the computation of higher-point instances becomes a nontrivial problem, see \cite{Hartl2010,Hartl2011}. However, a free-field description in even spacetime dimensions $D=2n$ can be found by representing the $\{\psi^\mu,S_a, S^\mu_b\}$-system via $n$ free bosons. These redefinitions are known as bosonization \cite{AlanKostelecky1987} and break the $SO(1,9)$ symmetry to an $SU(5)$ subgroup.

Let ${\bf H}$ denote an $SU(n)$ vector of free chiral bosons $\{H^j , \ j=1,2,\ldots,n \}$ subject to normalization $H^j(z) H^k(0)\sim - \delta^{jk}  \ln(z) + \ldots$, then its exponentials $\ee^{i{\bf p} \cdot {\bf H}}$ are conformal primaries of weight $\frac{1}{2} {\bf p}^2$ with OPEs\footnote{To simplify the notation, we neglect Jordan-Wigner cocycle factors \cite{Jordan1928,Frenkel1980a} in our discussion. These are additional algebraic objects accompanying the exponentials to ensure that $\ee^{\pm i H^j}$ and $\ee^{\pm i H^{k \neq j}}$ associated with different bosons anticommute. It suffices to remember that they are implicitly present and that the bosonized representation of $\psi^\mu$ still obeys fermi statistics.}
\begin{align}
\ee^{i {\bf p} \cdot {\bf H}}(z) \, \ee^{i {\bf q} \cdot {\bf H}}(0)  \sim  z^{{\bf p} \cdot {\bf q}} \, \ee^{i({\bf p } + {\bf q}) \cdot {\bf H}}(0) \ + \ \ldots \ .
\label{1,17}
\end{align}
The OPE among the worldsheet spinors, $\psi^\mu(z) \psi^\nu(0)\sim \eta^{\mu \nu} z^{-1} + \ldots$ can be reproduced from the dictionary
\beq
\psi^{\pm j} (z)\equiv \frac{1}{\sqrt{2}} \, \big(\psi^{2j-2}(z)  \pm \, i\psi^{2j-1} (z)\big)  \equiv  \ee^{\pm i  H^j(z)}  \ , \label{1,18}
\eeq
where $j\in \{1,2,\dots,n\}$. One can notice that $\psi^{\pm j}$ form the Cartan--Weyl basis for the fundamental representation of $SO(1,2n{-}1)$.

Spinor components of $SO(1,2n{-}1)$ can be labelled by their eigenvalues $\pm \frac{1}{2}$ under the $n$ simultaneously diagonalized Lorentz generators $\frac{1}{2}\gamma^{\mu \nu} $ which are most conveniently chosen as $\frac{1}{2}\gamma^{2i-2,2i-1}$ with $i=1,2,\ldots,n$ in the $SU(n)$ setting. This suggests to identify spinor indices with $n$-component lattice vectors $\left( \pm \tfrac{1}{2} ,\pm \tfrac{1}{2} ,\ldots ,\pm \tfrac{1}{2} \right)$ from the (anti-)spinor conjugacy classes of $SO(1,2n{-}1)$. The chiral irreducibles can be disentangled by counting the number of negative entries:
\begin{align}
S^{ {a} = \left( \pm \tfrac{1}{2} ,... ,\pm \tfrac{1}{2} \right) } &\leftrightarrow  \, \, \textrm{left-handed spinor} \, \, \leftrightarrow {a} \ \textrm{has an even number of '$-$' signs} \\
S_{ {a} = \left( \pm \tfrac{1}{2} ,... ,\pm \tfrac{1}{2} \right) } &\leftrightarrow  \textrm{right-handed spinor} \leftrightarrow {a} \ \textrm{has an odd number of '$-$' signs} \ .
\label{1,37}
\end{align}
Given this dictionary between spinor indices and lattice vectors, we can make the bosonization of spin fields more precise: The $S_a,S^a$ are represented as an exponential of bosons ${\bf H}$ contracted into a vector $a$ in the weight lattice of (the Lie algebra of) $SO(1,2n{-}1)$:
\begin{align}
S_{a}(z),S^{a}(z)\  \equiv\   \ee^{i {a} \cdot {\bf H}(z)},\ \   a \ \in\  \left\{ (a^1,a^2, \dots, a^n)
\ \big\vert\  a^j= \pm \tfrac{1}{2},\ j=1,\dots, n \right\} \ .
\label{1,38}
\end{align}
Accordingly, vector indices $\mu$ are identified with lattice vectors $(0,\ldots,0,\pm1,0,\ldots,0)$ of $SO(1,2n{-}1)$ from the vector conjugacy class with one nonzero entry $\pm 1$ such that (\ref{1,18}) can be written as $\psi^\mu = \ee^{i\mu \cdot {\bf H}}$.

Bosonization of $\psi^{\mu}$ and $S_{a},S^a$ allows us to relate other conformal primaries to their bosonized expressions, in particular the excited spin fields $S^{\mu}_a$ at the subleading order of the OPE $\psi^{\mu}$ and $S_a$ in (\ref{psiSOPE})\footnote{Obviously, the $n2^n$ bosonized fields of the form \eqref{eq;excitedb} do no exhaust the $(2n-1)2^n$ independent components of the excited spin field $S^\mu_a$ in a spin-3/2 representation of $SO(1,9)$. Still, the $n2^n$ components in \eqref{eq;excitedb} are sufficient to infer the Lorentz-covariant correlators in the next subsections.}:
\begin{align}\label{eq;excitedb}
S^{\mu}_{a}(z) \, \big|_{\mu_j + a_j = \pm \frac{3}{2}}  = \ee^{i ({a}+\mu) \cdot {\bf H}(z)}  =    \ee^{\pm \frac{i}{2} H^1 \pm \dots \pm \frac{i}{2} H^{j-1}  \pm \frac32 iH^j \pm \frac{i}{2} H^{j+1} \pm \dots \pm \frac{i}{2}H^{n}}(z)\ .
\end{align}
Therefore, in the bosonization scheme, $S^{\mu}_a$ can be taken as independent primaries involving a factor of $\ee^{\pm \frac32 iH^j}$ which capture the gamma-traceless components of the composite operators $\sim \psi^\mu \psi_\nu \gamma^\nu_{ab} S^b$.

The Cartan--Weyl basis has the remarkable advantage that entries of gamma- and charge conjugation matrices can be written as delta functions for the lattice vectors of $SO(1,2n{-}1)$ associated with the vector- and the spinor indices. Up to a complex phase (which can in principle be determined by keeping track of all the cocycles \cite{Jordan1928,Frenkel1980a}), one has
\beq
\delta_a^b \sim \delta(a+b) \ , \ \ \ \ \ \ \eta^{\mu \nu} \sim \delta(\mu+\nu) \ , \ \ \ \ \ \  \gamma^\mu_{ab} \sim \sqrt{2} \delta(\mu+a+b) \ .
\label{1,39}
\eeq
The relations in \eref{1,39} admit a derivation of the covariant OPE (\ref{psiSOPE}) in bosonized language, see appendix \ref{APPOPE} for details.

\subsection{Loop level correlators from bosonization}\label{sec;loop}

Correlation functions involving free bosons are well known on surfaces of arbitrary genus \cite{Atick1987b}, and their genus-one instances are given by \cite{Atick1987c}
\begin{align}\label{eq;220}
\cor{\prod_{j=1}^{N} \ee^{iq_j H(z_j)}}_\nu = \frac{1}{\tht{1}'(0)}   \delta(\sum_{j=1}^{N} q_j)  \tht{\nu}(\sum_{k=1}^{N} q_k z_k) \prod^N_{l<m} E(z_l,z_m)^{q_l q_m}  \ .
\end{align}
The Jacobi theta functions $\theta_\nu$ and the prime form $E(z_l,z_m)$ are defined in (\ref{thetaZ}), (\ref{thetaY}) and (\ref{thetaV}), and we again normalize the correlator such as to absorb the partition function of two worldsheet supermultiplets $X^\mu,\psi^\mu$. A general account on bosonization at nonzero genus including the role of spin structures can be found in \cite{AG1, AG2, AG3}, also see \cite{Clavelli:1990qd} for bosonization of odd-spin structure amplitudes.

For a given choice of the weight vectors $\mu,a,b$, one-loop correlators of the fields $\{\psi^\mu,S_a ,S^\mu_b\}$ can be straightforwardly reduced to products of the free-field correlator (\ref{eq;220}) by virtue of their bosonization (\ref{1,18}), (\ref{1,38}) and (\ref{eq;excitedb}). Once a sufficient number of such ``component'' results is available, they can be combined into Lorentz covariant expressions. The idea is to make an ansatz for the correlator with all admissible Lorentz tensors involving products of $\eta^{\mu\lambda}, \delta_a^b,\gamma^\mu_{ab}$ whose index structure is compatible with the $\{\psi^\mu,S_a,S^\mu_b\}$ insertions. Each linearly independent Lorentz tensor in the ansatz is accompanied by a spin-structure dependent function of the insertion points $z_i$ and the modular parameter $\tau$ which remains to be determined.

Then, for each component result computed via (\ref{eq;220}), one can use the delta-function representations (\ref{1,39}) of $\eta^{\mu\nu}, \delta_a^b$ and $\gamma^\mu_{ab}$ to identify the tensor structures compatible with the given choice of lattice vectors. Each choice yields an equation among the unknown functions of $z_j$ and $\tau$ along with the Lorentz tensors in the ansatz. In \cite{Atick:1986rs, Schlotterer2010, Hartl2011}, this procedure is applied to construct higher-point correlation functions involving $\psi^\mu,S_a$, some of which are reviewed in section \ref{sec;sa4}.

Given that the delta-function representation (\ref{1,39}) of $\eta^{\mu\nu}, \delta_a^b$ and $\gamma^\mu_{ab}$ is only fixed up to complex phases, covariant OPEs such as (\ref{psiSOPE}) and
\begin{align}
S_a(z) S^b(0) &\sim \frac{\delta_a^b}{z^{5/4}}\ +\ \frac{\gamma^{\mu\nu}{}_{a}{}^{b}\, \psi_{\nu} \psi_{\mu}(0) }{4\, z^{1/4}} + \ldots \\
S_a(z) S_b^\mu(0) &\sim \frac{\gamma^\nu_{ ab}\, \psi_{\nu} \psi^{\mu}(0)}{\sqrt{2}\, z^{5/4}}   + \ldots  \label{moreOPE}\\
\psi^\mu \psi^\nu(z) \psi^\lambda \psi^\rho(0) &\sim \frac{ \eta^{ \lambda [\nu} \eta^{\mu] \rho}  }{z^2} + \frac{ \eta^{\lambda [\nu} \psi^{\mu]} \psi^{ \rho}(0) - \eta^{\rho [\nu} \psi^{\mu]} \psi^{ \lambda}(0) }{z} + \ldots
\end{align}
are required to determine the signs in correlators, where we remind of the antisymmetrization conventions $\eta^{ \lambda [\nu} \eta^{\mu] \rho}= \eta^{ \lambda \nu} \eta^{\mu \rho} - \eta^{ \lambda \mu} \eta^{\nu \rho}$.

\subsubsection{Three-point example}

Since the two-point correlator $\langle S_a(z_A) S^\mu_b(z_B) \rangle_\nu$ of primary fields with different conformal weights vanishes, the simplest example involving an excited spin field reads
\beq
\cor{\psi^{\lambda}\psi^{\rho}(z_1) \, S_a(z_A) \, S_b^{\mu}(z_B)}_{\nu}
= (\eta^{\mu\rho}\,\gamma^{\lambda}_{ab}-\eta^{\mu\lambda}\, \gamma^\rho_{ab})\,r_\nu(z_1,z_A,z_B) \ ,
\label{eq;223}
\eeq
with some function $r_\nu$ of $z_1,z_A, z_B$ and $\tau$. The tensor on the right-hand side is uniquely determined by the antisymmetry of $\psi^{\lambda}\psi^{\rho}=-\psi^{\rho}\psi^{\lambda}$ and the irreducibility condition (\ref{notrace}) of the excited spin field $S^{\mu}_b$ which forbids a ``gamma-trace'' $\sim \gamma^\mu_{ab}$ as well as the corresponding three-form $\gamma^{\lambda \rho \mu}_{ab}$. By choosing the weight vectors to be
\begin{align}
\lambda &\rightarrow (-1,0,0,0,0) \ , \ \ \ \ \rho \rightarrow(0,+1,0,0,0)  \ , \ \ \ \ \mu \rightarrow(+1,0,0,0,0)  \notag \\
a &\rightarrow \tfrac{1}{2}(-,-,-,-,-)  \ , \ \ \ \ \ \ b \rightarrow \tfrac{1}{2}(+,-,+,+,+) \ ,\label{choiceA}
\end{align}
one can assemble the function in $r_\nu$ in \eqref{eq;223} via five copies of (\ref{eq;220}):
\begin{align}
\pm \sqrt{2} \, r_\nu(z_1,z_A,z_B) &= \langle \ee^{-iH^1(z_1)} \ee^{-\frac{i}{2} H^1(z_A)} \ee^{\frac{3i}{2}H^1(z_B)}\rangle_\nu  \,
\langle \ee^{iH^2(z_1)} \ee^{-\frac{i}{2} H^2(z_A)} \ee^{-\frac{i}{2}H^2(z_B)}\rangle_\nu \notag \\
& \ \ \ \ \ \times \prod_{j=3}^5 \langle   \ee^{-\frac{i}{2} H^j(z_A)} \ee^{\frac{i}{2}H^j(z_B)}\rangle_\nu
\label{eq;exam}\\
&=  \frac{ \theta_{\nu}(\frac32 z_B - z_1 - \frac12 z_A  ) \theta_{\nu} (z_1-\frac12 (z_A+z_B) )  \theta^3_{\nu}(\frac12 (z_B-z_A))}{  \theta'_1(0)^5  \, E(z_1, z_B)^{2}E(z_A,z_B)^{5/4}}  \notag
\end{align}
The factor of $\sqrt{2}$ on the left-hand side stems from the normalization (\ref{1,39}) of the gamma-matrices in the Cartan--Weyl basis.
By adjoining the correlator (\ref{eq;exams}) of the superghosts, the above results combine to
\begin{align}
\scor{\psi^{\lambda}\psi^{\rho}(z_1) \, S_a(z_A) \, S_b^{\mu}(z_B)}_{\nu} &= \eta^{\mu [\rho}\,\gamma^{\lambda]}_{ab} \,
\frac{  \theta_{\nu}(\frac12 (z_{A1}{+}z_{B1})) \theta_{\nu}(z_{1B} {+} \frac12 z_{AB} ) \theta^2_{\nu}(\frac12 z_{AB})}{\sqrt{2} \, \theta'_1(0)^4 \,  E_{1B}^{2}E_{AB}} \notag \\
&\equiv
\frac{ \theta^4_{\nu}(\frac12 z_{AB} )}{\sqrt{2} \, \theta'_1(0)^4 \,  E_{1A}E_{1B}E_{AB}} \times \eta^{\mu [\rho}\,\gamma^{\lambda]}_{ab} \,t^\nu_1 \, \newt^\nu_1
\ , \label{complete}
\end{align}
with shorthands $z_{ij}=z_i-z_j$ and $E_{ij}=E(z_i,z_j)$, where the sign can be fixed via Jordan--Wigner cocycles or the covariant OPE (\ref{moreOPE}). In passing to the second line of (\ref{complete}), we have introduced an additional building block
\beq
\newt^\nu_j \equiv  \frac{E_{jA} \theta_\nu(z_{jB}+\tfrac{1}{2}z_{AB}) }{E_{jB} \theta_\nu(\tfrac{1}{2} z_{AB})}
\label{defnewt}
\eeq
which extends the definitions of $T^\nu_{ij}$ and $t^\nu_i$ in (\ref{defT}) to account for the $z$-dependence from an excited spin field.

\subsubsection{Four-point example}

As an example with several viable tensor structures, we consider
\begin{align}
&\scor{\psi^{\mu_1}\psi^{\nu_1}(z_1) \psi^{\mu_2}\psi^{\nu_2}(z_2) S_a(z_A) S_b^{\lambda}(z_B)}_{\nu} =   \gamma^{[\mu_1}_{ab} \eta^{\nu_1][\mu_2}  \eta^{\nu_2] \lambda}  R^1_\nu(z) \label{compl} \\
&+  \gamma^{[\mu_2}_{ab} \eta^{\nu_2][\mu_1}  \eta^{\nu_1] \lambda} R^2_\nu(z)  +  \gamma^{\mu_1\nu_1[\mu_2}_{ab} \eta^{\nu_2] \lambda}   R^3_\nu(z)  +   \gamma^{\mu_2\nu_2[\mu_1}_{ab} \eta^{\nu_1] \lambda}   R^4_\nu(z) \ ,  \notag
\end{align}
with $R^j_\nu(z) \equiv R^j_\nu(z_1,z_2,z_A,z_B)$, where $R^1_\nu \leftrightarrow R^2_\nu$ and $R^3_\nu \leftrightarrow R^4_\nu$ are related to each other by exchange of $z_1$ and $z_2$. In order to see that four tensor structures are sufficient to express the correlator in question, one can verify that the tensor product of the Lorentz representations of $\psi^{\mu_1}\psi^{\nu_1}$, $\psi^{\mu_2}\psi^{\nu_2}$, $S_a$, $S_b^{\lambda}$ contains precisely four scalars.

Starting from $\lambda \rightarrow (1,0,0,0,0)$, one can isolate $R^3_\nu(z)$ through the choice
\begin{align}
\mu_1 &\rightarrow(0,0,-1,0,0)  \ , \ \ \ \ \nu_1 \rightarrow(0,0,0,-1,0)  \ , \ \ \ \ \mu_2 \rightarrow(0,1,0,0,0)  \label{choiceB}  \\
\nu_2 &\rightarrow(-1,0,0,0,0) \ , \ \ \ \
a \rightarrow \tfrac{1}{2}(-,-,+,+,-)  \ , \ \ \ \ \ \ b \rightarrow \tfrac{1}{2}(+,-,+,+,+) \ \notag
\end{align}
of lattice vectors, which specializes (\ref{compl}) to
\begin{align}
&\pm \, 2\sqrt{2} \, R_\nu^3(z) = \langle \ee^{-\phi/2}(z_A) \ee^{\phi/2}(z_B)  \rangle_\nu
\prod_{j=3}^4 \cor{ \ee^{-iH^j(z_1) }  \ee^{\frac{i}{2} H^j(z_A)} \ee^{\frac{i}{2} H^j (z_B)} }_\nu
\notag \\
& \ \ \times  \cor{\ee^{-iH^1(z_2)} \ee^{-\frac{i}{2} H^1(z_A)} \ee^{\frac{3i}{2} H^1 (z_B)} }_\nu  \cor{ \ee^{iH^2(z_2) }  \ee^{-\frac{i}{2} H^2(z_A)} \ee^{-\frac{i}{2} H^2 (z_B)} }_\nu
\cor{    \ee^{-\frac{i}{2} H^5(z_A)} \ee^{\frac{i}{2} H^5 (z_B)} }_\nu
\notag \\
&= \frac{ \theta_\nu(\tfrac{1}{2}(z_{A1}+z_{B1}) )^2 \theta_\nu(\tfrac{1}{2}(z_{A2}+z_{B2})) \theta_\nu(z_{2B} +\tfrac{1}{2}z_{AB}) }{ \tht{1}'(0)^4\, E_{1A} E_{1B} E_{2B}^2 } \ .\label{spec1}
\end{align}
The three powers of $\sqrt{2}$ stem from the product of three gamma-matrices in (\ref{compl}) along with $R^3_\nu$.

Likewise, combinations of $R^1_\nu$ and $R^3_\nu$ can be addressed via $\lambda \rightarrow (1,0,0,0,0)$ and
\begin{align}
\mu_1 &\rightarrow(0,0,1,0,0)  \ , \ \ \ \ \nu_1 \rightarrow(0,\mp 1,0,0,0)  \ , \ \ \ \ \mu_2 \rightarrow(0,\pm1,0,0,0)  \label{choiceC}  \\
\nu_2 &\rightarrow(-1,0,0,0,0) \ , \ \ \ \
a \rightarrow \tfrac{1}{2}(-,-,-,-,-)  \ , \ \ \ \ \ \ b \rightarrow \tfrac{1}{2}(+,+,-,+,+) \ ,\notag
\end{align}
which specializes (\ref{compl}) to
\begin{align}
& \langle \ee^{-\phi/2}(z_A) \ee^{\phi/2}(z_B)  \rangle_\nu   \cor{   \ee^{-iH^1(z_2)} \ee^{-\frac{i}{2} H^1(z_A)} \ee^{\frac{3i}{2} H^1 (z_B)}    }_\nu
\cor{  \ee^{\mp iH^2(z_1)} \ee^{\pm iH^2(z_2)} \ee^{-\frac{i}{2} H^2(z_A)} \ee^{\frac{i}{2} H^2 (z_B)}    }_\nu \notag \\
&\ \ \ \ \times  \cor{   \ee^{iH^3(z_1)} \ee^{-\frac{i}{2} H^3(z_A)} \ee^{-\frac{i}{2} H^3 (z_B)}    }_\nu \prod_{j=4}^5
\cor{    \ee^{-\frac{i}{2} H^j(z_A)} \ee^{\frac{i}{2} H^j (z_B)}    }_\nu \label{spec2} \\
&= \frac{ \theta_\nu(\tfrac{1}{2}z_{AB})  \theta_\nu(z_{2B} +\tfrac{1}{2}z_{AB})  \theta_\nu(\tfrac{1}{2}(z_{A1}+z_{B1}) )}{ \tht{1}'(0)^4\, E_{12} E_{AB} E_{1A} E_{1B} E_{2B}^2 }
\times \left\{ \begin{smallmatrix}
E_{1A} E_{2B} \theta_\nu(z_{12}+\tfrac{1}{2}z_{AB}) &: \ \nu_1=(0,-1,0,0,0) \\
E_{1B} E_{2A} \theta_\nu(z_{21}+\tfrac{1}{2}z_{AB}) &: \ \nu_1=(0,+1,0,0,0)
\end{smallmatrix} \right. \ .
\notag
\end{align}
Both $\gamma^{[\mu_1}_{ab} \eta^{\nu_1][\mu_2}  \eta^{\nu_2] \lambda}$ and $\gamma^{\mu_1\nu_1[\mu_2}_{ab} \eta^{\nu_2] \lambda}$ are non-zero for the lattice vectors in (\ref{choiceC}), but they exhibit different symmetry properties under exchange of $\mu_2$ and $\nu_1$: Since $\gamma^{[\mu_1}_{ab} \eta^{\nu_1][\mu_2}  \eta^{\nu_2] \lambda}$ is symmetric under $\mu_2 \leftrightarrow \nu_1$, its coefficient must be the sum of the two expressions in (\ref{spec2}) related by exchange of $\mu_2$ and $\nu_1$. The difference of the two expressions in (\ref{spec2}) in turn reproduces the coefficient (\ref{spec1}) of the tensor $\gamma^{\mu_1\nu_1[\mu_2}_{ab} \eta^{\nu_2] \lambda}$ with manifest antisymmetry in $\mu_2 \leftrightarrow \nu_1$, as
can be verified through the Fay trisecant identity \cite{Faybook}
\begin{align}
& E_{12}E_{AB} \theta_\nu(\tfrac{1}{2} (z_{1}+z_{2} - z_{A}-z_B) + z_0) \theta_\nu(\tfrac{1}{2} (z_{1}+z_{2} - z_{A}-z_B )-z_0)  \notag \\
&= E_{1A} E_{2B} \theta_\nu(\tfrac{1}{2} z_{12} {+} \tfrac{1}{2} z_{AB} {+} z_0) \theta_\nu(\tfrac{1}{2} z_{12} {+} \tfrac{1}{2} z_{AB} {-} z_0)
\label{Faytrisc} \\
& \ \ \ - E_{1B} E_{2A} \theta_\nu(\tfrac{1}{2} z_{12} {-} \tfrac{1}{2} z_{AB} {+} z_0) \theta_\nu(\tfrac{1}{2} z_{12} {-} \tfrac{1}{2} z_{AB} {-} z_0)  \notag
\end{align}
at $z_0 \rightarrow \frac{1}{2} z_{12}$. After assembling the above results and fixing the signs through covariant OPEs, the correlator of interest is given by
\begin{align}
&\scor{\psi^{\mu_1}\psi^{\nu_1}(z_1) \psi^{\mu_2}\psi^{\nu_2}(z_2) S_a(z_A) S_b^{\lambda}(z_B)}_{\nu} =\frac{\tht{\nu}(\tfrac12 z_{AB})^{4}}{2\sqrt{2} \, \tht{1}'(0)^4\, E_{1A} E_{1B} E_{2A} E_{2B}}\notag\\
& \ \ \ \ \ \ \ \times\bigg[   \gamma^{\mu_1\nu_1[\mu_2}_{ab} \eta^{\nu_2] \lambda} \, \newt^\nu_2 (t^{\nu}_{1})^{2} t^{\nu}_{2}
+  \gamma^{[\mu_1}_{ab} \eta^{\nu_1][\mu_2}  \eta^{\nu_2] \lambda}  \,  \newt^\nu_2  T_{12}^{\nu} t^{\nu}_{1}
+ (1\leftrightarrow2)  \bigg] \ .  \label{moreex}
\end{align}
The functions $t^\nu_i,T_{jk}^\nu$ and $\newt^\nu_l$ are defined in (\ref{defT}) and (\ref{defnewt}), respectively, and the notation $+ (1\leftrightarrow2) $ instructs to add the image of the previous two terms under $(z_1,\mu_1,\nu_1) \leftrightarrow (z_2,\mu_2,\nu_2)$.

\subsubsection{Five-point example}

The same strategy gives rise to five permutation-inequivalent functions of the punctures in the
correlator with three currents $\psi^{\mu_i}\psi^{\nu_i}$,
\begin{align}
&\scor{\psi^{\mu_1}\psi^{\nu_1}(z_1) \psi^{\mu_2}\psi^{\nu_2}(z_2) \psi^{\mu_3}\psi^{\nu_3}(z_3) S_a(z_A) S_b^{\lambda}(z_B)}_{\nu} = \frac{E_{AB} \tht{\nu}(\tfrac12 z_{AB})^{4}}{4\sqrt{2} \,  \tht{1}'(0)^4 \, \prod_{j=1}^{3} E_{jA} E_{jB}}\notag\\
& \ \ \ \ \times \bigg[ \Big\{ \gamma^{\mu_1 \nu_1 \mu_2 \nu_2 [\mu_3}_{ab}\eta^{ \nu_3] \lambda}
\, \newt^\nu_3\,  (t^{\nu}_{1} t^{\nu}_{2})^{2} t^{\nu}_{3}   + (\eta^{\nu_1 [\mu_2}\eta^{  \nu_2] \mu_1}-\eta^{\mu_1 [\mu_2}\eta^{  \nu_2] \nu_1})\gamma^{ [\mu_3}_{ab}\eta^{ \nu_3] \lambda}
\, \newt^\nu_3 \  (T_{12}^{\nu})^2 \  t^{\nu}_{3}\notag\\
&\ \ \ \ \ \ \ \ \ \ \ \ \ \ \ \ \ \ + (\eta^{ \mu_2 [\nu_1}\gamma^{\mu_1] \nu_2 [\mu_3}_{ab}\eta^{ \nu_3] \lambda}  -\eta^{ \nu_2 [\nu_1}\gamma^{\mu_1] \mu_2 [\mu_3}_{ab}\eta^{ \nu_3] \lambda} )
\,  \newt^\nu_3\, T_{12}^{\nu} \ t^{\nu}_{1} t^{\nu}_{2} t^{\nu}_{3} + \te{cyc}(1,2,3) \Big\} \notag\\
&\ \ \ \ \ \ \ \ \ + \Big\{ ( \gamma^{ \mu_1 \mu_2 \nu_2 }_{ab} \eta^{ \nu_1[\mu_3} \eta^{\nu_3] \lambda }-\gamma^{ \nu_1 \mu_2 \nu_2 }_{ab} \eta^{ \mu_1[\mu_3} \eta^{\nu_3] \lambda })
\, \newt^\nu_3 \, T_{13}^{\nu} \ t^{\nu}_{1} (t^{\nu}_{2})^{2}  \notag\\
&\ \ \ \ \ \ \ \ \ \ \ \ \ \ \ \ \ \ + \gamma^{ [\nu_2 }_{ab} \eta^{  \mu_2] [\mu_1}\eta^{\nu_1 ][\mu_3}\eta^{ \nu_3] \lambda}
\,  \newt^\nu_3\, T_{12}^{\nu} T_{13}^{\nu} \ t^{\nu}_{2} + \te{perm}(1,2,3) \Big\}\bigg] \ ,
\label{anotherex}
\end{align}
which can be determined by suitable choices of the lattice vectors $\lambda,\mu_i,\nu_i,a,b$ along the lines of the previous examples. Note that the explicit correlators in (\ref{complete}), (\ref{moreex}) and (\ref{anotherex}) are sufficient to capture the contributions of the excited spin field to open-string integrands (\ref{totalcorr}) with two fermions and $n\leq 3$ bosons.

\subsection{$n$-point spin-field correlators with an excited spin field}\label{sec;loopX}

Similar to the discussion in section \ref{sec;sa4}, the above examples of correlators involving $S_a(z_A) S^{\lambda}_b(z_B)$ strongly suggest their generalization to $n$ insertions of $\psi^{\mu_j} \psi^{\nu_j}$: After stripping off the overall prefactor of
\begin{align}
&\scor{ \prod_{j=1}^{n} \psi^{\mu_j} \psi^{\nu_j}(z_j) S_a(z_A) S^{\lambda}_b(z_B)}_{\nu}  =  \frac{\sqrt{2} \,E_{AB}^{n-2} \, \tht{\nu}(\tfrac12 z_{AB})^{4}}{2^{n} \, \tht{1}'(0)^4 \,  \prod_{j=1}^nE_{jA}E_{jB}}  \sum_{i}
({\cal L}_{(i)})^{[\mu_i \nu_i],\lambda}_{ab}   \Phi_\nu^{(i)}(z) \ ,
\label{eq;grule}
\end{align}
the remaining contributions are Lorentz tensors ${\cal L}_{(i)}$ with the index structure of the left-hand side and spin-structure dependent functions $\Phi_\nu^{(i)}(z)$ of $z_1,z_2,\ldots,z_n,z_A,z_B$ that obey the following rules: First, the tensors ${\cal L}_{(i)}$ are antisymmetric in all pairs $\mu_i \leftrightarrow \nu_i$ and cannot involve the vector index of the excited spin field $S^\lambda_b$ on a gamma-matrix to account for its irreducibility constraint. Second, ${\cal L}_{(i)}$ is a sum of products of a single odd-rank form $\gamma^{\rho_1\rho_2\ldots \rho_{2k+1}}_{ab}$ accompanied by $n{-}k$ factors of $\eta^{\cdot \cdot }$, and each summand has a prefactor $\pm1$ given the choice of normalization in (\ref{eq;grule}).

Most importantly, each Lorentz tensor ${\cal L}_{(i)}$ in (\ref{eq;grule}) can be translated into its accompanying function $\Phi_\nu^{(i)}(z)$ through the following dictionary (where $\supset$ is understood as ``contains a factor of''),
\begin{subequations} \label{superrules}
	\begin{align}
	{\cal L}_{(i)} \supset \gamma_{ab}^{\ldots \mu_j \ldots} \ {\rm or} \ \gamma_{ab}^{\ldots \nu_j \ldots} \ \ &\Rightarrow  \ \ \Phi_\nu^{(i)}(z) \supset t^{\nu}_{j}
	\label{superrule1} \\
	{\cal L}_{(i)} \supset \eta^{\mu_j \mu_k} \ {\rm or} \ \eta^{\nu_j \nu_k} \ {\rm or} \ \eta^{\mu_j \nu_k} \ \ &\Rightarrow  \ \ \Phi_\nu^{(i)}(z) \supset T^\nu_{jk}
	\label{superrule3} \\
	{\cal L}_{(i)} \supset \eta^{\mu_j \lambda} \ {\rm or} \ \eta^{\nu_j \lambda} \ \ &\Rightarrow  \ \ \Phi_\nu^{(i)}(z) \supset \newt^\nu_j  \label{superrule2}  \, ,
	\end{align}
\end{subequations}
see (\ref{defT}) and (\ref{defnewt}) for the building blocks $t^{\nu}_{j},\newt^\nu_j$ and $T^\nu_{jk}$. While the first two rules (\ref{superrule1}) and (\ref{superrule3}) tie in with those for two unexcited spin fields, see (\ref{superrule4}) and (\ref{superrule5}), the additional vector index of the excited spin field is addressed by (\ref{superrule2}).

The summation range $\sum_i$ in (\ref{eq;grule}) involves all Lorentz tensors ${\cal L}_{(i)}$ that can be obtained from partitions of the antisymmetrized pairs of indices $[\mu_1 \nu_1],[\mu_2 \nu_2],\ldots, [\mu_n \nu_n]$ into a form $(\gamma^{\ldots })_{a b}$, products of $\eta^{\cdot \cdot }$ and an additional $\eta^{\cdot \lambda}$ associated with the excited spin field.

Similar to the rules of section \ref{sec;nunex} to determine the signs in the correlator with unexcited spin fields, the idea is to start with a reference $(2n{+}1)$-form $\gamma^{\mu_1 \nu_1 \mu_2 \nu_2 \ldots \mu_n \nu_n \lambda}$. Pairs of indices which enter the given product of $\eta^{\cdot \cdot }$ must be moved to neighboring positions $\gamma^{\ldots \mu_i \mu_j \ldots }$, $\gamma^{\ldots \nu_i \nu_j \ldots}$, $\gamma^{\ldots \mu_i \nu_j \ldots}$ with $i<j$ (and {\rm not} $i>j$) or $\gamma^{\ldots \mu_i \lambda \ldots }$,  $\gamma^{\ldots \nu_i \lambda \ldots}$ (with $\lambda$ on the right of $\mu_i,\nu_i$) before transferring them to the metric tensors. For instance, the positive sign of $\eta^{\mu_1 \mu_2} \eta^{\nu_1\nu_2} \gamma^{\nu_3} \eta^{\mu_3\lambda}$ in (\ref{anotherex}) can be seen by rearranging $\gamma^{\mu_1 \nu_1 \mu_2 \nu_2 \mu_3 \nu_3 \lambda} = (-1)^2 \gamma^{\mu_1 \mu_2 \nu_1 \nu_2 \nu_3  \mu_3 \lambda}$ before transferring the pairs $\mu_1 \mu_2$, $\nu_1 \nu_2$, $\mu_3 \lambda$ to the $\eta^{\cdot \cdot}$ and converting $ \gamma^{\mu_1 \mu_2 \nu_1 \nu_2 \nu_3  \mu_3 \lambda} \rightarrow  \gamma^{\nu_3 }$.

\subsection{A standard form for spin sums}
\label{stform}

In view of the ultimate goal of this work to sum the above correlators (\ref{eq;moregrule}) and (\ref{eq;grule}) over the spin structures $\nu=1,2,3,4$, we identify a prototype spin sum from the dictionaries (\ref{moresuperrules}) and (\ref{superrules}). First, the prefactors of (\ref{eq;moregrule}) and (\ref{eq;grule}) along with the $\nu$-dependent minus sign in the amplitude prescription (\ref{eq;sa}) suggest to introduce the shorthands
\beq
Z_{\nu}(y)\equiv \frac{(-1)^{\nu+1} \tht{\nu}(y)^4 }{\tht{1}'(0)^4} \ , \ \ \ \ \ \
y \equiv \frac{1}{2} z_{AB}
\ ,
\label{Zandy}
\eeq
where $Z_\nu(y)$ may be interpreted as a partition function of $X^\mu$ and $\psi^\mu$ with twisted boundary conditions. All the $\nu$-dependence in the building blocks $t^{\nu}_{j},\newt^\nu_j$ and $T^\nu_{jk}$ for $\varphi_\nu^{(i)}(z) $ and $\Phi_\nu^{(i)}(z) $ in (\ref{moresuperrules}) and (\ref{superrules}) occurs via products of ratios $\frac{ \theta_{\nu}(x\pm y)}{\theta_\nu(y)}$, with $x$ representing some $z_{ij}$ with $i,j\in \{1,2,\ldots, n,A,B\}$. It is particularly convenient to gather such ratios of $\theta_\nu$ functions via
\beq
F_{\nu}(x,y) \equiv \frac{\tht{1}'(0) \tht{\nu}(x+y)}{\tht{1}(x) \tht{\nu}(y)} = \frac{ \tht{\nu}(x+y)}{E(x) \tht{\nu}(y)} \ ,
\label{Fnunu}
\eeq
which generalizes the Kronecker--Eisenstein series in (\ref{thetaO}) to even spin structures with $F_{\nu=1}(x,y)= F(x,y)$ and exhibits the following symmetry property,
\beq
F_{\nu}(-x,-y)=-F_{\nu}(x,y) \ .
\label{symprop}
\eeq
More precisely, the building blocks of the above spin-field correlators can be expressed in terms of the function $F_{\nu}(x,y) $ by means of
\begin{align}
T_{jk}^\nu &= \frac{E_{jA} E_{kB} F_\nu(z_{jk},y) + E_{jB} E_{kA} F_\nu(z_{jk},-y)}{E_{AB}} \notag  \\
t^\nu_j &= E_{Bi} F_\nu(z_{Bi},y) \ , \ \ \ \ \ \  \newt^\nu_j = E_{jA} F_\nu(z_{jB},y) \ .  \label{TFdict} 
\end{align}
Then, the most general spin sum we will be concerned with in the next section can be brought into the standard form
\begin{align}
\ssum \left[ \begin{array}{c}
\! x_1,x_2,\ldots,x_M \! \\
x_{M+1},x_{M+2},\ldots,x_N
\end{array}\right]
&\equiv
\sum_{\nu=1}^4 Z_\nu(y)   \prod_{i=1}^{M}F_{\nu}(x_i,y)   \prod_{j=M+1}^{N}   F_{\nu}(x_j,-y)   \notag \\
\sum_{i=1}^{M}x_i+  \sum_{j=M+1}^{N}   x_j&=0
\ , \label{standform}
\end{align}
which generalizes the prototype spin sum (\ref{thetaL}) for bosonic one-loop amplitudes. The first arguments $x_j,x_k$ of the above $F_\nu$ will always add up to zero after suitable application of (\ref{symprop}), and one can easily show that
\begin{align}\label{flip}
\ssuma{x_{1},x_{2},\dots,x_{M}}{x_{M+1},\dots,x_{N}}
=
(-1)^{N}
\ssuma{-x_{M+1},\dots,-x_{N}}{-x_{1},-x_{2},\dots,-x_{M}} \ .
\end{align}
In many cases, the complexity of the spin sum (\ref{standform}) governed by $M$ and $N$ can be reduced by expressing products $t^\nu_j t^\nu_k$ with $j\neq k$ through a single factor of $F_\nu(z_{jB},y) $ instead of an iteration of  (\ref{TFdict}):
\beq
t^\nu_j t^\nu_k = \frac{E_{jA} E_{kB} F_\nu(z_{jk},y) - E_{jB} E_{kA} F_\nu(z_{jk},-y)}{E_{AB}} \ , \ \ \ \ j\neq k \ .
\label{lesscomplex}
\eeq
Note that (\ref{lesscomplex}) is a consequence of the Fay trisecant identity (\ref{Faytrisc}).

\subsubsection{Examples with unexcited spin fields}

Let us give the simplest examples of spin-field correlators rewritten in terms of the standard spin sum (\ref{standform}) with
building blocks (\ref{Zandy}) and (\ref{Fnunu}): In presence of unexcited spin fields, the correlators (\ref{thetaR}), (\ref{thetaQ1}) and  (\ref{thetaQ2}) translate into
\begin{align}
\sum_{\nu=1}^{4}(-1)^{\nu+1} \scor{ S_a(z_A) S^b(z_B) }_\nu &=   \sum_{\nu=1}^{4} \frac{ \delta_a^b \, Z_\nu(y)}{E_{AB} } = \frac{ \delta_a^b }{E_{AB} } \ssum \left[ \begin{array}{c}
\! - \! \\
-
\end{array}\right]
\label{simpex} \\
\sum_{\nu=1}^{4} (-1)^{\nu+1}  \scor{\psi^{\lambda} \psi^{\rho}(z_1) S_a(z_A) S^b(z_B) }_\nu &=   \sum_{\nu=1}^{4} \frac{ \gamma^{\lambda \rho}{}_a{}^b \, E_{B1} }{2E_{1A}}  \,  Z_\nu(y) \,  F_{\nu}(z_{B1},y) F_{\nu}(z_{1B},-y)  \notag\\
& =   \frac{ \gamma^{\lambda \rho}{}_a{}^b \,  E_{B1}}{2E_{1A}}\,
\ssuma{z_{B1}}{z_{1B}}\label{eq;330}
\end{align}
as well as
\begin{align}
&\sum_{\nu=1}^{4}  (-1)^{\nu+1}  \scor{\psi^{\mu_1} \psi^{\nu_1}(z_1) \psi_{\mu_2}\psi_{\nu_2}(z_2)S_a(z_A) S^b(z_B) }_\nu  =(\gamma^{\mu_1 \nu_1}{}_{ \mu_2\nu_2})_a{}^b   \tfrac{E_{AB} E_{1B} E_{2B} }{4 E_{1A} E_{2A}}  \ssuma{z_{B1},z_{B2}}{z_{1B},z_{2B}}\, \notag\\
&\phantom{=\ } - \delta_{[\mu_2}^{  [\nu_1} \gamma^{\mu_1]}{}_{ \nu_2]}{}_a{}^b
\bigg\{
\tfrac{E_{2B}}{4 E_{2A}} \ssuma{z_{B1},z_{12}}{z_{2B}}
+
\tfrac{E_{1B}}{4 E_{1A}} \ssuma{z_{B1}}{z_{2B},z_{12}}
\bigg\}
\notag\\
&\phantom{=\ } - \delta^{\nu_1}_{ [\mu_2}  \delta_{\nu_2] }^{ \mu_1}  \delta_a^b
\bigg\{ \tfrac{1}{4E_{AB}} \ssuma{z_{12},z_{21}}{-}
+ \tfrac{1}{4E_{AB}} \ssuma{-}{z_{12},z_{21}}
\notag\\
&\phantom{=\  - \delta^{\nu_1}_{ [\mu_2}  \delta_{\nu_2] }^{ \mu_1}  \delta_a^b
	\bigg\{}
+ \tfrac{E_{1A} E_{2B}}{4E_{AB}E_{1B} E_{2A}} \ssuma{z_{12}}{z_{21}}
+ \tfrac{E_{1B} E_{2A}}{4E_{AB}E_{1A} E_{2B}} \ssuma{z_{21}}{z_{12}}
\bigg\} \notag\\
&=
\frac{1}{2 E_{AB}}\ssuma{z_{12},z_{21}}{-}
\left\{
(\gamma^{\mu_1 \nu_1}{}_{ \mu_2\nu_2})_a{}^b
-
\delta^{\nu_1}_{ [\mu_2}  \delta_{\nu_2] }^{ \mu_1}  \delta_a^b
\right\}
\notag\\
&\ \ \ -
\frac{E_{1A} E_{2B} }{4 E_{1B} E_{2A} E_{AB}} \ssuma{z_{12}}{z_{21}}
\left\{
(\gamma^{\mu_1 \nu_1}{}_{ \mu_2\nu_2})_a{}^b
+
\delta_{[\mu_2}^{  [\nu_1} \gamma^{\mu_1]}{}_{ \nu_2]}{}_a{}^b
+
\delta^{\nu_1}_{ [\mu_2}  \delta_{\nu_2] }^{ \mu_1}  \delta_a^b
\right\}
\notag\\
&\ \ \ -
\frac{E_{1B} E_{2A}}{4 E_{1A} E_{2B} E_{AB}} \ssuma{z_{21}}{z_{12}}
\left\{
(\gamma^{\mu_1 \nu_1}{}_{ \mu_2\nu_2})_a{}^b
-
\delta_{[\mu_2}^{  [\nu_1} \gamma^{\mu_1]}{}_{ \nu_2]}{}_a{}^b
+
\delta^{\nu_1}_{ [\mu_2}  \delta_{\nu_2] }^{ \mu_1}  \delta_a^b
\right\} \ .
\label{237a}
\end{align}
The last expression follows from \eqref{flip} and \eqref{lesscomplex}, and the generalization to three insertions of $\psi_{\mu_j}\psi_{\nu_j}(z_j)$ can be found in appendix \ref{sec;moreB}.

\subsubsection{Examples with an excited spin field}

In presence of excited spin fields, the expressions (\ref{complete}) and (\ref{moreex}) for the simplest correlators give rise to
\begin{align}\label{eq;332}
\sum_{\nu=1}^{4} (-1)^{\nu+1}\scor{\psi^{\lambda}\psi^{\rho}(z_1) \, S_a(z_A) \, S_b^{\mu}(z_B)}_{\nu} &= - \frac{ \eta^{\mu [\rho}\,\gamma^{\lambda]}_{ab} }{\sqrt{2} \, E_{AB}}\, \sum_{\nu=1}^{4} Z_\nu(y) \, F_\nu(z_{1B},y) F_\nu(z_{B1},y)   \notag\\
& = - \frac{ \eta^{\mu [\rho}\,\gamma^{\lambda]}_{ab} }{\sqrt{2} \, E_{AB}} \, \ssuma{z_{1B},z_{B1}}{-}
\end{align}
as well as
\begin{align}
&\sum_{\nu=1}^{4} (-1)^{\nu+1} \scor{\psi^{\mu_1}\psi^{\nu_1}(z_1) \psi^{\mu_2}\psi^{\nu_2}(z_2) S_a(z_A) S_b^{\lambda}(z_B)}_{\nu} \notag\\
&=
-
\gamma^{[\mu_1}_{ab} \eta^{\nu_1][\mu_2}  \eta^{\nu_2] \lambda}   \Big\{\tfrac{1 }{2\sqrt{2} \, E_{AB}} \ssuma{z_{B1},z_{2B},z_{12}}{-}
+ \tfrac{ E_{1B} E_{2A} }{2\sqrt{2} \, E_{AB}E_{1A} E_{2B} } \ssuma{z_{B1},z_{2B}}{z_{12}} \Big\}\notag\\
& \ \ \
+
\gamma^{\mu_1\nu_1[\mu_2}_{ab} \eta^{\nu_2] \lambda}   \tfrac{E_{1B} }{2\sqrt{2} E_{1A}} \ssuma{z_{B1},z_{B2},z_{2B}}{z_{1B}}  + (z_1,\mu_1,\nu_1) \leftrightarrow  (z_2,\mu_2,\nu_2)
\notag\\
&=
-
\tfrac{1 }{2\sqrt{2} \, E_{AB}} \ssuma{z_{B1},z_{2B},z_{12}}{-}
\left\{
\gamma^{[\mu_1}_{ab} \eta^{\nu_1][\mu_2}  \eta^{\nu_2]  \lambda}
+    \gamma^{\mu_1\nu_1[\mu_2}_{ab} \eta^{\nu_2] \lambda}
\right\}
\notag\\
&\ \ \
-
\tfrac{ E_{1B} E_{2A} }{2\sqrt{2} \, E_{AB}E_{1A} E_{2B} } \ssuma{z_{B1},z_{2B}}{z_{12}}
\left\{
\gamma^{[\mu_1}_{ab} \eta^{\nu_1][\mu_2}  \eta^{\nu_2]  \lambda}
-    \gamma^{\mu_1\nu_1[\mu_2}_{ab} \eta^{\nu_2] \lambda}
\right\}
\notag\\
& \ \ \   +
(z_1,\mu_1,\nu_1) \leftrightarrow  (z_2,\mu_2,\nu_2) \ . \label{mostex}
\end{align}
The last expression again follows from \eqref{symprop} and \eqref{flip}, and the generalization to three insertions of $\psi^{\mu_j}\psi^{\nu_j}(z_j)$ can be found in appendix \ref{sec;moreD}.

From the discussion in the next section, one can find that most of the spin sums in (\ref{simpex}) to (\ref{mostex}) vanish, except for the case with $\ssuma{z_{B1},z_{2B},z_{12}}{-} $. The latter leads to the non-vanishing four-point amplitude among two bosons and two fermions which has been first computed in \cite{Atick:1986rs}.


\section{Evaluating spin sums in two-fermion amplitudes} \label{sec;5}

In this section, we present a method to evaluate the prototype spin sum (\ref{standform}) for two-fermion amplitudes in terms of the doubly-periodic functions $f^{(n)}$ in (\ref{thetaO}). While the simplest case in (\ref{simpex}) 
\beq
\ssum \left[ \begin{array}{c}
	\! - \! \\
	-
\end{array}\right] =
\sum_{\nu=1}^4 Z_\nu(y) = \sum_{\nu=1}^4 \frac{(-1)^{\nu+1} \tht{\nu}(y)^4 }{\tht{1}'(0)^4} = 0
\label{simplst}
\eeq
due to $ \scor{ S_a(z_A) S^b(z_B)}_\nu$ can be dealt with via Riemann identities \cite{Mumford1983a},
\begin{align}
&\sum_{\nu=1}^{4} (-1)^{\nu+1} \tht{\nu}(z_1) \tht{\nu}(z_2) \tht{\nu}(z_3) \tht{\nu}(z_4) = \tht{1}(z'_1) \tht{1}(z'_2) \tht{1}(z'_3) \tht{1}(z'_4) \notag \\
&\ \ \ \ z'_1 \equiv \frac12 (z_1+z_2+z_3+z_4),\ \ \ \  z'_2 \equiv \frac12 (z_1+z_2-z_3-z_4) \label{prop;rie} \\
&\ \ \ \ z'_3 \equiv \frac12 (z_1-z_2-z_3+z_4),\ \ \ \  z'_4 \equiv \frac12 (z_1-z_2+z_3-z_4)\ , \notag
\end{align}
additional factors of $F_\nu(x_j,\pm y)$ in (\ref{Fnunu}) require further techniques which will be developed in this section for arbitrary multiplicities.

In contrast to the worldsheet bosons $X^\mu$, the $\{\psi^\mu,S_a,S^\mu_b\}$-system does not exhibit any zero modes shared between the left- and right-movers and therefore yields meromorphic correlation functions. At the same time, the interplay of different spin structures guarantees doubly-periodic expressions for the spin-summed correlators \cite{Atick1987c}. Accordingly, all the explicit final expressions for the latter are invariant under collective interchange of the doubly-periodic $f^{(n)}$ and their meromorphic truncations $g^{(n)}$ defined along the lines of (\ref{thetaO}),
\beq
F(z,\alpha) =\frac{ \theta_1'(0) \theta_1(z+\alpha)}{\theta_1(z) \theta_1(\alpha)}
\equiv \sum_{n =0}^{\infty} \alpha^{n-1} g^{(n)}(z) \ ,
\label{217h}
\eeq
e.g.\ $g^{(0)}(z)=1$ and $g^{(1)}(z) = \partial \log \theta_1(z)$. The freedom to collectively interchange $f^{(n)} \leftrightarrow g^{(n)}$ is inherited from the elliptic functions $V_w(z_1,z_2,\ldots,z_n)$ defined in (\ref{thetaH}) which capture the spin-summed correlators of bosonic vertex operators  and the subsequent results on their fermionic counterparts. Since various intermediate steps only manifest meromorphicity and obscure double periodicity, we will first derive expressions for fermionic spin sums in terms of $g^{(n)}$ and leave the freedom to globally replace $g^{(n)}\rightarrow f^{(n)}$ for the last step.

\subsection{General strategy}
\label{sec;51}

The evaluation of the prototype spin sum (\ref{standform}) can be organized into four steps:

\subsubsection{Reducing the factors of $F_\nu$}
\label{step1}

As a first step, we start from the spin-structure dependent Fay identity
\beq
F_{\nu}(x_1,y_1) F_{\nu}(x_2,y_2)
= F(x_1,y_1{+}y_2) F_{\nu}(x_2{-}x_1,y_2) + F(x_2,y_1{+}y_2) F_{\nu}(x_1{-}x_2,y_1)
\label{217a}
\eeq
to successively convert factors of $F_\nu(x_i,\pm y)$ into Kronecker--Eisenstein series $F(z,\alpha)$ in (\ref{217h}) and to thereby simplify the dependence on $\nu$. This process requires the following corollary of (\ref{217a}) with $x_1\neq -x_2$
\begin{align}
&F^{(0,k_1)}_{\nu}(x_1,y) F^{(0,k_2)}_{\nu}(x_2,y)
\label{217c}\\
&=
\sum_{l=0}^{k_2} 	(-1)^{l}(k_1{+}l)!
\left(\begin{array}{c}
k_2\\
l
\end{array}
\right)
\big[g^{(k_1+l+1)}(x_1) - (-1)^{k_1+l+1} g^{(k_1+l+1)}(x_2)
\big]
F^{(0,k_2-l)}_{\nu}(x_1{+}x_2,y)
\notag\\
&\phantom{=} - \sum_{l=0}^{k_2} \sum_{m=0}^{k_1+l}
\left(\begin{array}{c}
k_2\\
l
\end{array}
\right)
\frac{(-1)^{k_1-m} (k_1{+}l)!}{(m{+}1)!} \, g^{(k_1+l-m)}(x_2) \, F^{(0,m+k_2-l+1)}_{\nu}(x_1{+}x_2,y)\notag
\end{align}
for derivatives
\beq
F^{(k,l)} (x,y) \equiv \frac{\partial^{k}}{\partial x^{k}} \frac{\partial^{l}}{\partial y^{l}}F(x,y)  \ , \ \ \ \
F_\nu^{(k,l)} (x,y) \equiv \frac{\partial^{k}}{\partial x^{k}} \frac{\partial^{l}}{\partial y^{l}}F_\nu(x,y) \ .
\label{217b}
\eeq
The right-hand side of (\ref{217c}) naturally introduces the functions $g^{(n)}(x)$ in (\ref{217h}), for instance
$F_\nu(x_1,y) F_\nu(x_2,y) = (g^{(1)}(x_1) {+}g^{(1)}(x_2)) F_\nu(x_1{+}x_2,y)$  $- F^{(0,1)}_\nu(x_1{+}x_2,y) $.

In this way, one can always arrive at a simplified form of a generic spin sum
\begin{align}\label{eq;firstly}
&\ssuma{x_1,x_2,\dots,x_M}{x_{M+1},\dots x_{N}}= \! \! \sum_{k_1,k_2 \geq0}  \! \!{\cal R}_{k_1,k_2}(g^{(n)}_{pq}) \sum_{\nu=1}^{4} Z_{\nu}(y)  F^{(0,k_1)}_{\nu}(x,y) F^{(0,k_2)}_{\nu}(-x,-y) \ ,
\end{align}
where $x$ is a linear combination of $z_{ij}$, and ${\cal R}_{k_1,k_2}(g^{(n)}_{pq}) $ denote polynomials in $g^{(n)}$ functions due to the right-hand side of (\ref{217c}) at arguments $z_{pq}$, $p,q\in \{1,2,\ldots,n,A,B\}$.

Given that the applicability of (\ref{217c}) is tied to $x_1\neq -x_2$, we will need a separate corollary of (\ref{217a})
\begin{align}
&F_{\nu}^{(0,k_1)}(x,y) F_{\nu}^{(0,k_2)}(-x,y) =(-1)^{k_2}(k_1{+}k_2)!\partial  g^{(k_1+k_2+1)}(x)
\label{217d} \\
& \ \ \
- \sum_{l=0}^{k_2} (-1)^l
\sum_{m=0}^{k_1+l}
\left(\begin{array}{c}
k_2\\
l
\end{array}
\right)
\frac{(k_1{+}l)! }{(m{+}1)!}   \, g^{(k_1+l-m)}(x) \, F^{(0,m+k_2-l+1)}_{\nu}(0,y) \notag
\end{align}
for the degenerate case, where subsets of the first-row arguments $x_1,x_2,\ldots,x_M$ of $\ssuma{x_1,x_2,\dots,x_M}{x_{M+1},\dots x_{N}}$ (or subsets of the second-row entries $x_{M+1},\dots x_{N}$) add up to zero. As we will see, one can largely infer such degenerate spin sums from appropriate limits of the generic case, where the $x_j$ obey no constraint other than $ \sum_{i=1}^{M}x_i+  \sum_{j=M+1}^{N}   x_j = 0$. In such generic situations, the only need for (\ref{217d}) arises from cases with $M=N$, i.e.\ spin sums of the form $\ssuma{x_1,x_2,\dots,x_M}{-}$. Then, the derivatives $\partial g^{(k_1+k_2+1)}(x)$ in the first line of (\ref{217d}) without a $\nu$-dependent factor of $F^{(0,k)}_{\nu}(0,y)$ do not contribute to the spin sums because of (\ref{simplst}).

The central identities (\ref{217a}), (\ref{217c}) and (\ref{217d}) are proven in appendix \ref{proof1}.


\subsubsection{Merging $F_\nu$ at arguments $y$ and $-y$}
\label{step2}

The second step is to combine the spin-structure dependent terms in (\ref{eq;firstly}) via
\begin{align}
&\sum_{\nu=1}^{4} Z_{\nu}(y) F^{(0,k_1)}_{\nu}(x,y) F^{(0,k_2)}_{\nu}(-x,-y) \label{217f}\\
&=\sum_{\nu=1}^{4} Z_{\nu}(y) \bigg[ \sum_{l=1}^{k_2} (-1)^{k_2-l}
\left(	\begin{array}{c}	k_2\\l	\end{array}\right)F^{(0,l)}_{\nu}(0,-y) F^{(0,k_1+k_2-l)}(x,2 y)		\notag\\
&\ \ \ \ \ \ \ \ \ \ \ \ \ \ \ \ \ - \sum_{l=1}^{k_1} (-1)^{k_2} \left(	\begin{array}{c}	k_1\\
l\end{array}	\right)F^{(0,l)}_{\nu}(0,y) F^{(0,k_1+k_2-l)}(x,2 y)\bigg]\ , \notag
\end{align}
which is proven in appendix \ref{proof1A} and only holds under the spin sum. In particular, specialization to $k_1=k_2=0$ gives rise to $\sum_{\nu=1}^{4} Z_{\nu}(y) F_{\nu}(x,y) F_{\nu}(-x,-y) = 0$.

Application of (\ref{217f}) leads to the simpler expression
\begin{align}
\ssuma{x_1,x_2,\dots,x_M}{x_{M+1},\dots x_{N}}
&= \sum_{j>0} \tilde{{\cal R}}_j(g^{(n)}_{pq}, \partial^m E_{pq}) \sum_{\nu=1}^{4} Z_{\nu}(y)  F^{(0,j)}_{\nu}(0,y)  \ ,\label{271g}
\end{align}
with only a single factor of $F^{(0,j)}_{\nu}$ on the right-hand side, where $\tilde{{\cal R}}_j(g^{(n)}_{pq}, \partial^m E_{pq})$ are again polynomials in $g^{(n)}$ functions but also involve rational functions of prime forms and their derivatives. The appearance of prime forms can be seen from $F(x,2y) = \frac{ E(x+z_{AB}) }{E(x)E(z_{AB})}$ as well as the factors of $F^{(0,k_1+k_2-l)}(x,2 y)$ on the right-hand side of (\ref{217f}). Note that the divergent case of $F^{(0,j)}_{\nu}(0,y) $ at $j=0$ is absent since the sums over $ F^{(0,l)}_{\nu}(0,\pm y)$ on the right-hand side of (\ref{217f}) start at $l=1$.

While degenerate spin sums $\ssuma{x_1,x_2,\dots,x_M}{-}$ cannot be cast into the intermediate form (\ref{eq;firstly}), the
merging of $\nu$-dependent factors via (\ref{217c}) and (\ref{217d}) also leads to expressions of the form (\ref{271g}): The additional merging identity (\ref{217d}) special to degenerate spin sums involves no spin-structure dependence other than $F^{(0,j)}_{\nu}(0,y)$ on its right-hand side.


\subsubsection{Leftover spin sums}
\label{step3}

As detailed in appendix \ref{proof2}, the leftover spin sums
\beq
M_j \equiv \sum_{\nu=1}^{4} Z_{\nu}(y)  F^{(0,j)}_{\nu}(0,y)
\label{leftover}
\eeq
in (\ref{271g}) with $j\geq 1$ evaluate to a factor of $E_{AB}=E(2y)$ accompanied by combinations of Weierstrass functions
\begin{align}
\wp(z) \equiv -\partial^2 \log \theta_1(z) + \frac{ \theta_1'''(0) }{3 \theta_1'(0)} 
= (f^{(1)}(z))^2 - 2 f^{(2)}(z)
\label{defweier}
\end{align}
and their derivatives such as
\begin{align}
M_1 &= 0\ , \ \ \ \ M_2 = 4 E_{AB}\ , \ \ \ \ M_3 = 0\ ,\ \ \ \ M_4 = 48 E_{AB} \wp(2y) \label{stf1} \\
M_5 &= 240 E_{AB} \wp'(2 y) \ , \ \ \ \ M_6 = 288 E_{AB} ( 2\wp(2y)^2 +  3 \wp''(2y))  \ . \notag
\end{align}
By (\ref{defweier}), the Weierstrass functions can always be written in terms of Eisenstein series ${\rm G}_k$ and $f^{(1)},f^{(2)}$ at argument $z_{AB}=2y$. For $M_{j\geq 6}$, additional simplifications arise from the differential equations of the Weierstrass function, e.g.
\begin{align}
M_6 &= 2880 E_{AB} (2 \wp(2y)^2 - 9 {\rm G}_{4} ) \ .  \label{stf2} 
\end{align}
The general systematics of $M_j$ as well as additional explicit examples can be found in appendix \ref{proof2}. We emphasize that the prime form and the Weierstrass functions in the final expressions for (\ref{leftover}) were found to occur at argument $2y=z_A-z_B$.


\subsubsection{Cleaning up the prime forms}
\label{step4}

In the simplest cases, the rational dependence of $\tilde{{\cal R}}_j$ in (\ref{271g}) on prime forms cancels the factor of $E_{AB}$ from $M_j$ and additional ratios of prime forms accompanying a given spin sum $\ssuma{\ldots}{\ldots}$. Whenever this is not manifestly the case, one can apply the following identities (see appendix \ref{proof3} for the proof) to compensate the prime forms in $\tilde{{\cal R}}_j$,
\begin{align}\label{eq;clean}
&\frac{\partial^{m+n-1}}{\partial z ^{m+n-1}}\underset{j,k=1,\dots,n}{\det} [zF(x_j-y_k,z)]\bigg\vert_{z=0} =\notag\\
&\left(\begin{array}{c}
m + n-1\\
n-1
\end{array}
\right)
\sum_{l=0}^{m}
\frac{m!}{l!}
F^{(0,l)}\bigg(\sum_{p=1}^{n-1}(x_p- y_{p}), x_{n}-y_{n}\bigg)  \\
&\times\prod_{q=2}^{n-1} F\bigg(\sum_{r=1}^{q-1}(x_{r}- y_{r}), x_{q}-y_{q}\bigg)
g^{(m-l)}(x_n-y_n)\frac{\prod^n_{j<k} E(x_j,x_k) E(y_k,y_j)}{\prod^n_{j\ne k} E(x_j,y_k)}  \ , \notag
\end{align}
with $x_1,x_2,\ldots,x_n \in \mathbb C$ and $y_1,y_2,\ldots,y_n  \in \mathbb C$ subject to  $x_j\neq y_k \ \forall \ j,k=1,2,\ldots,n$, and $n\geq 2$. The determinant on the left-hand side refers to the $n\times n$ matrix defined by its entries $zF(x_j-y_k,z) = \sum_{l=0}^{\infty} z^l g^{(l)}(x_j - y_k)$. The simplest choice $m=0, \ n=2$ with $(x_1,x_2,y_1,y_2) = (z_i,z_A,z_j,z_B)$ specializes (\ref{eq;clean}) to
\begin{align}
g^{(1)}(z_{ij}) + g^{(1)}(z_{AB}) - g^{(1)}(z_{iB}) - g^{(1)}(z_{Aj})
&= \frac{ E_{iA}E_{jB}}{ E_{iB} E_{jA}} F(z_{ij},z_{AB}) \ ,
\label{clean;sp}
\end{align}
which will be applied to a concrete spin sum in section \ref{sec;5ptpt}.


\subsubsection{Comments}
\label{step5}

One can also mix the steps \ref{step1} and \ref{step2} to reduce the factors of $F_\nu$ by means of
\begin{align}
&F^{(0,k_1)}_{\nu}(x_1,y) F^{(0,k_2)}_{\nu}(x_2,-y)
= \sum_{l=0}^{k_2} (-1)^{k_2-l}\left(\begin{array}{c} \! k_2 \! \\  \!  l  \!  \end{array}\right) F^{(0,l)}_{\nu}(x_1{+}x_2,-y) F^{(0,k_1+k_2-l)}(x_1,2 y)\notag\\
&\ \ \ \ \ \ \ \ \ \ \ \ \ \ \ \ \ \ \ \ \ \ \ \ \ - \sum_{l=0}^{k_1} (-1)^{k_2} \left(\begin{array}{c}k_1\\l\end{array}\right)
F^{(0,l)}_{\nu}(x_1{+}x_2,y) F^{(0,k_1+k_2-l)}(-x_2,2 y) \label{eq;c124}
\end{align}
with $x_1+x_2 \ne 0$ which can be proven along the same lines as (\ref{217c}). This can yield alternative representations of the rational functions ${\cal R}_{k_1,k_2}$ in (\ref{eq;firstly}) and different situations for step \ref{step4}. 

\subsection{Worked out examples}
\label{sec;52}

While the procedure of the previous section can be applied to evaluate spin sums of arbitrary multiplicity,
we shall now present its simplest applications arising in three- to five-point amplitudes. In order to compactly
track intermediate steps of the subsequent calculations, we slightly generalize the notation in \eqref{standform} to
\begin{align}\label{eq;newnot}
\ssuma{\overset{(k_{1})}{x_1},\overset{(k_{2})}{x_2},\dots,\overset{(k_{M})}{x_M}}{\overset{(k_{M+1})}{x_{M+1}},\overset{(k_{M+2})}{x_{M+2}},\dots, \overset{(k_{N})}{x_N}} =
\sum_{\nu=1}^4 Z_\nu(y)   \prod_{i=1}^{M}F^{(0,k_i)}_{\nu}(x_i,y)   \prod_{j=M+1}^{N}   F^{(0,k_j)}_{\nu}(x_j,-y)\  .
\end{align}

\subsubsection{Three points}

Given that the spin sum in the fermionic two-point amplitude vanishes by \eqref{simplst}, the simplest non-trivial examples of the procedure in section \ref{sec;51} arise from three-point amplitudes involving one boson and two fermions. The spin sums in the relevant correlators \eqref{eq;330} and \eqref{eq;332} turn out to vanish: either by \eqref{217f} at $k_1=k_2=0$ or by the special case $F_\nu(x,y) F_{\nu}(-x,y)= \partial g^{(1)}(x) - F_\nu^{(0,1)}(0,y)$ of \eqref{217d},
\begin{subequations}\label{eq;general3}
	\begin{align}
	\ssuma{x}{-x} &= 0\label{eq;general1}\\
	\ssuma{x,-x}{-}  &= -M_{1} = 0 \ . \label{eq;general2}
	\end{align}
\end{subequations}

\subsubsection{Four points}
\label{sec;4pt}

The four-point amplitude with two fermions is built from the correlators \eqref{237a} and \eqref{mostex}, where the former vanishes by \eqref{eq;general3}. The first nonvanishing spin sums arise from the correlator \eqref{mostex} involving the following inequivalent topologies
\begin{align}\label{eq;4ptsums}
\ssuma{x_{1}, x_{2}, -x_{1} - x_{2}}{-},\ \ \ \ssuma{x_{1}, x_{2}}{-x_{1}-x_{2}}\ .
\end{align}
The specialization of \eqref{217c} to $(k_1,k_2) = (0,0)$ with $z+w \ne 0$ yields
\beq
\ssuma{\vec{x}_1,z,w}{\vec{x}_2} =  \left(g^{(1)}(z) +g^{(1)}(w)\right) \ssuma{\vec{x}_1,z+w}{\vec{x}_2}-\ssuma{\vec{x}_1,\overset{(1)}{z+w}}{\vec{x}_2} \ , \label{eq;4ptid1}
\eeq
where we use the notation of \eqref{eq;newnot}, and $\vec{x}_1,\vec{x}_2$ denote arbitrary (and possibly empty) collections of additional entries. The instances of the second term relevant to (\ref{eq;4ptsums}) follow from \eqref{217f} and the $(k_1,k_2)=(0,1)$ instance of \eqref{217d},
\begin{subequations}\label{eq;4ptids}
	\begin{align}
	\ssuma{\overset{(1)}{z}}{-z} &= - F(z,2y) M_1(y) = 0 \label{eq;4ptid2}
	\\
	\ssuma{\overset{(1)}{z},-z}{-} &= -\frac{1}{2} M_2(y) = -2 E_{AB}  \ .\label{eq;4ptid3}
	\end{align}
\end{subequations}
By applying these identities to \eqref{eq;4ptsums}, one finds
\begin{subequations}\label{eq;426}
	\begin{align}
	\ssuma{x_{1},x_{2},-x_{1} - x_{2}}{-}
	&\overset{\eqref{eq;4ptid1}}{=}
	\left(g^{(1)}(x_{1}) +g^{(1)}(x_{2})\right)
	\ssuma{x_{1}+x_{2},-x_{1}- x_{2}}{-}
	\notag\\
	&\phantom{\overset{\eqref{eq;4ptid1}}{=}\ }
	- \ssuma{ \overset{(1)}{x_{1} + x_{2}},-x_{1} -x_{2}}{-}\overset{\eqref{eq;4ptid3}\&\eqref{eq;general3}}{=} 2 E_{AB}
	\label{eq;426aa}\\
	\ssuma{x_{1},x_{2}}{-x_{1}-x_{2}}
	&\overset{\eqref{eq;4ptid1}}{=}
	\left(g^{(1)}(x_{1}) +g^{(1)}(x_{2})\right)
	\ssuma{x_{1}+x_{2}}{-x_{1}-x_{2}}
	\notag\\
	&\phantom{\overset{\eqref{eq;4ptid1}}{=}\ }
	- \ssuma{ \overset{(1)}{x_{1}+x_{2}}}{-x_{1}-x_{2}}
	\overset{\eqref{eq;4ptid2}\& \eqref{eq;general3}}{=} 0\ .
	\label{eq;426bb}
	\end{align}
\end{subequations}

\subsubsection{Five points}
\label{sec;5ptpt}

As will be shown in this section, the five-point correlator given in appendix \ref{sec;moreD}
yields non-trivial dependence on the punctures $\sim g^{(1)}_{ij}$ after spin sums. We compute the following spin sums
\begin{align}\label{eq;5ptsums}
&\ssuma{x_{1},x_{2},x_{3},x_4}{-}\ , \ \ \  \
\ssuma{x_{1},x_{2},x_{3}}{ x_4} \ , \ \ \ \
\ssuma{x_{1},x_{2}}{ x_{3},x_4} \ , \ \ \ \ \sum_{j=1}^4 x_j=0
\end{align}
and later on infer $\ssuma{x_{1},x_{2}, -x_{2}}{-x_{1}}, \ \ssuma{x_{1}, x_{2}}{-x_{1}, -x_{2}}, \ \ssuma{x_{1},-x_{1}}{ x_{2},-x_{2}}$ as well as $\ssuma{x_{1},-x_{1},x_{2},-x_2}{-}$ as degenerate cases for some $x_i \rightarrow -x_j$. As a starting point, we generalize \eqref{eq;4ptid1} by the following useful corollaries of (\ref{217c}) and (\ref{217d}) with $z,w \ne 0$,
\begin{subequations}\label{eq;5ptids}
	\begin{align}
	\ssuma{\vec{x}_1,\overset{(1)}{z},w}{\vec{x}_2}
	&=
	\left(g^{(2)}(z)-g^{(2)}(w) \right)
	\ssuma{\vec{x}_1,z+w}{\vec{x}_2}
	\label{eq;5ptid1}\\
	& \ \ +
	g^{(1)}(w) \ssuma{\vec{x}_1,\overset{(1)}{z+w}}{\vec{x}_2}
	-
	\frac{1}{2} \ssuma{\vec{x}_1,\overset{(2)}{z+w}}{\vec{x}_2} \notag \\
	\ssuma{\vec{x}_1,z,-z}{\vec{x}_2}
	&=
	\partial g^{(1)}(z) \ssuma{\vec{x}_1}{\vec{x}_2}
	-
	\ssuma{\vec{x}_1,\overset{(1)}{0}}{\vec{x}_2} \ ,\label{eq;5ptid1-0}
	\end{align}
\end{subequations}
where $\vec{x}_1,\vec{x}_2$ denote arbitrary (and possibly empty) collections of additional entries. Moreover, \eqref{217d} and \eqref{217f} along with $M_1=M_3=0$ and $M_2= 4E_{AB}$ imply that
\begin{subequations}\label{moreeq;5ptids}
	\begin{align}
	\ssuma{\overset{(1)}{z},\overset{(1)}{-z}}{-} &= \ssuma{\overset{(1)}{z}}{\overset{(1)}{-z}}  = 0\label{eq;5ptid2} \\
	\ssuma{\overset{(2)}{z}}{-z} &= -4 E_{AB} F(z,z_{AB}) \ .\label{eq;5ptid3}
	\end{align}
\end{subequations}
By repeatedly applying these identities to \eqref{eq;5ptsums}, one arrives at:
\begin{subequations} \label{many5pt}
	\begin{align}
	&\ssuma{x_{1},x_{2},x_{3},-x_{1}-x_{2}-x_{3}}{-}
	\notag\\
	&\overset{\eqref{eq;4ptid1}}{=}
	\left(g^{(1)}(x_{1}) +g^{(1)}(x_{2})\right)
	\ssuma{x_{1} + x_{2},x_{3},-x_{1}-x_{2}-x_{3}}{-}
	\notag\\
	&\phantom{\overset{\eqref{eq;4ptid1}}{=}\ }
	- \ssuma{ \overset{(1)}{x_{1} + x_{2}},x_{3},-x_{1}-x_{2}-x_{3}}{-}
	\notag\\
	&\overset{\eqref{eq;4ptid1}\&\eqref{eq;426}}{=}
	\left(g^{(1)}(x_{1}) +g^{(1)}(x_{2})\right)
	2 E_{AB}
	+
	\ssuma{\overset{(1)}{x_{1} + x_{2}},\overset{(1)}{-x_{1} - x_{2}}}{-}
	\notag\\
	&\phantom{\overset{\eqref{eq;426}\&\eqref{eq;4ptid1}}{=}}
	-\left( g^{(1)}(x_{3} ) + g^{(1)}(-x_{1} -x_{2}-x_{3}) \right) \ssuma{\overset{(1)}{x_{1} + x_{2}},-x_{1} - x_{2}}{-}
	\notag\\
	&
	\overset{\eqref{eq;4ptid3}\& \eqref{eq;5ptid2}}{=}
	2 E_{AB}
	\left(g^{(1)}(x_{1}) +g^{(1)}(x_{2}) + g^{(1)}(x_{3} ) + g^{(1)}(-x_{1} -x_{2}-x_{3})\right)
	\label{eq;primeAA}
	\end{align}
	\begin{align}
	&\ssuma{x_{1},x_{2},x_{3}}{ -x_{1}-x_{2}-x_{3}}
	\notag\\
	&\overset{\eqref{eq;4ptid1}}{=}
	\left(g^{(1)}(x_{1}) +g^{(1)}(x_{2})\right)
	\ssuma{x_{1} + x_{2},x_{3}}{ -x_{1}-x_{2}-x_{3}}
	- \ssuma{ \overset{(1)}{x_{1} + x_{2}},x_{3}}{ -x_{1}-x_{2}-x_{3}}
	\notag\\
	&
	\overset{\eqref{eq;426}\&\eqref{eq;5ptid1}}{=}
	-\left(
	g^{(2)}(x_{1} + x_{2}) - g^{(2)}(x_{3})\right)
	\ssuma{x_{1} + x_{2} + x_{3}}{- x_{1} - x_{2} - x_{3}}
	\notag\\
	&\phantom{\overset{\eqref{eq;426}\&\eqref{eq;5ptid1}}{=}}
	-g^{(1)}(x_{3})
	\ssuma{\overset{(1)}{x_{1} + x_{2} + x_{3}}}{-x_{1} - x_{2} - x_{3}}
	+\frac{1}{2}
	\ssuma{\overset{(2)}{x_{1} + x_{2} + x_{3}}}{-x_{1} - x_{2} - x_{3}}
	\notag\\
	&\overset{\eqref{eq;general1},\eqref{eq;4ptid2}\& \eqref{eq;5ptid3}}{=}
	-2 E_{AB}F(x_{1} + x_{2} + x_{3},z_{AB})\label{eq;prime}
	\end{align}
	\begin{align}
	&\ssuma{x_{1},x_{2}}{ x_{3},-x_{1}-x_{2}-x_{3}}
	\notag\\
	&
	\overset{\eqref{eq;4ptid1}}{=}
	\left(g^{(1)}(x_{1}) +g^{(1)}(x_{2})\right)
	\ssuma{x_{1} + x_{2}}{ x_{3},-x_{1}-x_{2}-x_{3}}
	- \ssuma{ \overset{(1)}{x_{1} + x_{2}}}{ x_{3},-x_{1}-x_{2}-x_{3}}
	\notag\\
	&
	\overset{\eqref{eq;4ptid1}\& \eqref{eq;426}}{=}
	-\left(
	g^{(1)}(x_{3}) +g^{(1)}(-x_{1}-x_{2}-x_{3})
	\right)
	\ssuma{\overset{(1)}{x_{1} + x_{2}}}{-x_{1}-x_{2}-x_{3}}
	\notag\\
	&\phantom{\overset{\eqref{eq;426}\& \eqref{eq;4ptid1}}{=}\ }
	+ \ssuma{\overset{(1)}{x_{1} + x_{2}}}{\overset{(1)}{-x_{1}-x_{2}-x_{3}}}
	\overset{\eqref{eq;4ptid2}\& \eqref{eq;5ptid2}}{=}0
	\label{eq;primeCC}
	\end{align}
\end{subequations}
Note that (\ref{eq;primeAA}) exemplifies the freedom to collectively redefine $g^{(n)} \rightarrow f^{(n)}$ within spin sums (\ref{standform}) since the non-meromorphic terms in $f^{(1)}(z)=g^{(1)}(z) + 2\pi i \frac{ \Im(z)}{\Im(\tau)}$ drop out from
\beq
\ssuma{x_{1},x_{2},x_{3},x_4}{-}=	2 E_{AB} \sum_{j=1}^4 f^{(1)}(x_{j}) \ , \ \ \ \ \sum_{j=1}^4 x_j=0 \ .
\eeq
Moreover, the possibility to eliminate prime forms as discussed in section \ref{step4}
applies to \eqref{eq;prime}: By virtue of (\ref{clean;sp}), one can simplify the following combination of
prime forms seen in the expression
\eqref{eq;moreexex} for the correlator $ \scor{ \prod_{j=1}^3\psi^{\mu_{j}} \psi^{ \nu_{j}}(z_j) S_{a} (z_A) S_{b}^\lambda(z_B)}_\nu$,
\begin{align}
\frac{E_{iB} E_{jA}}{ E_{AB}  E_{iA}  E_{jB} }
&\ssuma{z_{Bi},z_{kB},z_{jk}}{z_{ij}} = -2\frac{E_{iB} E_{jA}}{   E_{iA}  E_{jB} } F(z_{ji}, z_{AB}) \notag
\\
&=
-2
\left(
g^{(1)}(z_{ji}) + g^{(1)}(z_{AB}) - g^{(1)}(z_{jB}) - g^{(1)}(z_{Ai})
\right) \label{rewrit} \\
&=
-2
\left(
f^{(1)}(z_{ji}) + f^{(1)}(z_{iA}) + f^{(1)}(z_{AB})  + f^{(1)}(z_{Bj})
\right)  \ , \notag
\end{align}
where we have exploited the cancellation of $\Im(z_i)$ in the last step.

Finally, appropriate limits $x_i \rightarrow - x_j$ of (\ref{many5pt}) straightforwardly yield
\begin{align}
\ssuma{x_{1},x_{2}, -x_{2}}{-x_{1}} &= -2 E_{AB} F(x_{1},z_{AB})     \label{eq;primeBB} \\
\ssuma{x_{1}, x_{2}}{- x_{1}, - x_{2}} &= \ssuma{x_{1},-x_{1}}{ x_{2},-x_{2}} = \ssuma{x_{1},-x_{1},x_{2},-x_2}{-} = 0 \ , \notag
\end{align}
where the first line may be rewritten along the lines of (\ref{rewrit}).

Results for spin sums due to $n\geq 4$ insertions of $\psi^\mu \psi^\nu(z)$ will be given in the following subsections as well as
appendix \ref{sec;moremoremore}.

\subsection{Higher-multiplicity spin sums}
\label{sec;59}

As reviewed in section \ref{sec;sa5}, the elliptic functions in (\ref{thetaH}),
\beq
V_w(i_1,i_2,\ldots,i_n) \equiv V_w(z_{i_1i_2},z_{i_2i_3},\ldots,z_{i_{n-1}i_n} ,z_{i_n i_1})  \ ,
\label{thetaHprime}
\eeq
are a convenient language to compactly describe the spin sums over bosonic one-loop correlators.
They manifest doubly-periodicity and meromorphicity through the freedom to globally interchange
the functions $g^{(n)}$ and $f^{(n)}$ in their generating series. Accordingly, we will now rewrite the
results of the previous section and state various generalizations to higher multiplicity in terms of (\ref{thetaHprime}).

\subsubsection{Spin sum with all entries in the first line}

For spin sums of the form $\ssuma{z_{12},z_{23},\dots,z_{n1}}{-}$,
the results of (\ref{eq;general2}), (\ref{eq;426aa}) and (\ref{eq;primeAA}) are equivalent to
\begin{align}
\ssuma{z_{12},z_{21}}{-} &=  0 \notag
\\
E_{AB}^{-1}\, \ssuma{z_{12},z_{23},z_{31}}{-} &=  2 \label{pattern1} \\
E_{AB}^{-1}\,\ssuma{z_{12},z_{23},z_{34},z_{41}}{-} &= 2 V_1(1,2,3,4) \ ,\notag
\end{align}
and they generalize to
\begin{align}
E_{AB}^{-1}\,\ssuma{z_{12},z_{23},\dots,z_{51}}{-} &= 2 V_2(1,2,\dots,5) - 2 V_2(A,B)\notag\\
E_{AB}^{-1}\,\ssuma{z_{12},z_{23},\dots,z_{61}}{-} &= 2 V_3(1,2,\dots,6)-2 V_2(A,B) V_1(1,2,\dots,6)  + 2 \partial_{z_A} V_2(A,B) \notag\\
E_{AB}^{-1}\,\ssuma{z_{12},z_{23},\dots,z_{71}}{-} &= 2 V_4(1,2,\dots,7) - 2 V_2(A,B) V_2(1,2,\dots,7) \notag\\
&\  + 2 \partial_{z_A} V_2(A,B) V_1(1,2,\dots,7) - \frac{4}{3} \partial_{z_A}^{2} V_2(A,B) + 4 {\rm G}_4 \notag\\
E_{AB}^{-1}\,\ssuma{z_{12},z_{23},\dots,z_{81}}{-} &=  2 V_5(1,2,\dots,8) - 2 V_2(A,B) V_3(1,2,\dots,8)\notag\\
&\  + 2 \partial_{z_A} V_2(A,B) V_{2}(1,2,\dots,8) - \frac{4}{3} \partial_{z_A}^{2} V_2(A,B) V_{1}(1,2,\dots,8) \notag\\
&\ + \frac{2}{3} \partial_{z_A}^{3} V_2(A,B)
+4 {\rm G}_4 V_1(1,2,\dots,8)  \label{pattern2}  \\
E_{AB}^{-1}\,\ssuma{z_{12},z_{23},\dots,z_{91}}{-} &=
2 V_6(1,2,\dots,9)
- 2 V_2(A,B) V_4(1,2,\dots,9)\notag\\
&\   +2 \partial_{z_A} V_2(A,B) V_{3}(1,2,\dots,9) - \frac{4}{3} \partial_{z_A}^{2} V_2(A,B) V_{2}(1,2,\dots,9)\notag\\
&\ + \frac{2}{3} \partial_{z_A}^{3} V_2(A,B) V_{1}(1,2,\dots,9)
- \frac{4}{15} \partial_{z_A}^{4} V_{2}(A,B) \notag\\
&\   +4 {\rm G}_4 V_2(1,2,\dots,9) - 6 {\rm G}_4 V_{2}(A,B) + 18 {\rm G}_6 \ , \notag
\end{align}
with admixtures of the holomorphic Eisenstein series ${\rm G}_k$ defined in (\ref{thetaK}).

\subsubsection{Spin sum with entries in both lines}

In more general situations with entries in both lines of the spin sum, the results of (\ref{eq;426bb}), (\ref{eq;prime}) and (\ref{eq;primeCC}) can be aligned into
\begin{align}
\ssuma{z_{12},z_{23}}{z_{31}}  &= 0
\notag \\
\frac{E_{1A} E_{4B}}{E_{1B} E_{4A} E_{AB}} \ssuma{z_{12},z_{23},z_{34}}{z_{41}}
&=
2 V_{1}(1,B,A,4)
\label{pattern3} \\
\frac{E_{1A} E_{5B}}{E_{1B} E_{5A} E_{AB}} \ssuma{z_{12},z_{23},z_{34},z_{45}}{z_{51}}
&=
- 2 V_{2}(1,B,A,5,4,3,2) + 2 V_{2}(1,5,4,3,2)
\notag\\
&\ \ \ \, + 2 V_{2}(A,B)
\ ,  \notag
\end{align}
as well as
\begin{align}
\ssuma{z_{12},z_{23}}{z_{34},z_{41}}  &= 0  \label{pattern4}
\\
\frac{E_{1A} E_{4B}}{E_{1B} E_{4A} E_{AB}}\ssuma{z_{12},z_{23},z_{34}}{z_{45},z_{51}}
&=
2 V_{2}(5,1,B,A,4) - 2 V_{2}(A,B) \ . \notag
\end{align}
Note that the expressions for $\ssuma{z_{12},z_{23},\dots,z_{n1}}{-}$ in (\ref{pattern1}) and (\ref{pattern2}) exhibit stable patterns in the coefficients of $V_{n-3} (1,2,\dots,n)$ and $\partial_{z_A}^{k} V_2(A,B) V_{n-k-5}(1,2,\ldots,n)$,
\begin{align}
&E_{AB}^{-1}\,\ssuma{z_{12},z_{23},\dots,z_{n1}}{-} =
2 V_{n-3}(1,2,\dots,n)
- 2 V_2(A,B) V_{n-5}(1,2,\dots,n)\notag\\
&\ \ \ \ \ \ \ \ \ \ \ \ \ \ \ \ \ \ \ \ \ \ \ \ \ \ \ \ \ \ \ \ \ \ \ \ \ \ \ +2 \partial_{z_A} V_2(A,B) V_{n-6}(1,2,\dots,n)+ \ldots
\end{align}
with higher derivatives of $V_2(A,B)$ and Eisenstein series in the ellipsis. It would be interesting to investigate generalizations of (\ref{pattern3}) in view of similar systematics.

\subsection{Examples of spin-summed correlators}
\label{sec;53}

In this section, we assemble the expressions for various spin-summed correlators in fermionic one-loop amplitudes.

\subsubsection{Two unexcited spin fields}
\label{sec;53a}

The vanishing of the spin sums in (\ref{simpex}) to (\ref{237a}) immediately propagates to
\begin{align}
\sum_{\nu=1}^{4}(-1)^{\nu+1} 	\scor{   S_a(z_A) S^b(z_B)}_\nu &= 0 \notag \\
\sum_{\nu=1}^{4}(-1)^{\nu+1} 	\scor{ \psi^{\mu_1\nu_1}(z_1) S_a(z_A) S^b(z_B)}_\nu &= 0\label{237d} \\
\sum_{\nu=1}^{4}(-1)^{\nu+1} 	\scor{ \psi^{\mu_1\nu_1}(z_1) \psi^{\mu_2\nu_2}(z_2) S_a(z_A) S^b(z_B)}_\nu &= 0 \ . \notag
\end{align}
The first non-vanishing spin sums occur in the five-point correlator (\ref{5ptco}) such that
\begin{align}
&\sum_{\nu=1}^{4}(-1)^{\nu+1}  \scor{
	\psi^{\mu_{1}}\psi^{\nu_{1}}(z_{1})
	\psi^{\mu_{2}}\psi^{\nu_{2}}(z_{2})
	\psi^{\mu_{3}}\psi^{\nu_{3}}(z_{3})
	S_{a}(z_A)S^b(z_B) }_\nu \notag \\
&= (\gamma^{\mu_1\nu_1\mu_2\nu_2\mu_3\nu_3})_{a}{}^{b} \ h^{(0)}_{\emptyset} \ + \
\eta^{\nu_1\mu_2} (\gamma^{\mu_1\nu_2\mu_3\nu_3})_{a}{}^{b} \ h^{(0)}_{[12]}
\notag\\
&\ \ \ +\eta^{\mu_1\nu_2} \eta^{\nu_1\mu_2} (\gamma^{\mu_3\nu_3})_{a}{}^{b} \ h^{(0)}_{(12)}  +\eta^{\nu_1\mu_2}\eta^{\nu_2\mu_3}(\gamma^{\mu_1\nu_3})_{a}{}^{b} \ h^{(0)}_{12,23}
\notag\\
&\ \ \ +\ \eta^{\nu_1\mu_2}\eta^{\nu_2\mu_3}\eta^{\mu_1\nu_3} \delta_{a}^{b} \ h^{(0)}_{[123]} + {\rm permutations}
\label{237c}
\end{align}
with
\begin{align}
h^{(0)}_{\emptyset} = h^{(0)}_{12,23} = - h^{(0)}_{(12)} = \frac12  \ ,\ \ \ \ \ \ h^{(0)}_{[12]} = h^{(0)}_{[123]} = 0 \ ,
\label{237e}
\end{align}
or equivalently
\begin{align}
&\sum_{\nu=1}^{4}(-1)^{\nu+1}  \scor{
	\psi^{\mu_{1}}\psi^{\nu_{1}}(z_{1})
	\psi^{\mu_{2}}\psi^{\nu_{2}}(z_{2})
	\psi^{\mu_{3}}\psi^{\nu_{3}}(z_{3})
	S_{a}(z_A)S^b(z_B) }_\nu \notag \\
&= \frac{1}{2} \Big\{ (\gamma^{\mu_1\nu_1\mu_2\nu_2\mu_3\nu_3})_{a}{}^{b}
- \big[ (\eta^{\mu_1\nu_2} \eta^{\nu_1\mu_2} -\eta^{\mu_1\mu_2} \eta^{\nu_1\nu_2})(\gamma^{\mu_3\nu_3})_{a}{}^{b} + {\rm cyc}(1,2,3) \big] \cr
& \ \ \  \ \ \ \ \
+ \big[ \eta^{\mu_2 [\nu_1}(\gamma^{\mu_1][\nu_3})_{a}{}^{b}  \eta^{\mu_3]\nu_2}
-  \eta^{\nu_2 [\nu_1}(\gamma^{\mu_1][\nu_3})_{a}{}^{b}  \eta^{\mu_3]\mu_2}
+ {\rm cyc}(1,2,3) \big] \Big\} \ .
\label{equiva1}
\end{align}
The spin sums in the corresponding six-point correlator evaluate to
\begin{align}
&4\, \sum_{\nu=1}^{4}(-1)^{\nu+1} \scor{
	\psi^{\mu_{1}}\psi^{\nu_{1}}(z_{1})
	\psi^{\mu_{2}}\psi^{\nu_{2}}(z_{2})
	\psi^{\mu_{3}}\psi^{\nu_{3}}(z_{3})
	\psi^{\mu_{4}}\psi^{\nu_{4}}(z_{4})
	S_{a}(z_A)S^{b}(z_B)
}_\nu
\notag\\
&=(\gamma^{\mu_1\nu_1\mu_2\nu_2\mu_3\nu_3\mu_4\nu_4})_{a}{}^{b}\ h^{(1)}_{\emptyset} +\eta^{\nu_1\mu_2}(\gamma^{\mu_1\nu_2\mu_3\nu_3\mu_4\nu_4})_{a}{}^{b}\ h^{(1)}_{[12]}
\notag\\
&\phantom{=\ }+\ \eta^{\mu_1\nu_2}\eta^{\nu_1\mu_2}(\gamma^{\mu_3\nu_3\mu_4\nu_4})_{a}{}^{b}\ h^{(1)}_{(12)}
+
\eta^{\nu_1\mu_2}\eta^{\nu_2\mu_3}(\gamma^{\mu_1\nu_3\mu_4\nu_4})_{a}{}^{b} \ h^{(1)}_{12,23} \notag\\
&\phantom{=\ }+\eta^{\nu_1\mu_2}\eta^{\nu_3\mu_4}(\gamma^{\mu_1\nu_2\mu_3\nu_4})_{a}{}^{b} \ h^{(1)}_{[12],[34]}
+\eta^{\mu_1\nu_2}\eta^{\nu_1\mu_2}\eta^{\nu_3\mu_4}(\gamma^{\mu_3\nu_4})_{a}{}^{b} \ h^{(1)}_{(12),[34]}
\notag\\
&\phantom{=\ }+\eta^{\nu_1\mu_2}\eta^{\nu_2\mu_3}\eta^{\mu_1\nu_3}(\gamma^{\mu_4\nu_4})_{a}{}^{b} \ h^{(1)}_{[123]}
+\eta^{\nu_1\mu_2}\eta^{\nu_2\mu_3}\eta^{\nu_3\mu_4}(\gamma^{\mu_1\nu_4})_{a}{}^{b} \ h^{(1)}_{12,23,34}
\notag\\
&\phantom{=\ }+ \eta^{\mu_1\nu_2}\eta^{\nu_1\mu_2}\eta^{\nu_3\mu_4}\eta^{\mu_3\nu_4} \delta_{a}^{b} \ h^{(1)}_{(12),(34)} + \eta^{\nu_1\mu_2}\eta^{\nu_2\mu_3}\eta^{\nu_3\mu_4}\eta^{\mu_1\nu_4} \delta_{a}^{b} \ h^{(1)}_{(1234)} \notag \\
&\phantom{=\ }
+ {\rm permutations}\label{237f}
\end{align}
with doubly-periodic functions $h^{(1)}_{\ldots} \equiv h^{(1)}_{\ldots}(z_j,z_A,z_B) $ given by
\begin{subequations} \label{237g}
	\begin{align}
	h^{(1)}_{\emptyset}(z_j,z_A,z_B)\  &=\   \sum_{i=1}^4 V_1(i,A,B)\\
	h^{(1)}_{[12]}(z_j,z_A,z_B)\  &=\   V_{1}(1,2,A,B) - V_{1}(2,1,A,B)   \\
	h^{(1)}_{(12)}(z_j,z_A,z_B)\  &=\  \sum_{i=1}^{2} V_1(i,A,B)-\sum_{i=3}^{4}V_1(i,A,B)\\
	h^{(1)}_{12,23}(z_j,z_A,z_B)\  &=\ - V_{1}(2,A,4,B)\\
	h^{(1)}_{[12],[34]}(z_j,z_A,z_B)\  &=\ 0\\
	h^{(1)}_{(12),[34]}(z_j,z_A,z_B)\  &=\ -  V_1(3,4,A,B) - V_1(3,4,B,A)\\
	h^{(1)}_{[123]}(z_j,z_A,z_B)\  &=\ -2 V_1(1,2,3)\\
	h^{(1)}_{12,23,34}(z_j,z_A,z_B)\  &=\  \sum_{i=1}^{3}  \bigg[V_{1}(i,i{+}1,A,B)+V_{1}(i,i{+}1,B,A)\bigg]\\
	h^{(1)}_{(12),(34)}(z_j,z_A,z_B)\  &=\ - \sum_{i=1}^{4} V_1(i,A,B)\\
	h^{(1)}_{(1234)}(z_j,z_A,z_B)\  &=\ \frac{1}{2} \big[   V_1(1,2,A,B)-V_1(1,2,B,A) + {\rm cyc}(1,2,3,4) \big]
	\ .
	\end{align}
\end{subequations}
The analogous seven-point correlator can be found in appendix \ref{sec;moremoremoreA}, and the sum over permutations in (\ref{237f}) can be reconstructed from section \ref{sec;sa4}.

\subsubsection{One excited spin field}
\label{sec;53b}

Again, the vanishing of the relevant spin sums leads to
\begin{align}
\sum_{\nu=1}^{4}(-1)^{\nu+1} \scor{ \psi^{\mu_{1}}\psi^{\nu_{1}}(z_{1})S_{a}(z_A)S^{\lambda}_{b}(z_B)}_\nu = 0 \ ,
\end{align}
resulting in a vanishing three-point amplitude. The first non-vanishing spin-summed correlator with an excited spin field requires two insertions of $\psi^{\mu_{i}}\psi^{\nu_{i}}(z_{i})$,
\begin{align}
&\frac{1}{\sqrt{2}} \sum_{\nu=1}^{4}(-1)^{\nu+1} \scor{
	\psi^{\mu_{1}}\psi^{\nu_{1}}(z_{1})
	\psi^{\mu_{2}}\psi^{\nu_{2}}(z_{2})
	S_{a}(z_A)S^{\lambda}_{b}(z_B)
}_\nu  \label{psumA}  \\
&= (\gamma^{\mu_1\nu_1\mu_2})_{ab} \eta^{ \nu_2\lambda} H^{(0)}_{\emptyset}
+\eta^{\nu_1\mu_2}(\gamma^{\mu_1})_{ab}\eta^{ \nu_2 \lambda} H^{(0)}_{12} + {\rm permutations} \ , \notag
\end{align}
with
\begin{align}
H^{(0)}_{\emptyset} =  H^{(0)}_{12}= - \frac12 \ ,
\end{align}
or equivalently
\begin{align}
& \sum_{\nu=1}^{4}(-1)^{\nu+1} \scor{
	\psi^{\mu_{1}}\psi^{\nu_{1}}(z_{1})
	\psi^{\mu_{2}}\psi^{\nu_{2}}(z_{2})
	S_{a}(z_A)S^{\lambda}_{b}(z_B)
}_\nu \notag \\
&=  \frac{1}{\sqrt{2}} \big[ (\gamma^{\mu_1\nu_1[\mu_2})_{ab} \eta^{ \nu_2]\lambda}
+(\gamma^{[\mu_1})_{ab} \eta^{\nu_1][\mu_2} \eta^{ \nu_2] \lambda}  +(1\leftrightarrow 2) \big]   \label{equiva2}  \\
&=   \frac{1}{\sqrt{2}} \big[\eta^{\lambda [\nu_2} (\gamma^{\mu_2]} \gamma^{\mu_1 \nu_1})_{ab} + (1\leftrightarrow 2) \big]   \ . \notag
\end{align}
The corresponding five-point correlator
\begin{align}
&2\sqrt{2}\sum_{\nu=1}^{4}(-1)^{\nu+1} \scor{
	\psi^{\mu_{1}}\psi^{\nu_{1}}(z_{1})
	\psi^{\mu_{2}}\psi^{\nu_{2}}(z_{2})
	\psi^{\mu_{3}}\psi^{\nu_{3}}(z_{3})
	S_{a}(z_A)S^{\lambda}_{b}(z_B)
}_\nu
\notag\\
& = (\gamma^{\mu_1 \nu_1 \mu_2 \nu_2 \mu_3})_{ab}\eta^{\lambda \nu_3} \  H^{(1)}_{\emptyset}  + \eta^{ \nu_1 \mu_2}(\gamma^{\mu_1 \nu_2 \mu_3})_{ab}\eta^{\lambda \nu_3}  \ H^{(1)}_{[12]}\notag\\
&\ \ \ + \eta^{\nu_1 \mu_2}\eta^{\mu_1  \nu_2}(\gamma^{ \mu_3})_{ab}\eta^{\lambda \nu_3} \ H^{(1)}_{(12)} + (\gamma^{ \mu_1 \mu_2 \nu_2 })_{ab}\eta^{\nu_1 \mu_3}\eta^{\lambda \nu_3} \ H^{(1)}_{13} \notag\\
&\ \ \ + \eta^{ \mu_1 \mu_2}(\gamma^{ \nu_2 })_{ab}\eta^{\nu_1 \mu_3}\eta^{\lambda \nu_3}\ H^{(1)}_{12,13} + {\rm permutations}
\label{psumB}
\end{align}
involves the following doubly-periodic functions $H^{(1)}_{\ldots} \equiv H^{(1)}_{\ldots}(z_j,z_A,z_B) $:
\begin{subequations}
	\begin{align}
	H^{(1)}_{\emptyset}(z_j,z_A,z_B) &=- V_1(1, A, B) - V_1(2, A, B) \\
	H^{(1)}_{[12]}(z_j,z_A,z_B) &= -V_1(1, 2, A, B) - V_1(1, 2, B, A)\\
	H^{(1)}_{(12)}(z_j,z_A,z_B) &=-V_1(1, 2, A, B)+V_1(1, 2, B, A)\\
	H^{(1)}_{13}(z_j,z_A,z_B) &= -V_1(1, 3, A, 2, B)-V_1(1, 3, B, 2, A)-V_1(2, A, 3, B)\\
	H^{(1)}_{12,13}(z_j,z_A,z_B) &= - 2 V_1(1, 2, A, B, 3)-V_1(1, A, 2, B)\ .
	\end{align}
\end{subequations}
The analogous six-point correlator is presented in appendix \ref{sec;moremoremoreB}, and the sum over permutations in (\ref{psumA}) and (\ref{psumB}) can be reconstructed from (\ref{moreex}) and (\ref{anotherex}).

Note that (\ref{equiva1}) and (\ref{psumB}) yield an expression for the worldsheet integrand of the five-point amplitude (\ref{eq;sa}) in terms of the $f^{(1)}_{ij}$ functions with $i,j\in\{1,2,3,A,B\}$. It would be interesting to relate its factorization properties to the general considerations of \cite{Witten:2012bh, Sen:2014pia} on the distributions of superghost picture numbers at the boundary of (super-)moduli space.


\section{Conclusion and outlook}

In this work, we have studied the correlation functions of two fermionic and any number of bosonic vertex operators on the torus, with particular emphasis on the cancellations between different spin structures reflecting spacetime supersymmetry. These correlators form the worldsheet integrands for the respective massless one-loop amplitudes of the open RNS superstring, and their double copy yields closed-string amplitudes involving up to two Ramond--Ramond forms, gravitinos or dilatinos.

Among other things, the resulting fermionic RNS amplitudes are useful to test the equivalence with the pure-spinor formalism in more advanced situations. For example, the explicit correlators in section \ref{sec;53} and appendix \ref{sec;moremoremoreB} are suitable for comparison with the five- \cite{Mafra:2012kh} and six-point \cite{Mafra:2016nwr} results in pure-spinor superspace. 

Moreover, the $\tau \rightarrow i \infty$ limit of the present results extends the RNS ambitwistor-string setup \cite{Geyer:2015bja, Geyer:2015jch} to CHY formulae for one-loop SYM amplitudes with external fermions and the corresponding supergravity amplitudes. In particular, the tensor structure of our correlators at $\tau \rightarrow i \infty$ can be converted to explicit and local BCJ numerators using the techniques of \cite{He:2017spx}. Finally, we hope that our results are useful to study the forward-limit relations between ambitwistor-string correlators at different loop orders and the application of the gluing operators in \cite{Roehrig:2017gbt}.

While a detailed investigation of the resulting string and field-theory amplitudes is relegated to the future, the major novelties of this work are
\begin{itemize}
	\item[(i)] the one-loop correlation functions involving one excited spin field from the fermion vertex in the $+\frac{1}{2}$ picture and any number of Lorentz currents
	\item[(ii)] an algorithmic method to systematically perform and simplify the sum over spin structures for the one-loop integrand of two-fermion amplitudes
\end{itemize}
The $n$-point correlator (i) can be found in section \ref{sec;loopX}, and the mathematical techniques for the spin sums (ii) are presented in section \ref{sec;5}, see in particular subsection \ref{sec;53} and appendix \ref{sec;moremoremore} for explicit $n\leq 6$-point expressions.

A mild generalization of the techniques which led to the main results (i) and (ii) can be applied to one-loop correlators involving any number of fermion pairs. And we expect that several of the mathematical tools developed in this work are helpful for higher-genus amplitudes, for instance to extend the two-loop spin sums of \cite{DHoker:2005vch, Tsuchiya:2017joo} for bosonic external states to fermionic amplitudes.

On the one hand, the pure-spinor formalism bypasses the spin sums, gathers all component amplitudes into supersymmetric expressions and held the key to the first explicit three-loop calculation \cite{Gomez:2013sla}. On the other hand, the form of the RNS spin sums at genus one given in \cite{Tsuchiya:1988va, Broedel:2014vla} pinpointed the ubiquity of doubly-periodic functions $f^{(n)}(z,\tau)$ (see section \ref{sec;sa5}) in multiparticle correlators which is crucial to construct the latter from an ansatz in both RNS- and pure-spinor variables. Hence, we expect that explicit control over RNS spin sums provides valuable inspiration for the design of multiparticle correlators at higher genus and appropriate generalizations of the $f^{(n)}(z,\tau)$ functions.

Another kind of follow-up question concerns the extension of the present results to string compactifications with reduced supersymmetry, see e.g.\ \cite{Blumenhagen:2006ci} for a review. Higher-genus correlators involving two spin fields and an arbitrary number of NS fermions were found to be robust under dimensional reduction \cite{Hartl2011}, and the same is expected for excited spin fields, see \cite{Schlotterer:2010kk} for tree-level evidence. It remains to incorporate the fingerprints of the compactification geometry on the fermionic vertex operators where universal statements for a given number of supersymmetries can be made from \cite{Friedan:1985ge, Banks:1987cy, Banks:1988yz, Ferrara:1989ud}.

For bosonic one-loop amplitudes, the spin sums in half-maximally and quarter-maximally supersymmetric setups could be identified as specializations of maximally supersymmetric spin sums with two additional legs \cite{Berg:2016wux}. Upon extrapolation to external fermions, the spin-summed five- and six-point correlators in the maximally supersymmetric setup of this work should admit a similar map to spin summed three- and four-point correlators with reduced supersymmetry.

The resulting expressions for fermionic one-loop RNS amplitudes with reduced supersymmetry will provide helpful cross-checks and guidance to supersymmetrize their bosonic counterparts \cite{Bianchi:2006nf, Bianchi:2015vsa,Berg:2016wux}: They are important in comparing RNS results with one-loop amplitudes in the hybrid formalism with four or eight supercharges manifest \cite{Berkovits:1994wr, Berkovits:1996bf, Berkovits:1999im, Berkovits:1999in}. While one-loop hybrid amplitudes with
maximally supersymmetric multiplets in the loop have been computed in \cite{Berkovits:2001nv}, it remains to derive their generalizations to spectra with reduced supersymmetry.

\acknowledgments

We are grateful to Arnab Rudra and Massimo Taronna for enlightening discussions as well as Song He and Carlos Mafra for valuable comments on the draft. OS would like to thank Johannes Br\"odel, Carlos Mafra and Nils Matthes for fruitful collaboration on related topics. The research of OS was supported in part by Perimeter Institute for Theoretical Physics. Research at Perimeter Institute is supported by the Government of Canada through the
Department of Innovation, Science and Economic Development Canada and by the Province
of Ontario through the Ministry of Research, Innovation and Science.

\appendix

\section{OPEs and bosonization}
\label{APPOPE}

The bosonization technique discussed in section \ref{sec;21} renders the OPEs among $\psi^{\mu}$ and spin fields $S_a,S^b$ of $SO(D=2n)$ accessible to free-field methods. For example, (\ref{1,18}) and (\ref{1,38}) give rise to
\begin{align}\label{eq;bosonex}
\psi^\mu(z) \, S_a(0) =&  \ee^{i \mu \cdot {\bf H}(z)} \, \ee^{i a \cdot {\bf H}(0)}
\sim
z^{\mu \cdot a} \ee^{i (\mu + a)\cdot {\bf H}(0)} \bigg( 1 + z i \mu \cdot \partial {\bf H}(0) + \dots \bigg)\ .
\end{align}
Since $\mu = (0,\dots, 0,\pm 1, 0, \dots, 0)$ and $a = (\pm \tfrac{1}{2}, \pm \tfrac{1}{2}, \dots, \pm \tfrac{1}{2})$, the exponent $\mu \cdot a$ of $z$ is either $-\tfrac{1}{2}$ or $+\tfrac{1}{2}$. Therefore, one can split \eqref{eq;bosonex} into (up to the subleading order)
\begin{align}
\psi^\mu(z) \, S_a(0) \sim
\begin{cases}
\frac{1}{z^{1/2}}
\ee^{i (\mu + a ) \cdot \mathbf{H}(0)}
+ z^{1/2} i \mu \cdot \partial \mathbf{H} \ee^{i (\mu +a) \cdot \mathbf{H} (0)}  &\textrm{  if  } \mu \cdot a = - \frac12\\
z^{1/2}
\ee^{i (\mu + a ) \cdot \mathbf{H}(0)} &\textrm{  if  } \mu \cdot a = +\frac12\ .
\end{cases}
\end{align}
The subleading term $i \mu \cdot \partial \mathbf{H} \ee^{i (\mu + a) \cdot \mathbf{H}(0)}$ can be further decomposed into a primary and a descendant part with respect to the energy-momentum tensor $T(z) = -\tfrac{1}{2} \partial \mathbf{H} \cdot \partial \mathbf{H}$ of the bosonized system,
\begin{align}
i \mu \cdot \partial \mathbf{H} \ee^{i (\mu + a) \cdot \mathbf{H}(0)}
=
\frac{4}{D} i (\mu + a) \cdot \partial \mathbf{H} \ee^{i (\mu +a) \cdot \mathbf{H}(0)} + i \left( \frac{D-4}{D} \mu - \frac{4}{D} a \right) \cdot \partial \mathbf{H} \ee^{i (\mu + a) \cdot \mathbf{H}(0)} \ .
\end{align}
Thus, we have primary fields $S^{\mu}_{a}(z)$  defined by
\begin{align}
S^{\mu}_{a}(z)
&=
\de \! \left(  \mu {\cdot a} {+} \frac12 \right) \! \left( \frac{D{-}4}{D} \mu - \frac{4}{D} a \right)\! {\cdot}i \partial \mathbf{H} \ee^{i (\mu + a) \cdot \mathbf{H}(z)}
+\de \! \left(  \mu {\cdot} a {-} \frac12 \right) \!\ee^{i(\mu + a) \cdot \mathbf{H}(z)} \label{line12}
\end{align}
at the subleading order in the OPE (\ref{eq;bosonex}). Although the first term of (\ref{line12}) could in principle be used in section \ref{sec;loop} to evaluate components of the correlators involving $S^\mu_a$, we found the second term $\ee^{i(\mu + a) \cdot \mathbf{H}(z)}$ more convenient to extract the small number of required examples.

Moreover, if $\mu \cdot a = -\frac{1}{2}$, the resulting lattice vector $\mu + a=b$ refers to a spin field $S^b$ of opposite chirality. Therefore, the OPE \eqref{eq;bosonex} can be written as
\begin{align}
\psi^{\mu}(z) S_{a}(0) &\sim \sum_{b\in (\pm\tfrac12,\dots,\pm\tfrac12)} \frac{\delta (\mu +a -b)}{z^{1/2}} \left\{\ee^{i b \cdot \mathbf{H}(0)} + z \frac{4}{D} \partial \ee^{i b \cdot \mathbf{H}(0)}  \right\} + z^{1/2} S^{\mu}_a(0) \notag\\
&\equiv  \frac{\gamma^{\mu}_{ab}}{\sqrt{2} z^{1/2}} \left\{S^{b}(0) + z \frac{4}{D} \partial S^{b}(0)  \right\} + z^{1/2} S^{\mu}_a(0)\ .
\end{align}
In passing to the last line, we have used the definition \eqref{1,39} of gamma-matrices in the Cartan--Weyl basis, where the sign of $b$ is flipped by the contraction through the charge-conjugation matrix in $\gamma^{\mu}_{ab} S^b$. The computation above exemplifies how Lorentz covariance can be a posteriori restored in results obtained from bosonization. In \cite{Atick:1986rs, Schlotterer2010, Hartl2011}, this procedure is applied to construct higher-point correlation functions involving $\psi^\mu$ and $S_a$.


\section{Examples for the standard form of spin sums}\label{sec;more}
\newcommand{\cyc}{\textrm{cyclic}}

This appendix complements the discussion in section \ref{stform} by identifying
the standard form (\ref{standform}) of spin sums
in correlators with three insertions of $\psi^{\mu_{j}} \psi^{ \nu_{j}}(z_j)$. The evaluation of
the spin sums is addressed in section \ref{sec;5}.

\subsection{Unexcited spin fields}
\label{sec;moreB}

With two unexcited spin fields, the five-point correlator
\begin{align}
&8\,\sum_{\nu=1}^{4} (-1)^{\nu+1} \scor{\psi^{\mu_{1}} \psi^{ \nu_{1}}(z_1) \psi^{\mu_{2}}\psi^{ \nu_{2}}(z_2) \psi^{\mu_{3} } \psi^{\nu_{3}}(z_3) S_{a} (z_A) S^{b}(z_B)}_\nu
\notag\\
&=
(\eta^{ \nu_{1} [ \mu_{2} } \eta^{ \nu_{2}] [\mu_{3} } \eta^{  \nu_{3}]   \mu_{1} }
-\eta^{ \mu_{1} [ \mu_{2} } \eta^{ \nu_{2}] [\mu_{3} } \eta^{  \nu_{3}]   \nu_{1} }) \delta_{a}{}^{b} \
\xi^{(1)}(z_1,z_2,z_3,z_A,z_B)
\notag\\
& \ \ + \Big[   (\eta^{ \mu_{2}   [\nu_{1} } (\ga^{ \mu_{1}]   \nu_{2} \mu_{3} \nu_{3} } )_{a}{}^{b}
-\eta^{ \nu_{2}   [\nu_{1} } (\ga^{ \mu_{1}]   \mu_{2} \mu_{3} \nu_{3} } )_{a}{}^{b} ) \
\xi^{(2)}(z_1,z_2,z_3,z_A,z_B)
\label{5ptco}\\
& \ \  \ \  \ \ +
(\eta^{ \mu_{3}  [\nu_{2} } \eta^{\mu_{2}] [\nu_{1}   }  (\ga^{ \mu_{1}]   \nu_{3} } )_{a}{}^{b}
-\eta^{ \nu_{3}  [\nu_{2} } \eta^{\mu_{2}] [\nu_{1}   }  (\ga^{ \mu_{1}]   \mu_{3} } )_{a}{}^{b} ) \
\xi^{(3)}(z_1,z_2,z_3,z_A,z_B)
\notag\\
& \ \  \ \  \ \  +  (\eta^{\nu_1   [\mu_2} \eta^{\nu_2]  \mu_1} - \eta^{\mu_1   [\mu_2} \eta^{\nu_2]  \nu_1} )(\gamma^{\mu_3 \nu_3})_a{}^{b} \
\xi^{(4)}(z_1,z_2,z_3,z_A,z_B) + {\rm cyc}(1,2,3)
\Big] \notag \\
& \ \ +  (\ga^{ \mu_{1} \nu_{1} \mu_{2} \nu_{2} \mu_{3} \nu_{3} })_{a}{}^{b} \xi^{(5)}(z_1,z_2,z_3,z_A,z_B)
\notag
\end{align}
involves spin sums
\begin{subequations}\label{eq;morenonex}
	\begin{align}
	\xi^{(1)}(z_i,z_j,z_k,z_A,z_B) &=
	-\frac{1}{ E_{AB} }
	\ssuma{z_{ij},z_{jk},z_{ki}}{-}
	-\frac{E_{iA} E_{kB}}{ E_{AB} E_{iB} E_{kA}}
	\ssuma{z_{ij},z_{jk}}{z_{ki}}
	\notag\\
	&\! \! \! \! \! \! \! \! \! \! \! \! \! \! \! \! \! \! \! \!\phantom{=\ }
	-\frac{E_{jB} E_{kA}}{ E_{AB}  E_{jA}   E_{kB}}
	\ssuma{z_{ij},z_{ki}}{z_{jk}}
	-\frac{E_{iA} E_{jB}}{ E_{AB}  E_{iB} E_{jA}   }
	\ssuma{z_{ij}}{z_{jk},z_{ki}}
	\notag\\
	&\! \! \! \! \! \! \! \! \! \! \! \! \! \! \! \! \! \! \! \!\phantom{=\ }
	-\frac{E_{iB} E_{jA}}{ E_{AB} E_{iA} E_{jB}}
	\ssuma{z_{jk},z_{ki}}{z_{ij}}
	-\frac{E_{jA} E_{kB}}{ E_{AB}    E_{jB} E_{kA} }
	\ssuma{z_{jk}}{z_{ij},z_{ki}}
	\notag\\
	&\! \! \! \! \! \! \! \! \! \! \! \! \! \! \! \! \! \! \! \!\phantom{=\ }
	-\frac{E_{iB} E_{kA}}{ E_{AB} E_{iA}  E_{kB}}
	\ssuma{z_{ki}}{z_{ij},z_{jk}}
	-\frac{1}{ E_{AB} }
	\ssuma{-}{z_{ij},z_{jk},z_{ki}}
	\end{align}
	\begin{align}
	\xi^{(2)}(z_i,z_j,z_k,z_A,z_B) &= \frac{1}{ E_{AB} }
	\ssuma{z_{ij},z_{jk},z_{ki}}{-}
	-\frac{E_{iA} E_{kB}}{ E_{AB} E_{iB} E_{kA}}
	\ssuma{z_{ij},z_{jk}}{z_{ki}}
	\notag\\
	&\! \! \! \! \! \! \! \! \! \! \! \! \! \! \! \! \! \! \! \!\phantom{=\ }
	-\frac{E_{jB} E_{kA}}{ E_{AB}  E_{jA}   E_{kB}}
	\ssuma{z_{ij},z_{ki}}{z_{jk}}
	+\frac{E_{iA} E_{jB}}{ E_{AB}  E_{iB} E_{jA}   }
	\ssuma{z_{ij}}{z_{jk},z_{ki}}
	\notag\\
	&\! \! \! \! \! \! \! \! \! \! \! \! \! \! \! \! \! \! \! \!\phantom{=\ }
	+\frac{E_{iB} E_{jA}}{ E_{AB} E_{iA} E_{jB}}
	\ssuma{z_{jk},z_{ki}}{z_{ij}}
	-\frac{E_{jA} E_{kB}}{ E_{AB}    E_{jB} E_{kA} }
	\ssuma{z_{jk}}{z_{ij},z_{ki}}
	\notag\\
	&\! \! \! \! \! \! \! \! \! \! \! \! \! \! \! \! \! \! \! \!\phantom{=\ }
	-\frac{E_{iB} E_{kA}}{ E_{AB} E_{iA}  E_{kB}}
	\ssuma{z_{ki}}{z_{ij},z_{jk}}
	+\frac{1}{ E_{AB} }
	\ssuma{-}{z_{ij},z_{jk},z_{ki}}
	\end{align}
	\begin{align}
	\xi^{(3)}(z_i,z_j,z_k,z_A,z_B) &=
	- \frac{1}{ E_{AB} }
	\ssuma{z_{ij},z_{jk},z_{ki}}{-}
	+\frac{E_{iA} E_{kB}}{ E_{AB} E_{iB} E_{kA}}
	\ssuma{z_{ij},z_{jk}}{z_{ki}}
	\notag\\
	&\! \! \! \! \! \! \! \! \! \! \! \! \! \! \! \! \! \! \! \!\phantom{=\ }
	-\frac{E_{jB} E_{kA}}{ E_{AB}  E_{jA}   E_{kB}}
	\ssuma{z_{ij},z_{ki}}{z_{jk}}
	+\frac{E_{iA} E_{jB}}{ E_{AB}  E_{iB} E_{jA}   }
	\ssuma{z_{ij}}{z_{jk},z_{ki}}
	\notag\\
	&\! \! \! \! \! \! \! \! \! \! \! \! \! \! \! \! \! \! \! \!\phantom{=\ }
	-\frac{E_{iB} E_{jA}}{ E_{AB} E_{iA} E_{jB}}
	\ssuma{z_{jk},z_{ki}}{z_{ij}}
	+\frac{E_{jA} E_{kB}}{ E_{AB}    E_{jB} E_{kA} }
	\ssuma{z_{jk}}{z_{ij},z_{ki}}
	\notag\\
	&\! \! \! \! \! \! \! \! \! \! \! \! \! \! \! \! \! \! \! \!\phantom{=\ }
	-\frac{E_{iB} E_{kA}}{ E_{AB} E_{iA}  E_{kB}}
	\ssuma{z_{ki}}{z_{ij},z_{jk}}
	+\frac{1}{ E_{AB} }
	\ssuma{-}{z_{ij},z_{jk},z_{ki}}
	\end{align}
	\begin{align}
	\xi^{(4)}(z_i,z_j,z_k,z_A,z_B) &=
	\frac{E_{kB} }{ E_{kA} }
	\ssuma{z_{Bk},z_{ij}, z_{ji}}{z_{kB}}
	+
	\frac{E_{kB} E_{iA} E_{jB}}{  E_{iB} E_{jA}  E_{kA} }
	\ssuma{z_{Bk},z_{ij}}{z_{kB}, z_{ji}}
	\notag\\
	&\! \! \! \! \! \! \! \! \! \! \! \! \! \! \! \! \! \! \! \!\phantom{= \ }+
	\frac{E_{kB}  E_{iB} E_{jA} }{ E_{iA} E_{jB} E_{kA} }
	\ssuma{z_{Bk},z_{ji}}{z_{kB},z_{ij}}
	+
	\frac{E_{kB} }{  E_{kA} }
	\ssuma{z_{Bk}}{z_{kB},z_{ij}, z_{ji}}
	\end{align}
	\begin{align}
	\xi^{(5)}(z_i,z_j,z_k,z_A,z_B) &= \frac{1}{ E_{AB} }
	\ssuma{z_{ij},z_{jk},z_{ki}}{-}
	-\frac{E_{iA} E_{kB}}{ E_{AB} E_{iB} E_{kA}}
	\ssuma{z_{ij},z_{jk}}{z_{ki}}
	\notag\\
	&\! \! \! \! \! \! \! \! \! \! \! \! \! \! \! \! \! \! \! \! \phantom{=\ }
	-\frac{E_{jB} E_{kA}}{ E_{AB}  E_{jA}   E_{kB}}
	\ssuma{z_{ij},z_{ki}}{z_{jk}}
	+\frac{E_{iA} E_{jB}}{ E_{AB}  E_{iB} E_{jA}   }
	\ssuma{z_{ij}}{z_{jk},z_{ki}}
	\notag\\
	&\! \! \! \! \! \! \! \! \! \! \! \! \! \! \! \! \! \! \! \!\phantom{=\ }
	-\frac{E_{iB} E_{jA}}{ E_{AB} E_{iA} E_{jB}}
	\ssuma{z_{jk},z_{ki}}{z_{ij}}
	+\frac{E_{jA} E_{kB}}{ E_{AB}    E_{jB} E_{kA} }
	\ssuma{z_{jk}}{z_{ij},z_{ki}}
	\notag\\
	&\! \! \! \! \! \! \! \! \! \! \! \! \! \! \! \! \! \! \! \!\phantom{=\ }
	+\frac{E_{iB} E_{kA}}{ E_{AB} E_{iA}  E_{kB}}
	\ssuma{z_{ki}}{z_{ij},z_{jk}}
	-\frac{1}{ E_{AB} }
	\ssuma{-}{z_{ij},z_{jk},z_{ki}} \ .
	\end{align}
\end{subequations}
The notation $+{\rm cyc}(1,2,3)$ in (\ref{5ptco}) refers to cyclic permutations of both the Lorentz indices and the punctures including for instance $\{(z_1,\mu_1,\nu_1), (z_2,\mu_2,\nu_2) ,  (z_3,\mu_3,\nu_3) \}$ $\rightarrow \{ (z_2,\mu_2,\nu_2),(z_3,\mu_3,\nu_3),(z_1,\mu_1,\nu_1)\}$.

\subsection{Excited spin field}
\label{sec;moreD}

In case of an excited spin field, the five-point correlator
\begin{align}
&4 \sqrt{2} \sum_{\nu=1}^{4}(-1)^{\nu+1} \scor{\psi^{\mu_{1}} \psi^{ \nu_{1}}(z_1) \psi^{\mu_{2}} \psi^{ \nu_{2}}(z_2) \psi^{\mu_{3}} \psi^{ \nu_{3}}(z_3) S_{a} (z_A) S_{b}^\lambda(z_B)}_\nu
\notag\\
&= \Big[ (\gamma^{\mu_1\nu_1\mu_2\nu_2 [ \mu_3})_{a b} \eta^{\nu_3] \lambda} \
\Xi^{(1)}(z_1,z_2,z_3,z_A,z_B)
\notag\\
& \ \ \ \
+ ( \eta^{ \mu_2  [\nu_1} (\gamma^{\mu_1]  \nu_2 [ \mu_3})_{a b}
- \eta^{ \nu_2  [\nu_1} (\gamma^{\mu_1] \mu_2 [ \mu_3})_{a b} )\eta^{\nu_3] \lambda} \
\Xi^{(2)}(z_1,z_2,z_3,z_A,z_B)
\notag\\
& \ \ \ \
+  ( \eta^{ \mu_2  [\nu_1} \eta^{\mu_1]   \nu_2}-  \eta^{ \nu_2 \vert [\nu_1} \eta^{\mu_1]  \mu_2}) (\gamma^{ [ \mu_3})_{a b} \eta^{\nu_3] \lambda} \
\Xi^{(3)}(z_1,z_2,z_3,z_A,z_B) + {\rm cyc}(1,2,3) \Big]
\notag\\
& \ \ + \Big[
(\gamma^{\mu_1 \nu_1 [\mu_2})_{a b} \eta^{\nu_2] [\mu_3} \eta^{\nu_3] \lambda} \
\Xi^{(4)}(z_1,z_2,z_3,z_A,z_B)
\notag\\
& \ \ \ \ +
(\gamma^{[\mu_1})_{a b} \eta^{ \nu_1] [\mu_2} \eta^{\nu_2] [\mu_3} \eta^{\nu_3] \lambda} \
\Xi^{(5)}(z_1,z_2,z_3,z_A,z_B) + {\rm perm}(1,2,3)
\Big]
\label{5ptcoB}
\end{align}
involves spin sums
\begin{subequations}\label{eq;moreexex}
	\begin{align}
	\Xi^{(1)}(z_i,z_j,z_k,z_A,z_B) &=
	-
	\tfrac{1}{ E_{AB} }
	\ssuma{z_{Bi},z_{kB},z_{ij},z_{jk}}{-}
	+
	\tfrac{E_{jB} E_{kA}}{ E_{AB}    E_{jA}   E_{kB}}
	\ssuma{z_{Bi},z_{kB},z_{ij}}{z_{jk}}
	\notag\\
	&\! \! \! \! \! \! \! \! \! \! \! \! \! \! \! \! \! \! \! \! \phantom{=\ }
	+
	\tfrac{E_{iB} E_{jA}}{ E_{AB}  E_{iA}  E_{jB} }
	\ssuma{z_{Bi},z_{kB},z_{jk}}{z_{ij}}
	-
	\tfrac{E_{iB} E_{kA}}{ E_{AB}  E_{iA}   E_{kB}}
	\ssuma{z_{Bi},z_{kB}}{z_{ij},z_{jk}}\\
	\Xi^{(2)}(z_i,z_j,z_k,z_A,z_B) &=
	-
	\tfrac{1}{ E_{AB} }
	\ssuma{z_{Bi},z_{kB},z_{ij},z_{jk}}{-}
	+
	\tfrac{E_{jB} E_{kA}}{ E_{AB}    E_{jA}   E_{kB}}
	\ssuma{z_{Bi},z_{kB},z_{ij}}{z_{jk}}
	\notag\\
	&\! \! \! \! \! \! \! \! \! \! \! \! \! \! \! \! \! \! \! \!\phantom{=\ }
	-
	\tfrac{E_{iB} E_{jA}}{ E_{AB}  E_{iA}  E_{jB} }
	\ssuma{z_{Bi},z_{kB},z_{jk}}{z_{ij}}
	+
	\tfrac{E_{iB} E_{kA}}{ E_{AB}  E_{iA}   E_{kB}}
	\ssuma{z_{Bi},z_{kB}}{z_{ij},z_{jk}}
	\end{align}
	\begin{align}
	\Xi^{(3)}(z_i,z_j,z_k,z_A,z_B) &=
	\tfrac{1}{E_{AB}}\ssuma{z_{ij},z_{ji},z_{Bk},z_{kB}}{-}
	+
	\tfrac{E_{iA} E_{jB}}{E_{iB} E_{jA} E_{AB}}
	\ssuma{z_{ij},z_{Bk},z_{kB}}{z_{ji}}
	\notag\\
	&\! \! \! \! \! \! \! \! \! \! \! \! \! \! \! \! \! \! \! \!\phantom{=\  }
	+
	\tfrac{E_{iB} E_{jA}}{ E_{iA}  E_{jB} E_{AB} }
	\ssuma{z_{ji},z_{Bk},z_{kB}}{z_{ij}}
	+
	\tfrac{1}{E_{AB} }
	\ssuma{z_{Bk},z_{kB}}{z_{ij},z_{ji}}
	\\
	\Xi^{(4)}(z_i,z_j,z_k,z_A,z_B) &=
	-
	\tfrac{1}{ E_{AB} }
	\ssuma{z_{Bi},z_{kB},z_{ij},z_{jk}}{-}
	-
	\tfrac{E_{jB} E_{kA}}{ E_{AB}    E_{jA}   E_{kB}}
	\ssuma{z_{Bi},z_{kB},z_{ij}}{z_{jk}}
	\notag\\
	&\! \! \! \! \! \! \! \! \! \! \! \! \! \! \! \! \! \! \! \!\phantom{=\ }
	+
	\tfrac{E_{iB} E_{jA}}{ E_{AB}  E_{iA}  E_{jB} }
	\ssuma{z_{Bi},z_{kB},z_{jk}}{z_{ij}}
	+
	\tfrac{E_{iB} E_{kA}}{ E_{AB}  E_{iA}   E_{kB}}
	\ssuma{z_{Bi},z_{kB}}{z_{ij},z_{jk}}\\
	\Xi^{(5)}(z_i,z_j,z_k,z_A,z_B) &=
	-
	\tfrac{1}{ E_{AB} }
	\ssuma{z_{Bi},z_{kB},z_{ij},z_{jk}}{-}
	-
	\tfrac{E_{jB} E_{kA}}{ E_{AB}    E_{jA}   E_{kB}}
	\ssuma{z_{Bi},z_{kB},z_{ij}}{z_{jk}}
	\notag\\
	&\! \! \! \! \! \! \! \! \! \! \! \! \! \! \! \! \! \! \! \! \phantom{=\ }
	-
	\tfrac{E_{iB} E_{jA}}{ E_{AB}  E_{iA}  E_{jB} }
	\ssuma{z_{Bi},z_{kB},z_{jk}}{z_{ij}}
	-
	\tfrac{E_{iB} E_{kA}}{ E_{AB}  E_{iA}   E_{kB}}
	\ssuma{z_{Bi},z_{kB}}{z_{ij},z_{jk}}\ .
	\end{align}
\end{subequations}


\section{Simplifying the spin-structure dependence}
\label{proof1}

\subsection{The spin-structure dependent Fay identity}

This subsection is dedicated to a proof of the spin-structure dependent Fay identity (\ref{217a}) which generalizes the property \cite{Mumford1983a, Brown2011}
\begin{align}\label{eq;fay}
&F(x_1,y_1) F(x_2,y_2)= F(x_1,y_1{+}y_2) F(x_2{-}x_1,y_2) + F(x_2,y_1{+}y_2) F(x_1{-}x_2,y_1)
\end{align}
of the Kronecker--Eisenstein series (\ref{thetaO}). While (\ref{eq;fay}) immediately yields the $\nu=1$ version of (\ref{217a}) by $F(x,y)=F_{\nu=1}(x,y)$, its extension to even spin structures requires (quasi-)periodicity properties of the Jacobi theta functions $\tht{\nu}(x),\ \nu=1,2,3,4$ defined by (\ref{thetaZ}) and (\ref{thetaY}). The latter are quasi-periodic
\begin{subequations}
	\begin{align}
	\tht{1}(z) &= -\tht{1}(z+1) = -q^{1/2} \ee^{2i\pi z}\tht{1}(z+\tau)
	\\
	\tht{2}(z) &= -\tht{2}(z+1) = q^{1/2} \ee^{2i\pi z}\tht{2}(z+\tau)
	\\
	\tht{3}(z) &= \tht{3}(z+1) = q^{1/2} \ee^{2i\pi z}\tht{3}(z+\tau)
	\\
	\tht{4}(z) &= \tht{4}(z+1) = -q^{1/2} \ee^{2i\pi z}\tht{4}(z+\tau)
	\end{align}
\end{subequations}
with $q=\ee^{2\pi i \tau}$ and related to each other by the half-periodicity:
\begin{align}
\tht{1}(z) = - \tht{2}(z+\tfrac{1}{2}) = -i q^{1/8} \ee^{i\pi z}\tht{4}(z+\tfrac{\tau}{2}) =-i q^{1/8} \ee^{i\pi z} \tht{3}(z+\tfrac{1}{2} + \tfrac{\tau}{2})\ .
\end{align}
Then, the Kronecker--Eisenstein series $F_{\nu}(x,y)$ with spin structures $\nu = 1,2,3,4$ as defined in (\ref{Fnunu}) is related to $F(x,y)$ via
\begin{align}
F_{\nu}(x,y) = \ee^{i \pi \phi_{\nu}x} F(x,y+s_{\nu}) \ ,
\end{align}
where the shift in the second argument and the complex phase is determined by
\begin{align}
s_{\nu} = (0, \tfrac{1}{2}, \tfrac{1}{2} + \tfrac{\tau}{2}, \tfrac{\tau}{2})	\ , \ \ \ \ \ \
\phi_{\nu}=
\begin{cases}
\ 0,\ \ &\textrm{for } \nu = 1,2
\\
\ 1,\ \ &\textrm{for } \nu = 3,4\ .
\end{cases}
\end{align}
Finally, the proof of (\ref{217a}) follows from the Fay trisecant identity \eqref{eq;fay} and the quasi-periodicity of $F(x,y)$
\begin{align}
F_{\nu}(x_1,y_1) F_{\nu}(x_2,y_2) 	=& \ee^{i \pi \phi_{\nu}(x_1 + x_2)} F(x_1,y_1+s_{\nu}) F(x_2,y_2+s_{\nu})
\\
=& F(x_1, y_1 + y_2) \ee^{i \pi \phi_{\nu}(x_2 - x_1)} F(x_2-x_1,y_2 + s_{\nu})\notag\\
& + F(x_2,y_1 + y_2) \ee^{i \pi \phi_{\nu} (x_1 - x_2)} F(x_1-x_2,y_1 + s_{\nu})
\notag\\
=& F(x_1, y_1 + y_2) F_{\nu}(x_2-x_1,y_2) + F(x_2, y_1 + y_2) F_{\nu}(x_1-x_2,y_1)\ . \notag
\end{align}

\subsection{Merging products of $F_\nu$ at the same second argument $y$}

In order to prove (\ref{217c}), we note that the derivatives $F^{(0,k_1)}_{\nu}(x_1,y) F^{(0,k_2)}_{\nu}(x_2,y)$ in the notation of (\ref{217b}) can be rewritten as
\begin{align}
&F^{(0,k_1)}_{\nu}(x_1,y) F^{(0,k_2)}_{\nu}(x_2,y) = \frac{\partial^{k_1}}{\partial y^{k_1}} \frac{\partial^{k_2}}{\partial y'{}^{k_2}} \left[ F_{\nu}(x_1,y) F_{\nu}(x_2,y') \right]_{y'\rightarrow y}  \ .\label{eq;c126}
\end{align}
Then, (\ref{217c}) follows by inserting the corollary
\begin{align}
&F_{\nu}(x_1,y) F_{\nu}(x_2,y') = - F_{\nu}(x_1,y) F_{\nu}(-x_2,-y')\notag\\
&= F(x_1,y-y') F_{\nu}(x_1+x_2,y') - F(-x_2,y-y') F_{\nu}(x_1+x_2,y)\label{eq;c127}\ .
\end{align}
of the Fay identity (\ref{217a}) into \eqref{eq;c126}.

In order to prove (\ref{217d}), we first note that derivatives $F_{\nu}^{(0,k)}(x,y)$ with $k\geq 1$ are non-singular at $x=0$ since the residue of the simple pole of $F_{\nu}(x,y)$ at $x=0$ does not depend on $y$. A similar argument implies that $\lim_{x\rightarrow 0}x F_{\nu}(x,y) = 1$. Then, (\ref{217d}) can be obtained by taking $x_1 \rightarrow -x_2$ in (\ref{217c}).

\subsection{Merging products of $F_\nu$ at second argument $y$ and $-y$}
\label{proof1A}

In order to prove (\ref{217f}), we separate \eqref{eq;c124} into the following two contributions according to the number of derivatives w.r.t.\ the second argument of $F_\nu$:
\begin{align}
&\sum_{\nu=1}^{4} Z_{\nu}(y) F^{(0,k_1)}_{\nu}(x_1,y) F^{(0,k_2)}_{\nu}(x_2,-y)
\notag\\
&=(-1)^{k_2}\sum_{\nu=1}^{4} Z_{\nu}(y) \bigg[ F_{\nu}(x_1+x_2,-y) F^{(0,k_1+k_2)}(x_1,2 y)\notag\\
&\phantom{\sum_{\nu=1}^{4} Z_{\nu}(y) (-1)^{k_2}\bigg[}\ \ \ - F_{\nu}(x_1+x_2,y) F^{(0,k_1+k_2)}(-x_2,2 y) \bigg] \label{eq;c128a} \\
&\ \ \ + \sum_{\nu=1}^{4} Z_{\nu}(y) \bigg[\sum_{l=1}^{k_2}
\left(
\begin{array}{c}
k_2\\
l
\end{array}
\right)
(-1)^{k_2-l} F^{(0,l)}_{\nu}(x_1+x_2,-y) F^{(0,k_1+k_2-l)}(x_1,2 y)
\notag\\
&\phantom{\sum_{\nu=1}^{4} Z_{\nu}(y) (-1)}\ - \sum_{l=1}^{k_1}
\left(
\begin{array}{c}
k_1\\
l
\end{array}
\right)
(-1)^{k_2} F^{(0,l)}_{\nu}(x_1+x_2,y) F^{(0,k_1+k_2-l)}(-x_2,2 y) \bigg] \notag
\end{align}
Since $F_{\nu}^{(0,l)}(x,y)$ is non-singular at $x=0$ for $l\geq 1$, one can set $x_1 + x_2 \rightarrow 0$ or $x_2\rightarrow -x_1$ in the last two lines of \eqref{eq;c128a}, casting them into the form
\begin{align}
&\sum_{\nu=1}^{4} Z_{\nu}(y) \bigg[\sum_{l=1}^{k_2}
\left(
\begin{array}{c}
k_2\\
l
\end{array}
\right)
(-1)^{k_2-l} F^{(0,l)}_{\nu}(0,-y) F^{(0,k_1+k_2-l)}(x_1,2 y)
\notag\\
&\phantom{\sum_{\nu=1}^{4} Z_{\nu}(y)\bigg[}- \sum_{l=1}^{k_1}
\left(
\begin{array}{c}
k_1\\
l
\end{array}
\right)
(-1)^{k_2} F^{(0,l)}_{\nu}(0 , y) F^{(0,k_1+k_2-l)}(x_1,2 y)\bigg]\ .
\label{theRHS}
\end{align}
The first two lines on the right-hand side of \eqref{eq;c128a} in turn vanish as $x_1 + x_2 \rightarrow 0$, as one can check by combining
\begin{align}
\lim_{x\rightarrow 0} \bigg[ F_{\nu}(x,-y) - F_{\nu}(x,y) \bigg] &= \lim_{x\rightarrow 0}   \frac{1}{E(x)} \bigg[ \frac{\tht{\nu}(x{-}y)}{\tht{\nu}(-y)} -  \frac{\tht{\nu}(x{+}y)}{\tht{\nu}(y)} \bigg] = - \frac{2 \tht{\nu}'(y)}{\tht{\nu}(y)}
\end{align}
with the Riemann identity (\ref{prop;rie}):
\begin{align}
&\sum_{\nu=1}^{4} Z_{\nu}(y)\bigg[F_{\nu}(x_1{+}x_2,-y) F^{(0,k_1+k_2)}(x_1,2 y)- F_{\nu}(x_1{+}x_2,y) F^{(0,k_1+k_2)}(-x_2,2 y) \bigg]_{x_1 + x_2 = 0}\notag\\
&=- 2 F^{(0,k_1+k_2)}(x_1,2y) \sum_{\nu=1}^{4} \tht{\nu}(y)^{3} \theta'_{\nu}(y) = 0 \ .
\end{align}
Hence, the claim (\ref{217f}) follows from identifying (\ref{theRHS}) as the right-hand side of \eqref{eq;c128a} in the limit $x_2\rightarrow -x_1$.

\section{Evaluating the leftover spin sums}
\label{proof2}

The general method of section \ref{sec;51} reduces all the spin sums (\ref{standform}) in two-fermion amplitudes to a family of functions $M_j = M_j(z_A-z_B)$ with $j\in \mathbb N$ defined in (\ref{leftover}). In this appendix, we will present a recursive method to express $M_j$ with arbitrary $j\geq 1$ in terms of the prime form $E_{AB}$ and combinations of Weierstrass functions at argument $2y= z_A-z_B$.

\subsection{Properties of the Weierstrass function}

Given the Weierstrass $\wp$-function in (\ref{defweier}), we define
\beq
e_1 \equiv \wp(\tfrac{1}{2}) \ , \ \ \ \
e_2 \equiv \wp(-\tfrac{1}{2}-\tfrac{\tau}{2}) \ , \ \ \ \
e_3 \equiv \wp(\tfrac{\tau}{2})
\eeq
subject to
\begin{align}
e_1 + e_2 + e_3 = 0 \ .
\label{prop;e1e2e3}
\end{align}
The Weierstrass function obeys the addition theorem
\begin{align}
\wp(z+w) = \frac14 \bigg[ \frac{\wp'(z)- \wp'(w)}{\wp(z) - \wp(w)} \bigg]^{2} - \wp(z) - \wp(w)
\end{align}
which provides two useful corollaries,
\begin{align}
\wp(2 z) = \frac{1}{4}\left( \frac{\wp''(z)}{\wp' (z)} \right)^2 -2 \wp(z) \label{eq;addp}
\end{align}
as well as
\begin{align}
\wp(z+\tfrac{1}{2}) &= e_1 + \frac{(e_1-e_2)(e_1-e_3)}{\wp(z) - e_1}\notag\\
\wp(z-\tfrac{1}{2}-\tfrac{\tau}{2}) &= e_2 + \frac{(e_2-e_1)(e_2-e_3)}{\wp(z) - e_2}  \label{cor;wphalf} \\
\wp(z+\tfrac{\tau}{2}) &= e_3 + \frac{(e_3-e_1)(e_3-e_2)}{\wp(z) - e_3} \ . \notag
\end{align}
For $\nu=2,3,4$, the representation $(\wp(z) - e_\nu)^{1/2} = \frac{\tht{1}'(0) \tht{\nu+1}(z)}{\tht{1}(z) \tht{\nu+1}(0)}$ of the fermion Green function (\ref{thetaM}) implies
\begin{align}
F_{\nu}(x,0) = (\wp(x) - e_{\nu-1})^{1/2} \ , \ \ \ \ \ \ \nu=2,3,4 \ ,
\label{prop;Fnuwp}
\end{align}
which yields the following representation of the first derivative
\begin{align}
\wp'(x) &= -2 \sqrt{(\wp(x) - e_1) (\wp(x) - e_2)(\wp(x) - e_3)}\notag\\
& = -2 F_{2}(x,0) F_{3}(x,0) F_{4}(x,0) \ . \label{cor;Fnuwp}
\end{align}

\subsection{$M_j$ and Weierstrass functions}

In this subsection, we demonstrate that the above properties of the Weierstrass function (\ref{defweier}) can be used cast the spin sums $M_{j\geq 1}$ in (\ref{leftover}) into the form
\begin{align}
M_{j}(y) = \frac{E(2 y)}{\wp'(y)} \left\{ \wp^{(j-1)}(y) - \sum^{3}_{k=1}(\wp(y)-e_k)^2 \frac{\partial^{j-1}}{\partial y^{j-1}} \frac{1}{\wp(y) - e_k} +2 \delta_{j,1} \wp(y) \right\}\ .
\label{prop;d2}
\end{align}
We first note that $F^{(1,0)}(x,0)=  - \wp(x) - {\rm G}_2$
(with ${\rm G}_k$ defined in (\ref{thetaK})) such that
\begin{subequations}\label{eq;Z}
	\begin{align}
	F^{(0,j)}(0,y) &=  F^{(j,0)}(y ,0) = - \frac{\partial^{j-1}}{\partial y^{j-1}} [  \wp(y) + {\rm G}_2 ]\\
	F^{(0,j)}_{\nu \ne 1}(0,y)
	&=
	\begin{cases}
	- \frac{\partial^{j-1}}{\partial y^{j-1}} \big[  \frac{(e_1-e_2)(e_1-e_3)}{\wp(x) - e_1}+e_1 + {\rm G}_2 \big],\ \ \nu=2\\
	- \frac{\partial^{j-1}}{\partial y^{j-1}} \big[  \frac{(e_2-e_1)(e_2-e_3)}{\wp(x) - e_2}+ e_2 + {\rm G}_2 \big],\ \ \nu=3\\
	- \frac{\partial^{j-1}}{\partial y^{j-1}} \big[  \frac{(e_3-e_1)(e_3-e_2)}{\wp(x) - e_3} + e_3 + {\rm G}_2 \big],\ \ \nu=4
	\end{cases}   \ , \label{eq;d155b}
	\end{align}
\end{subequations}
where \eqref{eq;d155b} results from \eqref{cor;wphalf}. Then, by (\ref{prop;Fnuwp}) and (\ref{cor;Fnuwp}),
\begin{subequations}\label{eq;Fwp}
	\begin{align}
	Z_{1}(y) &=\frac{\tht{1}(y)^{4}}{\tht{1}'(0)^{4}} =  \frac{E(2y)}{2 F_{2}(y,0) F_{3}(y,0)F_{4}(y,0)} = -\frac{E(2y)}{\wp'(y)}\\
	Z_{\nu\ne 1}(y) &= \frac{(-1)^{\nu+1} \tht{\nu}(y)^{4}}{\tht{1}'(0)^{4}} = Z_{\nu}(0) Z_{1}(y) F_{\nu}(y,0)^{4}\notag\\
	&=
	\begin{cases}
	\displaystyle \frac{E(2y) (\wp(y)-e_1)^{2}}{\wp'(y) (e_1 - e_2) (e_1 -e_3)},\ \ \ \ \nu=2\\
	\displaystyle \frac{E(2y) (\wp(y)-e_2)^{2}}{\wp'(y) (e_2 - e_1) (e_2 -e_3)},\ \ \ \ \nu=3\\
	\displaystyle \frac{E(2y) (\wp(y)-e_3)^{2}}{\wp'(y) (e_3 - e_1) (e_3 -e_2)},\ \ \ \ \nu=4
	\end{cases} \ .
	\end{align}
\end{subequations}
Hence, combining \eqref{eq;Z} and \eqref{eq;Fwp} gives rise to
\begin{align}
&\sum_{\nu=1}^{4} Z_{\nu}(y) F^{(0,j)}_{\nu}(0,y) = \frac{E(2 y)}{\wp'(y)} \left\{ \wp^{(j-1)}(y) - \sum^{3}_{k=1}(\wp(y)-e_k)^2 \frac{\partial^{j-1}}{\partial y^{j-1}} \frac{1}{\wp(y) - e_k} \right\}\notag\\
&\ \ \  - \delta_{j,1} \frac{E(2 y)}{\wp'(y)} \bigg\{ \frac{ (\wp(y)-e_1)^{2}e_1}{ (e_1 - e_2) (e_1 -e_3)} + \frac{ (\wp(y)-e_2)^{2}e_2}{ (e_2 - e_1) (e_2 -e_3)} + \frac{ (\wp(y)-e_3)^{2}e_3}{ (e_3 - e_1) (e_3 -e_2)}  \bigg\} \ . \notag \\
\end{align}
Finally, the claim (\ref{prop;d2}) follows by simplifying the coefficient of $\delta_{j,1}$ via (\ref{prop;e1e2e3}).

\subsection{Initial conditions}
\label{sec:init}

By (\ref{prop;e1e2e3}), the instances of (\ref{prop;d2}) at $j=1,2$ imply
\begin{subequations} \label{initicond}
	\begin{align}
	M_{1}(y) &= \frac{E(2y)}{\wp'(y)} \bigg[ \wp(y) - \sum_{k=1}^{3} (\wp(y) - e_k) + 2 \wp(y) \bigg] = 0\\
	M_{2}(y) &= \frac{E(2y)}{\wp'(y)} \bigg[ \wp'(y) - \sum_{k=1}^{3} (\wp(y) - e_k)^{2}\frac{\partial}{\partial y}\frac{1}{\wp(y) - e_k} \bigg] \\
	&=  \frac{E(2y)}{\wp'(y)} \bigg[ \wp'(y) - \frac{\partial}{\partial y}\bigg(\sum_{k=1}^{3} (\wp(y) - e_k)\bigg) + 6 \wp'(y) \bigg] = 4 E(2y)\ . \notag
	\end{align}
\end{subequations}

\subsection{Recursive construction of $M_j$}

Given that the initial conditions for $M_j$ at $j=1,2$ have been settled in (\ref{initicond}), we now proceed to demonstrating that (\ref{prop;d2}) with $j>2$ is equivalent to the recursion
\begin{align}
M_{j}(y) = E(2 y) \frac{\partial}{\partial y} \left( \frac{M_{j-1}(y)}{E(2 y)} \right) -  Q_1(y) M_{j-1}(y) + 4 E(2 y) \sum_{l=0}^{j-3} Q_{j-2-l}^{(l)}(y)
\label{prop;measure}
\end{align}
with the following building blocks:
\beq
Q_k(y) \equiv \wp'(y) \frac{\partial^{k}}{\partial y^k} \frac{1}{\wp'(y)}
\ , \ \ \ \ \ \
Q^{(m)}_{k}(y) \equiv \frac{\partial^{m}}{\partial y^{m}} Q_{k}(y) \ .
\label{notationxxx}
\eeq
As a first step towards proving (\ref{prop;measure}), we note that, by (\ref{prop;e1e2e3}) and (\ref{cor;Fnuwp}),
\begin{align}
\sum_{k=1}^{3} \frac{1}{\wp(y) - e_k} = -2 \frac{\partial}{\partial y} \frac{1}{\wp'(y)} \ ,
\end{align}
which implies
\begin{align}
\sum_{k=1}^{3} \wp'(y) \frac{\partial^{l-1}}{\partial y^{l-1}} \frac{1}{\wp(y) - e_{k}} = -2 Q_{l}(y)\ .
\label{fromhere}
\end{align}
It then follows from (\ref{fromhere}) that, for $l>1$,
\begin{align}
&\sum_{k=1}^{3} (\wp(y) - e_k) \frac{\partial^{l}}{\partial y^{l}} \frac{1}{\wp(y) - e_{k}}\notag\\
&= \frac{\partial}{\partial y} \sum_{k=1}^{3} \bigg[ (\wp(y) - e_{k}) \frac{\partial^{l-1}}{\partial y^{l-1}} \frac{1}{\wp(y) - e_{k}} \bigg] - \sum_{k=1}^{3} \wp'(y) \frac{\partial^{l-1}}{\partial y^{l-1}} \frac{1}{\wp(y) - e_{k}} \label{equd22}\\
& = 2 \sum_{m=0}^{l-1} Q^{(m)}_{l-m}(y) \ ,
\notag
\end{align}
see (\ref{notationxxx}) for $Q^{(m)}_{l-m}(y)$. Now, as a consequence of (\ref{equd22}), we have
\begin{align}
&\frac{\wp'(y)}{E(2 y)}M_{j}(y) =    \wp^{(j-1)}(y) - \sum^{3}_{k=1}(\wp(y)-e_k)^2 \frac{\partial^{j-1}}{\partial y^{j-1}} \frac{1}{\wp(y) - e_k}    \notag \\
& =   \wp^{(j-1)}(y) - \frac{\partial}{\partial y}\sum^{3}_{k=1}(\wp(y)-e_k)^2 \frac{\partial^{j-2}}{\partial y^{j-2}} \frac{1}{\wp(y) - e_k}\notag\\
&\ \ \ \ \ + 2 \sum^{3}_{k=1} \wp'(y) (\wp(y)-e_k) \frac{\partial^{j-2}}{\partial y^{j-2}} \frac{1}{\wp(y) - e_k}   \\
& = \frac{\partial}{\partial y}\bigg\{ \wp^{(j-2)}(y) - \sum^{3}_{k=1}(\wp(y)-e_k)^2 \frac{\partial^{j-2}}{\partial y^{j-2}} \frac{1}{\wp(y) - e_k} \bigg\}+ 4 \wp'(y)  \sum_{l=0}^{j-3} Q^{(l)}_{j-2-l}(y) \notag \\
& = \frac{\partial}{\partial y}\bigg\{ \frac{\wp'(y)}{E(2 y)}M_{j-1}(y) \bigg\}+ 4 \wp'(y)  \sum_{l=0}^{j-3} Q^{(l)}_{j-2-l}(y) \notag
\end{align}
for $j>2$. Upon multiplication by $\frac{E(2y)}{\wp'(y)}$ and exploiting that $Q_{1}(y) = - \frac{\wp''(y)}{\wp'(y)}$, this lands us on the claim (\ref{prop;measure}).

\subsection{The recursion for $M_j$ in terms of Weierstrass derivatives}

Finally, we shall express the building blocks $Q_k^{(m)}$ seen in the recursion (\ref{prop;measure}) for $M_j$ in terms of the Weierstrass function and its derivatives at argument $2y = z_{AB}$. As a first step, we eliminate any $Q_k^{(m)}$ with $k\geq 2$ by repeated application of
\begin{align}
Q_k(y) = Q'_{k-1}(y) + Q_1(y) Q_{k-1}(y)\ ,
\label{lem2}
\end{align}
which can be proven as follows (with $Q_k = Q_k(y)$ and $\wp = \wp(y)$):
\begin{align}
Q_k = \wp'\frac{\partial^k}{\partial y^k} \frac{1}{\wp'} = \frac{\partial}{\partial y}\left( \wp' \frac{\partial^{k-1}}{\partial y^{k-1}} \frac{1}{\wp'} \right) - \wp'' \frac{\partial^{k-1}}{\partial y^{k-1}} \frac{1}{\wp'} = Q'_{k-1} + Q_1 Q_{k-1} \ .
\end{align}
Then, (\ref{prop;measure}) reduces to a polynomial in $Q_1(y) = - \frac{\wp''(y)}{\wp'(y)}$ and its derivatives. Such polynomials connect with the desired $\wp^{(k)}(2y)$ by means of
\begin{align}
\wp(2 y) = \frac{1}{12} \left\{ 2 Q'_1(y) + Q_1^2(y) \right\}
\label{lem1}
\end{align}
and its derivatives, e.g.\ $\wp'(2 y) = \frac{1}{12} \left\{ Q''_1(y) + Q_1(y)Q_1'(y) \right\}$. One can derive (\ref{lem1}) from the differential equation
\begin{align} \label{eq;wpdiff}
\wp'' (z) = 6 \wp(z)^{2}  - 30 {\rm G}_4
\end{align}
of the Weierstrass function which implies
\begin{align}
\wp^{(3)}(z) = 12 \wp'(z) \wp(z)
\end{align}
for the third derivative. Then, (\ref{lem1}) follows from (\ref{eq;addp}) along with the above expressions for
$\wp'' (z) $ and $\wp^{(3)}(z) $:
\begin{align}
\wp(2 y) &= -\frac{1}{4}\wp''(y) \frac{\partial}{\partial y} \frac{1}{\wp'(y)} - \frac{2 \wp'(y) \wp(y)}{\wp'(y)} \notag \\
& = -\frac{1}{4}\frac{\wp''(y)}{\wp'(y)} \wp'(y) \frac{\partial}{\partial y} \frac{1}{\wp'(y)} - \frac{ \wp^{(3)}(y)}{6 \wp'(y)} \notag \\
& = \frac{1}{4} \wp'(y) \frac{\partial}{\partial y} \frac{1}{\wp'(y)} Q_1(y) - \frac{1}{6} \left(\frac{\partial}{\partial y} \frac{ \wp''(y)}{ \wp'(y)} - \wp''(y)
\frac{\partial}{\partial y} \frac{1}{\wp'(y)} \right)\label{eq;4c}\\
& = \frac{1}{4}  Q_1(y)^2 - \frac{1}{6} \left( - \frac{\partial}{\partial y} Q_1 + Q_1(y)^2 \right) \notag \\
& = \frac{1}{12} \left\{ 2 Q'_1(y) + Q_1^2(y) \right\}\ . \notag
\end{align}
In summary, the recursion (\ref{prop;measure}) along with (\ref{lem2}) imply that any $M_j$ is given by $E(2y)$ multiplied by polynomials in $Q_1(y) = - \frac{\wp''(y)}{\wp'(y)}$ and its derivatives. We have checked up to order $j=8$ that these polynomials can be expressed in terms of $\wp(2 y) $ as given on the right-hand side of (\ref{lem1}) and its derivatives. It is reasonable to expect this to hold for all $M_j$ since the properties of $Z_\nu(y)$ and $F^{(0,j)}(0,y)$ under shifts of $y$ imply that $\frac{ M_j(y)}{E(2y)}$ is periodic under $y\rightarrow y+\frac{1}{2}$ and $y\rightarrow y+\frac{\tau}{2}$. Hence, $\frac{ M_j(y)}{E(2y)}$ be expressed through doubly-periodic functions at argument $2y$.


\subsection{Additional examples}

The above recursive method with \eqref{lem1} and \eqref{lem2} gives rise to the expressions for $M_{k\leq 6}$ in (\ref{stf1}) and
\begin{align}
M_7 &= 2688  \,E_{AB} (3 \wp(2 y) \wp'(2 y) + \wp^{(3)}(2y))\\
M_8 &= 192  \, E_{AB} (40 \wp^{(4)}(2 y)+204 \wp(2 y) \wp''(2 y)+165 \wp'(2 y)^2+36 \wp(2 y)^3 ) \ . \notag
\end{align}
This can be simplified to
\begin{align}
M_7 &= 40320 E_{AB} \wp(z_{AB}) \partial \wp(z_{AB})\\
M_8 &= 80640 \,E_{AB} \, \big[ 16 \wp(z_{AB})^{3} -141 {\rm G}_{4} \wp(z_{AB}) - 215 {\rm G}_{6} \big]\notag
\end{align}
with $\wp(z_{AB}) = (f^{(1)}(z_{AB}))^2 - 2 f^{(2)}(z_{AB})$, see (\ref{stf2}) for similar representations of~$M_{k\leq 6}$.


\section{Cleaning up prime forms}
\label{proof3}

In this appendix, we prove the central identity (\ref{eq;clean}) of section \ref{step4}.
Let $\tht{\nu}(z) \ne 0$ and consider complex variables $\{x_1,x_2,\ldots,x_n\}$ and $\{y_1,y_2,\ldots,y_n\}$, where $x_j$ and $y_k$ are pairwise different for $j,k=1,2,\ldots,n$. Then, the following Fay trisecant identities hold for $\nu=1,2,3,4$ \cite{Faybook},
\begin{align}\label{eq;faysX}
&\tht{\nu}\left(\sum_{j=1}^n (x_j-y_j)+z\right)\tht{\nu}(z)^{n-1} \frac{\prod_{j<k}^{n} E(x_j,x_k) E(y_k,y_j)}{\prod_{j,k=1}^{n} E(x_j,y_k)}
= \underset{j,k}{\det}\left[ \frac{\tht{\nu}(x_j-y_k+z)}{E(x_j,y_k)}\right] \ ,
\end{align}
where the determinant refers to the $n\times n$ matrix with entries $\frac{\tht{\nu}(x_j-y_k+z)}{E(x_j,y_k)}$. Multiplication with $\big(\frac{z}{\tht{1}(z)}\big)^n$ and setting $\nu=1$ gives rise to
\begin{align}
\frac{z^{n} \tht{1}\bigg(\sum_{p=1}^{n}(x_{p} -y_{p}) + z\bigg)}{\tht{1}(z)} \frac{\prod^n_{j<k} E(x_j,x_k) E(y_k,y_j)}{\prod_{j,k =1}^{n} E(x_j,y_k)} = \underset{j,k}{\det} \left[zF(x_j-y_k,z)\right]\ .
\end{align}
Then, inserting $1= \frac{\prod_{q=1}^{n-1} \tht{1}(\sum_{r=1}^{q} (x_r-y_r))}{\prod_{q=1}^{n-1} \tht{1}(\sum_{r=1}^{q} (x_r-y_r))}$ on the left-hand side and using the definition (\ref{thetaO}) of $F(x,y)$ yields the following lemma: Let $n>1$, then
\begin{align}\label{eq;cross}
&\underset{j,k}{\det} [zF(x_j-y_k,z)] =
z^{n-1} \left[  \prod_{q=2}^{n-1}
F\bigg(\sum_{r=1}^{q-1} (x_{r} - y_{r}), x_{q}-y_{q} \bigg)  \right]
\\
& \ \ \ \ \times F\bigg(\sum_{p=1}^{n-1}(x_p-y_p), x_n-y_n+z \bigg) z F(x_n-y_n,z)
\frac{\prod^n_{j<k} E(x_j,x_k) E(y_k,y_j)}{\prod^n_{j\ne k} E(x_j,y_k)} \ . \notag
\end{align}
Finally, the identity (\ref{eq;clean}) in the main text follows by applying $\frac{\partial^{m+n-1}}{\partial  z^{m+n-1}}$ to (\ref{eq;cross}) and taking $z \rightarrow 0$.


\section{Examples for spin-summed correlators}
\label{sec;moremoremore}

This appendix adds further examples of spin-summed correlators to section \ref{sec;53}.


\subsection{Unexcited spin-fields}
\label{sec;moremoremoreA}

The seven-point generalization of the correlators in section \ref{sec;53a} is given by
\begin{align}
&8 \, \sum_{\nu=1}^{4}(-1)^{\nu+1} \scor{\psi^{\mu_1}\psi^{\nu_1}(z_1) \psi^{\mu_2}\psi^{\nu_2}(z_2) \psi^{\mu_3}\psi^{\nu_3}(z_3) \psi^{\mu_4}\psi^{\nu_4}(z_4) \psi^{\mu_5}\psi^{\nu_5}(z_5) S_a(z_A) S^b(z_B)}_\nu\notag\\
&=(\gamma^{\mu_1\nu_1 \mu_2\nu_2\dots \mu_5\nu_5})_{a}{}^{b}h^{(2)}_{\emptyset}  + \eta^{\nu_1\mu_2}(\gamma^{\mu_1 \nu_2 \mu_3 \nu_3\mu_4 \nu_4 \mu_5\nu_5})_{a}{}^{b}h^{(2)}_{[12]} \notag\\
&+\eta^{\nu_1 \mu_2} \eta^{\mu_1 \nu_2}(\gamma^{\mu_3\nu_3 \mu_4 \nu_4 \mu_5\nu_5})_{a}{}^{b}h^{(2)}_{(12)} + \eta^{\nu_1 \mu_2} \eta^{\nu_2 \mu_3}(\gamma^{\mu_1 \nu_3 \mu_4 \nu_4 \mu_5\nu_5})_{a}{}^{b}h^{(2)}_{12,23} \notag\\
&+\eta^{\nu_1 \mu_2} \eta^{\nu_3 \mu_4}(\gamma^{\mu_1 \nu_2 \mu_3 nu_4 \mu_5\nu_5})_{a}{}^{b}h^{(2)}_{[12],[34]} + \eta^{\nu_1 \mu_2} \eta^{\mu_1 \nu_2} \eta^{\nu_3 \mu_4} (\gamma^{\mu_3 \nu_4 \mu_5\nu_5})_{a}{}^{b}h^{(2)}_{(12),[34]} \notag\\
&+ \eta^{\nu_1 \mu_2} \eta^{\nu_2 \mu_3} \eta^{\mu_1 \nu_3} (\gamma^{ \mu_4 \nu_4 \mu_5\nu_5})_{a}{}^{b}h^{(2)}_{[123]} + \eta^{\nu_1 \mu_2} \eta^{\nu_2 \mu_3} \eta^{\nu_3 \mu_4} (\gamma^{\mu_1 \nu_4 \mu_5\nu_5})_{a}{}^{b}h^{(2)}_{12,23,34}  \notag\\
\end{align}
with the following doubly-periodic functions $h^{(2)}_{\ldots} \equiv h^{(2)}_{\ldots}(z_j,z_A,z_B) $:
\begin{subequations}
	\begin{align}
	h^{(2)}_{\emptyset}(z_j,z_A,z_B)
	&=
	- \sum _{i=1}^5 \sum _{j=i+1}^5 V_2(i,A,j,B)
	- 16 V_2(A,B)\\
	h^{(2)}_{[12]}(z_j,z_A,z_B) &=  \sum _{i=3}^5 V_2(1,2,B,i,A) + 4 V_2(1,2,A,B) - (A \leftrightarrow B)\\
	h^{(2)}_{(12)}(z_j,z_A,z_B) &= \sum_{i=3}^{5}\sum_{j=i+1}^{5} V_2(i,B,j,A)
	- \sum_{i=1}^{2}\sum_{j=i+1}^{5} V_2(i,B,j,A)  - 4 V_2(1,2)\\
	h^{(2)}_{12,23}(z_j,z_A,z_B) &= \sum_{i=1,i\ne 2}^{5} V_2(2,A,i,B)-V_2(1,A,3,B)-V_2(4,A,5,B) \notag\\
	&\phantom{=\ }
	- 2 \left\{ V_2(2,A,1,3) + V_2(2,A,3,1) + (A \leftrightarrow B) \right\}\\
	h^{(2)}_{[12],[34]}(z_j,z_A,z_B) & =V_2(1,2,A,3,4,B) - (1 \leftrightarrow 2) -  (3 \leftrightarrow 4) + (1 \leftrightarrow 2, 3 \leftrightarrow 4) \\
	h^{(2)}_{(12),[34]}(z_j,z_A,z_B) & = \left(V_1(1,A,B)+V_1(2,A,B)-V_1(5,A,B)\right)
	\notag\\
	&\ \ \ \times
	\left( V_1(3,4,A,B) + V_{1}(3,4,B,A) \right)
	\notag\\
	&\ \ \
	-  V_2(3,4,A,B) + V_2(3,4,B,A)\\
	h^{(2)}_{[123]}(z_j,z_A,z_B) &= V_{2}(1,A,2,3,B) +  \sum_{i=4}^{5} V_2(1,2,A,i,B)+ \textrm{cyc}(1,2,3)   \notag\\
	&\ \ \ \,- (A \leftrightarrow B) \\
	h^{(2)}_{12,23,34}(z_j,z_A,z_B) & =
	- V_{2}(2,A,3,4,B) - V_{2}(3,A,1,2,B)
	-\sum_{i=1}^{3} V_2(i,i+1,A,5,B)
	\notag\\
	&\phantom{=\ } - (A \leftrightarrow B)\\
	h^{(2)}_{12,23,45}(z_j,z_A,z_B) & =
	- V_2(2,A,4,5,B)+V_2(2,B,4,5,A)
	\\
	h^{(2)}_{(12),(34)}(z_j,z_A,z_B) &=
	V_2(1,A,2,B)+V_2(3,A,4,B)+4 V_2(1,2)+4 V_2(3,4)
	\notag\\
	& \phantom{=\ }
	+\sum_{i=1}^{4}V_2(i,A,5,B)
	-\sum_{i=1}^{2} \sum_{j=3}^{4} V_2(i,A,j,B)
	\\
	h^{(2)}_{[123],[45]}(z_j,z_A,z_B) & = -2 V_1(1,2,3) \left( V_1(4,5,A,B) +  V_1(4,5,B,A) \right)
	\\
	h^{(2)}_{(1234)}(z_j,z_A,z_B) & =  V_2(1,A,3,B) + V_2(2,A,4,B) -\sum_{i=1}^{4}V_2(i,A,5,B)
	\notag\\
	& \phantom{=\ }
	-4 V_{2}(1,2,3,4) \\
	h^{(2)}_{(12),34,45}(z_j,z_A,z_B) & = V_2(1,B,4,A) + V_2(2,B,4,A) - V_2(1,B,2,A) - 4 V_2(1,2)
	\notag\\
	&\phantom{=\ }
	- V_2(3,B,4,A) - V_2(4,B,5,A) + V_2(3,B,5,A)\notag\\
	&\phantom{=\ }
	- 2 V_2(3,A,5,4) - 2 V_2(3,B,5,4)
	\end{align}
	\newpage
	\begin{align}
	h^{(2)}_{12,23,34,45}(z_j,z_A,z_B) & =  V_2(1,A,2,B)-V_2(2,A,4,B)+V_2(4,A,5,B)
	\notag\\
	& \phantom{=\ }
	-V_2(1,A,5,B)
	+2 V_2(1,2,3,4,5,A)+2 V_2(1,2,3,4,5,B)
	\\
	h^{(2)}_{[123],(45)}(z_j,z_A,z_B) & =
	\left( V_2(1,2,A,B) + {\rm cyc}(1,2,3) \right)  - V_1(1,2,3) \sum_{i=1}^{5}V_1(i,A,B)
	\notag\\
	&\phantom{=\ } - (A \leftrightarrow B) \\
	h^{(2)}_{(12345)}(z_j,z_A,z_B) & =
	\big( V_{2}(1,A,2,3,B) + V_{2}(1,A,3,4,B) + V_{2}(1,A,4,5,B)   \notag\\
	&\phantom{=\ }
	+ 4 V_2(1,2,A,B) + {\rm cyc} (1,2,3,4,5) \big) - (A \leftrightarrow B)\ .
	\end{align}
\end{subequations}

\subsection{One excited spin-fields}
\label{sec;moremoremoreB}
%
The six-point generalization of the correlators in section \ref{sec;53b} reads
\begin{align}
&4\sqrt{2} \sum_{\nu=1}^{4}(-1)^{\nu+1} \scor{\psi^{\mu_1}\psi^{\nu_1}(z_1) \psi^{\mu_2}\psi^{\nu_2}(z_2) \psi^{\mu_3}\psi^{\nu_3}(z_3) \psi^{\mu_4}\psi^{\nu_4}(z_4) S_a(z_A) S^{\lambda}_b(z_B)}_\nu\notag\\
&=  (\gamma^{\mu_1\nu_1 \mu_2\nu_2 \mu_3\nu_3 \mu_4 })_{ab} \eta^{\nu_4 \lambda} H^{(2)}_{\emptyset} + \eta^{\nu_1 \mu_2} (\gamma^{\mu_1\nu_2 \mu_3\nu_3 \mu_4 })_{ab} \eta^{\nu_4 \lambda} H^{(2)}_{[12]} \notag\\
&+ \eta^{\nu_1 \mu_2} \eta^{\mu_1\nu_2} (\gamma^{ \mu_3\nu_3 \mu_4 })_{ab} \eta^{\nu_4 \lambda} H^{(2)}_{(12)} + \eta^{\nu_1 \mu_2} \eta^{\nu_2 \mu_3} (\gamma^{\mu_1\nu_3 \mu_4 })_{ab} \eta^{\nu_4 \lambda} H^{(2)}_{12,23} \notag\\
&+ \eta^{\nu_1 \mu_2} \eta^{\nu_2 \mu_3} \eta^{\mu_1\nu_3} (\gamma^{ \mu_4 })_{ab} \eta^{\nu_4 \lambda} H^{(2)}_{[123]} +  (\gamma^{\mu_1 \mu_2\nu_2 \mu_3\nu_3 })_{ab} \eta^{\nu_1 \mu_4} \eta^{\nu_4 \lambda} H^{(2)}_{14} \notag\\
&+ \eta^{\nu_2 \mu_3} (\gamma^{\mu_1 \mu_2\nu_3 })_{ab} \eta^{\nu_1 \mu_4} \eta^{\nu_4 \lambda} H^{(2)}_{[23],14} + \eta^{\nu_2 \mu_3} \eta^{ \mu_2\nu_3} (\gamma^{\mu_1 })_{ab} \eta^{\nu_1 \mu_4} \eta^{\nu_4 \lambda} H^{(2)}_{(23),14} \notag\\
&+ (\gamma^{\nu_2 \mu_3\nu_3 })_{ab} \eta^{\mu_1 \mu_2}  \eta^{\nu_1 \mu_4}  \eta^{\nu_4 \lambda} H^{(2)}_{12,14} +  \eta^{\nu_2 \mu_3} (\gamma^{\nu_3 })_{ab} \eta^{\mu_1 \mu_2} \eta^{\nu_1 \mu_4} \eta^{\nu_4 \lambda} H^{(2)}_{12,23,14} \notag \\
&+ {\rm permutations} \label{hereY}\end{align}
with doubly-periodic functions $H^{(2)}_{\ldots} \equiv H^{(2)}_{\ldots}(z_j,z_A,z_B) $ given by
\begin{subequations}
	\begin{align}
	H^{(2)}_{\emptyset}(z_j,z_A,z_B) & = \big[ V_{2}(1,A,2,B) + {\rm cyc}(1,2,3) \big] -2 V_2(4,B) + 8 V_2(A,B) \\
	H^{(2)}_{[12]}(z_j,z_A,z_B) & =V_2(1,2,A,3,B) -2 V_2(1,2,A,B)  - (A\leftrightarrow B)\\
	H^{(2)}_{(12)}(z_j,z_A,z_B) & = \big[  V_2(1,B,2,A) + {\rm cyc}(1,2,3) \big] +4 V_2(1,2) + 2 V_2(4,B)\\
	H^{(2)}_{12,23}(z_j,z_A,z_B) & =- V_2(1,B,2,A)- V_2(2,B,3,A) + V_2(1,B,3,A)\notag\\
	&\phantom{=\ }
	- 2 V_2(1,A,3,2) - 2 V_2(1,B,3,2) - 2 V_2(4,B) \\
	H^{(2)}_{[123]}(z_j,z_A,z_B) & =- 2 V_2(2,3,A,B) - 2 V_2(1,2,A,B,3) - V_2(1,A,2,3,B)
	\notag\\
	&\phantom{=\ }
	- (A\leftrightarrow B) \\
	H^{(2)}_{14}(z_j,z_A,z_B) & = \Big[
	-2 V_{2}(1,4,A,B)
	- \sum_{i=2}^{3}  V_{2}(i,A,1,4,B)
	- (A \leftrightarrow B) \Big]
	\notag\\
	&\phantom{=\ \big\{}
	- \sum_{i=2}^{3} V_{2}(i,B,4,A)
	+ V_{2}(2,B,3,A)
	+ 2 V_{2}(4,B)
	\end{align}
	\newpage
	\begin{align}
	H^{(2)}_{[23],14}(z_j,z_A,z_B) & =
	\sum_{i=2}^{3}
	\left(
	V_{2}(1,A,i,B) + V_{2}(4,A,i,B)
	\right)
	-2 V_{2}(1,A,4,B)
	\notag\\
	&\phantom{=\ }
	- 2 V_{2}(2,A,3,B)
	-2 V_{2}(1,4,A,2,3,B) - 2 V_{2}(1,4,B,2,3,B)
	\notag\\
	&\phantom{=\ }
	+ V_{2}(4,A,2,3,B) - V_{2}(4,B,2,3,A)
	\\
	H^{(2)}_{(23),14}(z_j,z_A,z_B) & = \Big[
	-
	\sum_{i=2}^{3} \left(
	V_2(i,A,1,4,B)
	\right)
	-2
	V_{2}(1,4,A,B)
	-
	(A\leftrightarrow B) \Big]
	\notag\\
	&\phantom{=\  }
	- \sum_{i=2}^{3}V_{2}(4,A,i,B)
	+4 V_2(2,3) + V_2(2,B,3,A) -  2 V_2(4,B)\\
	H^{(2)}_{12,14}(z_j,z_A,z_B) & =
	-2
	V_2(2,A,1,4,B)
	+2
	V_2(3,A,1,4,B)
	+2
	V_2(1,B,2,A) \notag\\
	&\phantom{=\ }
	-V_2(3,A,1,2,B)+V_2(3,B,1,2,A)+4 V_2(4,B,2,1)  \\
	H^{(2)}_{12,23,14}(z_j,z_A,z_B) & =- \big(
	2 V_{2}(1, 2, A, B)
	+2 V_{2}(1, 4, B, A)
	+2 V_{2}(2, 3, A, B)
	\notag\\
	& \phantom{=\ -\big(}
	+V_{2}(1, A, 2, 3, B) - (A \leftrightarrow B)
	\big)
	-2 V_{2}(2, B, 4, A)
	\notag\\
	&\phantom{= \ }
	-2 V_{2}(1, 4, A, 3, 2)-2 V_{2}(1, 4, B, 3, 2)
	+2 V_{2}(2, A, 1, 4, B)
	\notag\\
	&\phantom{= \ }
	+2 V_{2}(4, A, 2, 1, B)+2 V_{2}(4, A, 3, 2, B)
	+2 V_{2}(4, B)\ .
	\end{align}
\end{subequations}


\end{document}